\documentclass[fleqn,usenatbib]{mnras}

\usepackage{amsfonts}

\usepackage[T1]{fontenc}
\usepackage{ae,aecompl}

\usepackage{grffile}

\usepackage{graphicx}	
\usepackage{amsmath}	
\usepackage{amssymb}	
\usepackage[position=top]{subfig}
\usepackage{float}

\newcommand\HII{{H\,\textsc{ii}}} 
\usepackage{etoolbox}
\makeatletter 
  \patchcmd{\NAT@citex}
    {\@citea\NAT@hyper@{%
      \NAT@nmfmt{\NAT@nm}%
      \hyper@natlinkbreak{\NAT@aysep\NAT@spacechar}{\@citeb\@extra@b@citeb}%
      \NAT@date}}
    {\@citea\NAT@nmfmt{\NAT@nm}%
    \NAT@aysep\NAT@spacechar\NAT@hyper@{\NAT@date}}{}{}

  \patchcmd{\NAT@citex}
    {\@citea\NAT@hyper@{%
      \NAT@nmfmt{\NAT@nm}%
      \hyper@natlinkbreak{\NAT@spacechar\NAT@@open\if*#1*\else#1\NAT@spacechar\fi}%
        {\@citeb\@extra@b@citeb}%
      \NAT@date}}
    {\@citea\NAT@nmfmt{\NAT@nm}%
    \NAT@spacechar\NAT@@open\if*#1*\else#1\NAT@spacechar\fi\NAT@hyper@{\NAT@date}}
    {}{}
\makeatother

\title[Radiation from early structure formation]{Global radiation signature from early structure formation}

\author[B. Liu et al.]{Boyuan Liu,$^{1}$\thanks{E-mail: boyuan@utexas.edu} Jason Jaacks,$^{1}$ Steven L. Finkelstein$^{1}$ and Volker Bromm$^{1}$\\
$^{1}$Department of Astronomy, University of Texas, Austin, TX 78705, USA}

\date{Accepted XXX. Received YYY; in original form ZZZ}

\pubyear{2019}

\begin{document}
\label{firstpage}
\pagerange{\pageref{firstpage}--\pageref{lastpage}}
\maketitle

\begin{abstract}
We use cosmological hydrodynamic zoom-in simulations to study early structure formation in two dark matter (DM) cosmologies, the standard CDM model, and a thermal warm DM (WDM) model with a particle mass of $m_{\chi}c^{2}=3$~keV. We focus on DM haloes with virial masses $M\sim 10^{10}\ M_{\odot}$. 
We find that the first star formation activity is delayed by $\sim 200$~Myr in the WDM model, with similar delays for metal enrichment and the formation of the second generation of stars. However, the differences between the two models in globally-averaged properties, such as star formation rate density and mean metallicity, decrease towards lower redshifts ($z\lesssim 10$). Metal enrichment in the WDM cosmology is restricted to dense environments, while low-density gas can also be significantly enriched in the CDM case. The free-free contribution from early structure formation at redshifts $z>6$ to the cosmic radio background (CRB) is $3_{-1.5}^{+13}$\% ($8_{-3.5}^{+33}$\%) of the total signal inferred from radio experiments such as ARCADE 2, in the WDM (CDM) model. The direct detection of the $\mathrm{H_{2}}$ emission from early structure formation ($z\gtrsim 7.2$), originating from the low-mass haloes explored here, will be challenging even with the next generation of far-infrared space telescopes, unless the signal is magnified by at least a factor of 10 via gravitational lensing or shocks. However, more massive haloes with $M\gtrsim 10^{12}\ M_{\odot}$ may be observable for $z\gtrsim 10$, even without magnification, provided that our extrapolation from the scale of our simulated haloes is valid. 
\end{abstract}

\begin{keywords}
cosmology: observations -- radio continuum: general -- cosmology: theory -- dark matter
\end{keywords}


\section{Introduction}
\label{s1}
In the standard cold dark matter (CDM) model, based on weakly interacting massive particles \citep[e.g.][]{jedamzik2009}, the first stars are predicted to form at redshifts $z\sim 20-30$ ($100-200$~Myr after the Big Bang) in minihaloes with masses $M\sim 10^{6}\ M_{\odot}$, and the first galaxies at redshifts $z\sim 10-15$ (cosmic times of $300-500$~Myr) in atomic cooling haloes of masses $M\gtrsim 10^{8}\ M_{\odot}$ (for reviews, see \citealt{bromm2009formation,bromm2011first,dayal2018}). The first stars and galaxies provide powerful diagnostics for early structure formation, through their radiation fields and metal enrichment, as well as their impact on the thermal history of the intergalactic medium (IGM). Although the direct observation of the first stars and galaxies is still challenging in the upcoming era of the {\it James Webb Space Telescope (JWST)}, we can infer their properties from their chemical, thermal and radiative footprints, such as the abundance patterns of low-mass, metal-poor stars \citep[e.g.][]{Karlsson2013,ji2015preserving}, and the absorption and emission of 21-cm radiation in the early IGM (reviewed in \citealt{barkana2016rise}). 

Recently, the Experiment to Detect the Global Epoch of Reionization Signature (EDGES) measured an absorption feature at $78$~MHz, which is attributed to the 21-cm absorption signal from primordial neutral hydrogen, illuminated by the Lyman-$\alpha$ (Ly$\alpha$) photons from first star formation \citep{nature}. Whether this is a true detection of the 21-cm absorption signal from the early Universe is still uncertain with concerns regarding the foreground model \citep{hills2018concerns,bowman2018reply}. If this signal is real, with its large absorption depth and flat profile, it cannot be explained in the framework of the standard CDM model \citep{witte2018edges}. Recalling the failure of the CDM model in predicting observed features in small-scale structures, such as the missing satellite, cusp-core, and too-big-to-fail problems (e.g. \citealt{strigari2007redefining,spekkens2005cusp,boylan2011too}), we should seriously explore alternative dark matter (DM) models, including self-interacting DM (e.g. \citealt{carlson1992self,rocha2013cosmological}), fuzzy DM (e.g. \citealt{hu2000fuzzy,woo2009high}), and warm DM (WDM, e.g. \citealt{gelmini2010inert,WDM2012}). Indeed, \citet{nature1} has argued that the stronger EDGES absorption signal implies a cooler IGM at redshift $z\sim 17$ than current theoretical predictions, which could be achieved by non-gravitational scattering between baryons and DM particles, e.g. with millicharged atomic DM \citep{millicharge2012}. An alternative interpretation posits a possible early radio background, in addition to the cosmic microwave background (CMB) \citep{feng2018enhanced}. 

This unique absorption feature can also place constraints on the global properties of the first stars and galaxies, such as the UV luminosity function (UVLF), UV luminosity density, average star formation efficiency, and star formation rate density (SFRD). For instance, \citet{Mirocha2018} found that the EDGES detection implies a constant star formation efficiency for DM haloes with masses $M\lesssim 10^{10}\ M_{\odot}$, resulting in a steepening of the UVLF at high redshifts.  
Based on the required Ly$\alpha$ photon field that couples the spin and gas temperatures, \citet{Madau2018} inferred that the high-redshift UV luminosity density is consistent with an extrapolation of UV measurements at lower redshifts. The timing of the EDGES signal can also place constraints on the mass of WDM particles \citep{Sitwell2014,Safar2018,Schneider2018}. 
These studies represent efforts to bridge observation and theory of early structure formation within idealized semi-analytical models. Although valuable insights can be obtained in this way, the inferred models are largely phenomenological, and need further validation from fundamental {\it ab initio} physics, which can be implemented in state-of-the-art cosmological hydrodynamic simulations. For example, \citet{jaacks2018legacy} simulated early structure formation in CDM cosmology with legacy star formation and feedback prescriptions for Population~III (Pop~III) and Population~II (Pop~II) stars, deriving an SFRD evolution consistent with that in \citet{Mirocha2018}.  

It is also interesting to study how early structure formation contributes to other observables, such as the cosmic radio background (CRB) and $\mathrm{H_{2}}$ emission in mid- and far-infrared (IR) bands, especially for different DM models. For the former, the ARCADE 2 experiment has measured the absolute temperature of the sky at frequencies 3, 8, 10, 30, and 90~GHz, using an open-aperture cryogenic instrument at balloon altitudes. This mission discovered an excess of $54\pm 6$~mK at 3.3~GHz, in addition to the CMB \citep{arcade}, which is confirmed by more recent measurements with the Long Wavelength Array \citep{dowell2018radio}. The observed CRB brightness temperature, above the CMB baseline contribution, can be modeled as $T_{\mathrm{excess}}\ [\mathrm{mK}]=(24100\pm 2100)(\nu_{\mathrm{obs}}/310\ \mathrm{MHz})^{-2.599\pm 0.036}$ \citep{arcade}. The signal is dominated by synchrotron emission, but there could exist a non-negligible free-free component \citep{kogut2011}. 
Besides, the average brightness (zeroth-moment) of the CRB cannot be explained by CMB spectral distortions or known radio sources \citep{seiffert2011, singal2010}, and the unusual smoothness of the CRB further indicates that it is unlikely to come from sources at $z\lesssim 5$ \citep{holder2013unusual}.
Therefore, currently undetectable high-redshift sources may contribute most of the unresolved CRB.

The $\mathrm{H_{2}}$ emission from Pop~III star formation was calculated with 1-D models by \citet{mizusawa2004h2,mizusawa2005}. Their predicted signal is unlikely to be observable even with the next generation of infrared space telescopes, such as the SPace Infrared telescope for Cosmology and Astrophysics (SPICA)\footnote{\url{http://www.ir.isas.jaxa.jp/SPICA/SPICA_HP/}} and the Origins Space Telescope (OST)\footnote{\url{https://asd.gsfc.nasa.gov/firs/}}, as detection would require an extremely high Pop~III star formation rate (SFR) of $\sim 10^{3}-10^{4}\ M_{\odot}\mathrm{\ yr^{-1}}$. However, it is only via 3-D simulations within a realistic cosmological context that one can obtain more robust predictions for the $\mathrm{H_{2}}$ emission from both Pop~III and Pop~II stars during early structure formation. In light of this, we carry out cosmological hydrodynamic zoom-in simulations with the \textsc{gizmo} code to study the radiation signature of high-redshift DM haloes with virial masses $M\sim 10^{10}\ M_{\odot}$. Considering standard CDM and a (thermal) WDM model with a particle mass of $m_{\chi}c^{2}=3$~keV, we here specifically focus on the free-free and $\mathrm{H_{2}}$ emissions. 

The paper is structured as follows. In Section~\ref{s2} we describe the numerical methods and tools used in simulations and post-processing. In Section~\ref{s3}, we assess the difference between the CDM and WDM models, in terms of the physical processes during early structure formation, such as ionization and heating/cooling, star formation, feedback effects, as well as metal enrichment. In Section \ref{s4}, we derive the global radiation signature of the simulated DM haloes, specifically their free-free and $\mathrm{H_{2}}$ emissions. Finally, Section~\ref{s5} contains a summary of our findings and a discussion of the overall implications. 

\section{Numerical methods}
\label{s2}

\begin{table*}
\centering
\caption{Simulation parameters used in this paper. In the first part (a), the co-moving size of a simulation box is shown in the form $l_{x}\times l_{y}\times l_{z}$, where $l_{k}$ is the length in a given dimension ($k=x,\ y,\ z$); $N_{p}$ is the number of gas and dark matter particles, which is not determined in advance for zoom-in simulations (thus labeled as `-'); $m_{\mathrm{DM}}$ and $m_{\mathrm{gas}}$ are the DM and gas (simulation) particle masses; $\epsilon_{\mathrm{grv}}$ is the co-moving gravitational softening length. Part~(b) shows the initial abundances (with respect to the reference nuclei) for select species, at the initial redshift $z_{i}=99$, from \citet{galli2013dawn}.}
\begin{tabular}{ccccccc}
\hline
(a) Run & Box size [$h^{-1} \mathrm{Mpc}$] & $N_{p}$ (DM, gas) & $m_{\mathrm{DM}}\ [M_{\odot}]$ & $m_{\mathrm{gas}}\ [M_{\odot}]$ & $\epsilon_{\mathrm{grv}}\ [h^{-1} \mathrm{kpc}]$ & SF scheme\\
\hline
Fiducial & $4\times 4\times 4$ & $2\times 128^{3}$ & $3.3\times 10^{6}$ & $6.0\times 10^{5}$ & 1.25 & -\\ 
Z\_Nsfdbk & $1.5\times 1.4\times 1.6$ & - & $5.22\times 10^{4}$ & $9.34\times 10^{3}$ & 0.2 & SINK \\
Z\_sfdbk & $1.5\times 1.4\times 1.6$ & - & $5.22\times 10^{4}$ & $9.34\times 10^{3}$ & 0.2 & P3L+P2L\\
\hline
(b) [Species/reference] & $\mathrm{[H^{+}/H]}$ & $\mathrm{[H^{-}/H]}$ & $\mathrm{[H_{2}/H]}$ & $\mathrm{[H_{2}^{+}/H]}$ & $\mathrm{[D/H]}$ & $\mathrm{[HD/H]}$\\
\hline
Abundance & $2.8\times 10^{-4}$ & $1\times 10^{-13}$ & $5\times 10^{-8}$ & $2\times 10^{-15}$ & $4.3\times 10^{-5}$ & $1\times 10^{-11}$\\
\hline
\end{tabular}
\label{t0}
\end{table*}

We use the \textsc{gizmo} code \citep{hopkins2015new} for our simulations, which adopts a Lagrangian meshless finite-mass (MFM) method to solve hydrodynamics equations, addressing many numerical problems encountered in previous methods, e.g. smoothed particle hydrodynamics (SPH) and adaptive mesh refinement (AMR). We start with the version of \citet{jaacks2018baseline}, which includes the primordial chemistry and cooling model from \citet{johnson2006}. This model identifies $\mathrm{H_{2}}$ and $\mathrm{HD}$ as the main molecular coolants in primordial gas in the low temperature regime ($T\lesssim 10^{3}$~K), while another molecular coolant, $\mathrm{LiH}$, has long been suspected to play an important role as well \citep[e.g.][]{bovino2011,galli2013dawn}, due to its high cooling efficiency per molecule. We have verified that the effect of $\mathrm{LiH}$ on the cooling of primordial gas is negligible because the abundance of $\mathrm{LiH}$ remains extremely low throughout, i.e. $[\mathrm{LiH/H}]\sim 10^{-19}-10^{-15}$ \citep{lithium}. Therefore, we do not further enlarge the chemical network and coolant set of \citet{jaacks2018baseline}. 

For the initial conditions and zoom-in procedure, we employ the \textsc{music} code \citep{hahn2011multi} to generate initial conditions for the CDM and WDM simulations, for the latter assuming a particle mass of $m_{\chi}c^{2}=3$~keV (labeled `CDM' and `WDM\_3\_keV', respectively). We use the parameterization of the WDM power spectrum by \citet{bode2001halo} for thermal-relic WDM. We perform post processing with the \textsc{yt} \citep{turk2010yt} and \textsc{caesar}\footnote{\url{http://caesar.readthedocs.io/en/latest/index.html}} software packages to identify DM haloes. In a second step, we trace the simulation particles in the pre-selected refinement zones back to their initial distributions, defining the Lagrangian regions as the smallest Cartesian boxes that enclose all selected particles. We also use \textsc{music} to generate the refined initial conditions for our zoom-in simulations, in which the spatial (mass) resolution in the Lagrangian regions of interest is enhanced by a factor of $2^{2}$ ($8^{2}$). 

The low-resolution fiducial simulation operates in a box of co-moving size $4\ h^{-1}\mathrm{Mpc}$, containing $128^{3}$ particles in both gas and DM, and employing {\it Planck} cosmological parameters \citep{planck}: $\Omega_{m}=0.315$, $\Omega_{b}=0.048$, $\sigma_{8}=0.829$, $n_{s}=0.966$, and $h=0.6774$. 
Both the fiducial and zoom-in simulations start from the same initial redshift $z_{i}=99$, at which the primordial chemical network is initialized according to the values in \citet{galli2013dawn}. In Table~\ref{t0}, we provide a summary of the simulation parameters.
\begin{figure*}
\centering
\includegraphics[width=2.0\columnwidth]{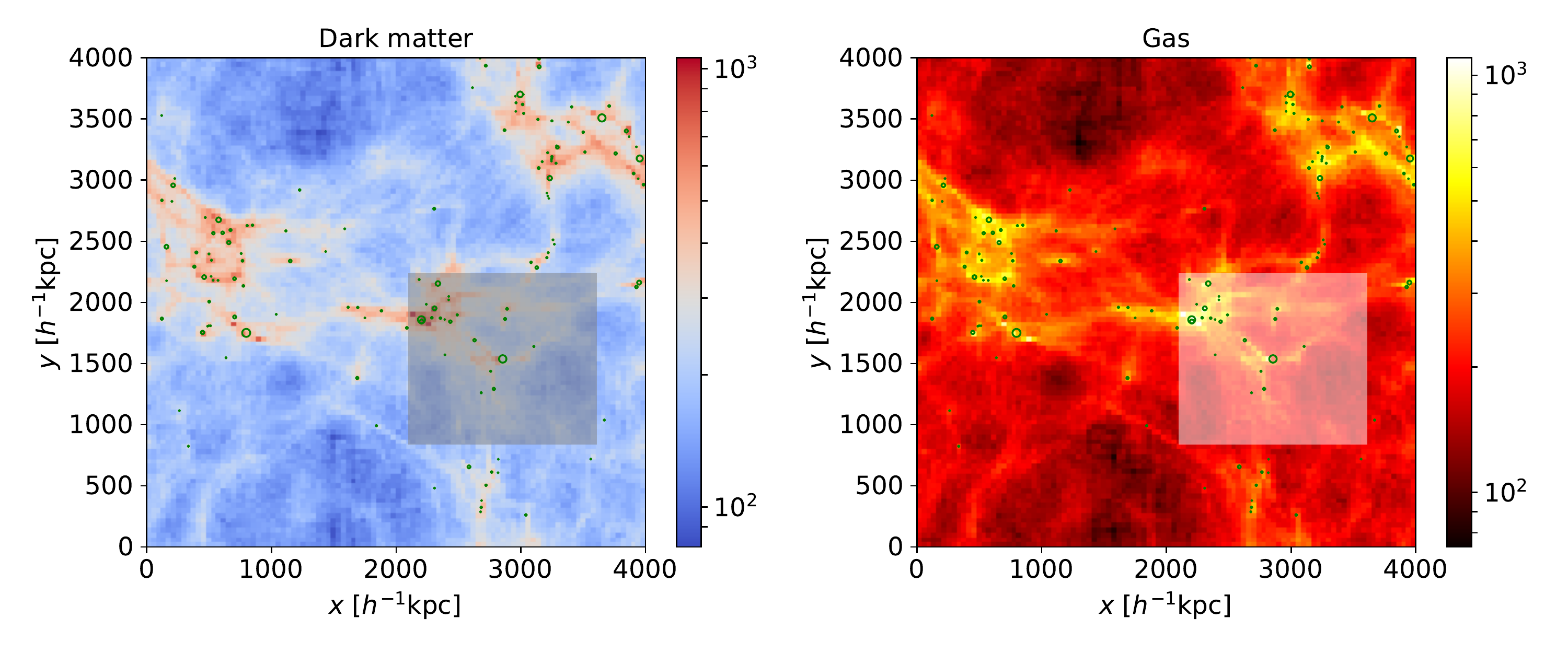}
\vspace{-15pt}
\caption{Cosmic web, in terms of projected distribution of simulation particles, at redshift $z=8.8$, for a WDM cosmology with a particle mass of $m_{\chi}c^{2}=3$~keV. The thickness of the slice is $4\ h^{-1}$~Mpc. Green circles denote DM haloes identified with the friends-of-friends (FOF) methodology, whose sizes scale with the corresponding virial masses. Colour corresponds to the number of simulation particles in each pixel. The zoom-in region is shaded. The structures in the top-left corner of the shaded area do not reside in the 3-D zoom-in region, and appear only in projection.}
\label{f1}
\end{figure*}
We run two sets of zoom-in simulations, Z\_Nsfdbk and Z\_sfdbk, employing different prescriptions for star formation (SF), resulting in correspondingly modified emission physics, as described below.

In Z\_Nsfdbk, the SF process is modeled with sink particles, which only interact gravitationally and do NOT generate any stellar feedback. Therefore, only the collisional ionization and heating of primordial gas by structure formation shocks are captured. The main purpose of using sink particles is to avoid simulating high-density regions, which is computationally expensive. When a gas particle has a local hydrogen number density above $n_{\mathrm{th}}=100\ \mathrm{cm^{-3}}$, all gas particles within the accretion radius $r_{\mathrm{acc}}=0.029\ \mathrm{kpc}$ will form a sink particle, if the condition $\nabla\cdot \mathbf{v}<0$ is also satisfied. Subsequently, gas particles within $r_{\mathrm{acc}}$ around any sink will be accreted onto it. If a gas particle is close to more than one sink, it will be accreted by the one to which it is most tightly bound. At each timestep, only the densest gas particle that meets the above requirements will be allowed to seed a new sink. 
Here the accretion radius is chosen to include the resolved mass, $m_{\mathrm{res}}=N_{\mathrm{ngb}}m_{\mathrm{gas}}\approx 3\times 10^{5}\ M_{\odot}$, at the threshold number density of $n=n_{\mathrm{th}}$, with $N_{\mathrm{ngb}}=32$ being the number of nearest neighbours used for the hydrodynamical update. The choice of $n_{\mathrm{th}}=100\ \mathrm{cm^{-3}}$ is consistent with the star formation criteria in \citet{jaacks2018baseline}, to ensure that the gas has reached a density such that cooling via molecular processes is efficient.

Throughout our simulations, the initial masses of sink particles are mostly above $m_{\mathrm{res}}\approx 3\times 10^{5}\ M_{\odot}$, and accretion events are quite rare. This implies that our sink creation scheme is very aggressive, such that most of the dense gas around newly-engendered sink particles is removed from the gas phase, and subsequent accretion will be unimportant. We can regard each sink particle as representing a single stellar population together with its associated interstellar medium (ISM), but the total mass of star forming gas may thus be overestimated. Note that this is a highly simplified model of star formation, aimed at testing our algorithms at intermediate densities, where computational cost is not yet prohibitive. 

The second scheme, Z\_sfdbk, provides improved physical realism, based on the legacy models for Pop~III and Pop~II star formation and feedback (P3L and P2L), developed previously \citep{jaacks2018legacy,jaacks2018baseline}. Within this model, a gas particle is turned into a stellar particle when a threshold density of $n_{\mathrm{th}}=100\ \mathrm{cm^{-3}}$ is reached, while the temperature remains at $T<10^{3}$~K. The particle will be assigned a Pop~III or Pop~II stellar population according to its metallicity $Z$. Specifically, Pop~II stars are formed when the metallicity exceeds a critical value, $Z>Z_{\mathrm{crit}}=10^{-4}\ Z_{\odot}$ \citep{safranek2010, schneider2011}. 
Each stellar population is modeled with individual star formation efficiencies, $\epsilon_{\mathrm{PopIII}}=0.05$ and $\epsilon_{\mathrm{PopII}}=0.1$, as well as separate choices for the initial mass function (IMF). The corresponding thermal, chemical and radiative feedback is `painted' onto nearby gas particles, and the SF activity is reflected in global radiation fields.

Locally, photo-ionization heating from Pop~III (Pop~II) stars is applied on-the-fly to the gas particles within the ionization front with $R_{\mathrm{ion}}=2\ (0.24)$~kpc around each newly-formed stellar particle for 3 (10)~Myr. Whenever a stellar population comes to the end of its lifetime, instantaneous thermal energy injection is applied to each gas particle within $R_{\mathrm{ion}}$, and the metals produced by supernovae (SNe) are equally distributed to the gas particles within the terminal radius $r_{\mathrm{final}}$ of shell expansion, which depends on the total energy released by SN events $E_{\mathrm{tot}}$. Globally, the Lyman-Werner (LW) background is derived from the combined Pop~III and Pop~II SFRD, and the UV background is modeled separately by the redshift-dependent photo-ionization rate $\zeta (z)$ from \citet{faucher2009new}. Note that the resolution of our zoom-in simulations is the same as in \citet{jaacks2018legacy}, so that our results are directly comparable to that earlier study.

In the following, we mainly present the results from Z\_sfdbk, for one sample zoom-in region, in which a dominant DM halo forms (the target halo, henceforth), reaching a virial mass of $1.13\times 10^{10}\ M_{\odot}$ ($1.06\times 10^{10}\ M_{\odot}$) at redshift $z=8.5$, for WDM (CDM) cosmology\footnote{In our zoom-in simulations, the structure formation histories and radiation properties of other DM haloes in the mass range $10^{9}-10^{10}\ M_{\odot}$ are similar to those of the target halo, and thus, not shown. }. For illustration, Fig.~\ref{f1} shows the initial extent of the zoom-in region with an initial co-moving volume of $V_{Z,c}=1.5\times 1.4\times 1.6\ h^{-3}\mathrm{Mpc^{3}}\approx 11\ \mathrm{Mpc^{3}}$, with respect to the cosmic web of the parent WDM simulation. Below, we will also discuss select results from Z\_Nsfdbk for comparison.

\section{Early structure formation}
\label{s3}
The early structure formation in WDM models has been studied with semi-analytic models and numerical simulations (e.g. \citealt{yoshida2003early,oshea2006,gao2007lighting,2016MNRAS.463.3848B,dayal2017reionization,2018MNRAS.477.2886L,10.1093/mnras/stz766}). In cosmological hydrodynamic simulations for a WDM cosmology with a particle mass of $m_{\chi}c^{2}=3$~keV, \citet{yoshida2003early} found that the formation of star forming clouds is delayed by $\sim 60$~Myr, and suppressed in number by about two orders of magnitude, which leads to much less efficient early ionization of the IGM, compared with that in the CDM model calibrated to the initial WMAP data release. On the other hand, \citet{dayal2017reionization} illustrated in a semi-analytical framework that despite of the delay in the start of reionization, WDM models (with $m_{\chi}c^{2}=1.5$, 3, and 5~keV) can produce plausible ending redshifts ($z\simeq 5.5$) with higher escape fractions and gas accretion rates. Similarly, \citet{2016MNRAS.463.3848B} found that the build-up of ionizing sources is faster in sterile neutrino WDM cosmologies, as they are formed in more massive haloes compared with the CDM case. The same trend is also seen in the simulation of \citet{2018MNRAS.477.2886L} for an effective WDM model under the ETHOS framework \citep{PhysRevD.93.123527,2016MNRAS.460.1399V}. These studies show that current observations of the electron scattering optical depth and UVLF (e.g. \citealt{planck,finkelstein2015,mcleod2016z,livermore2017directly}) are equally compatible with CDM and WDM models.

Based on simulations of a WDM model with $m_{\chi}c^{2}=3$~keV, \citet{gao2007lighting} argued that the first stars in WDM cosmologies, in the absence of small-scale perturbations, will form in filaments of masses $\sim 10^{7}\ M_{\odot}$, where fragmentation occurs at high densities. As a result, fragmentation of such dense filaments can cause bursts of star formation and produce stellar mass functions quite different from that in the CDM case. A recent study by \citet{10.1093/mnras/stz766} also found that star formation, although delayed, tends to be more rapid and violent in more gas-rich filaments (see their Fig.~9), for a ETHOS model with a power spectrum similar to that of the thermal WDM model with $m_{\chi}c^{2}=3$~keV.
In general, all these studies have shown that early structure formation and the concomitant processes (e.g. star formation and ionization) in WDM models are delayed and shifted to more massive, and thus luminous, objects. 
Here, we focus on the thermal, star formation and metal enrichment histories during early structure formation, comparing WDM and CDM cosmologies. In this section, we evaluate the physics that drives these histories, while deferring the discussion of the resulting radiation signature to the next section.

\begin{figure*}
\subfloat[WDM, $z=15.8$]{\includegraphics[width= 1\columnwidth]{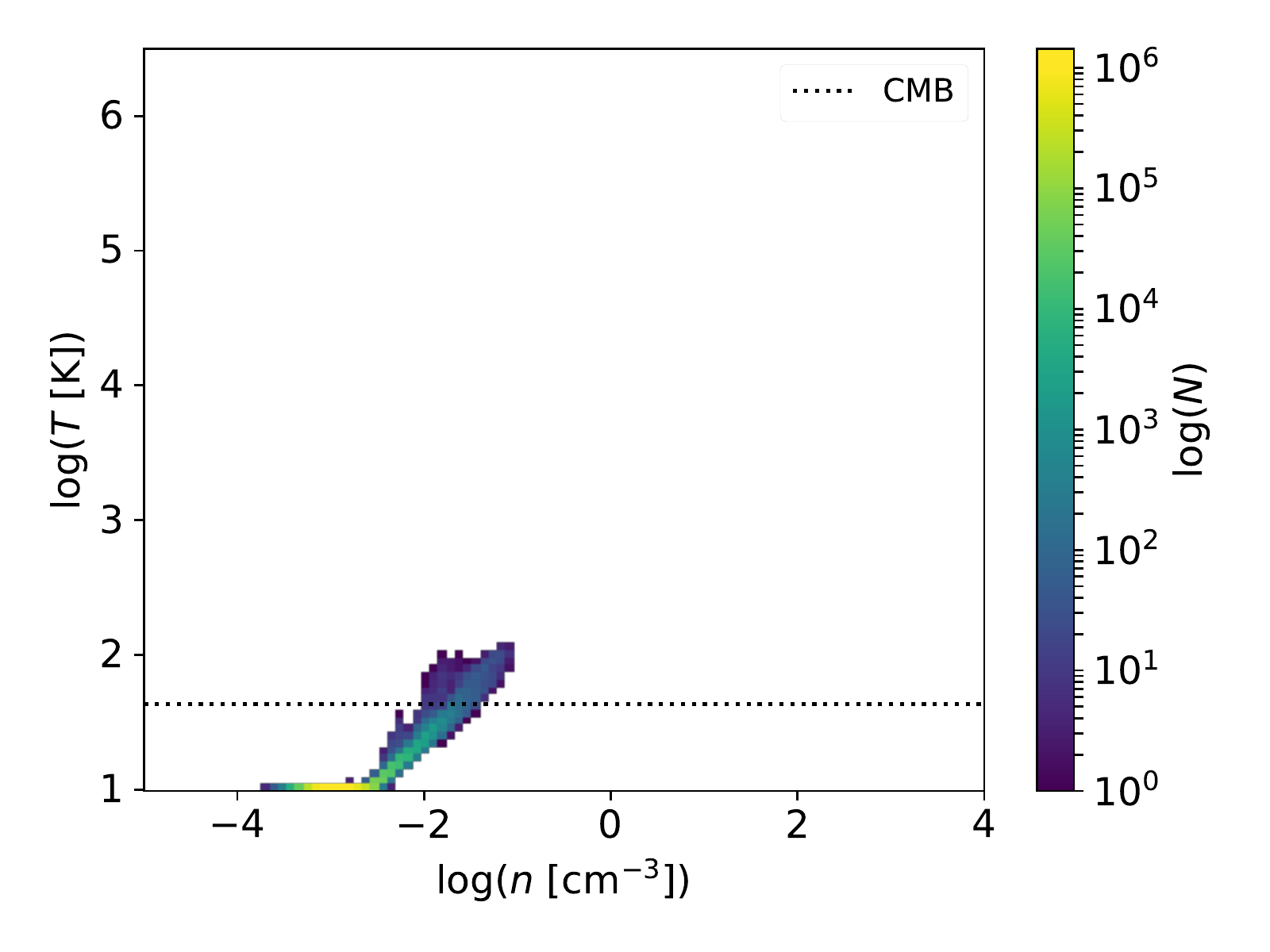}}
\hfill
\subfloat[CDM, $z=15.8$]{\includegraphics[width= 1\columnwidth]{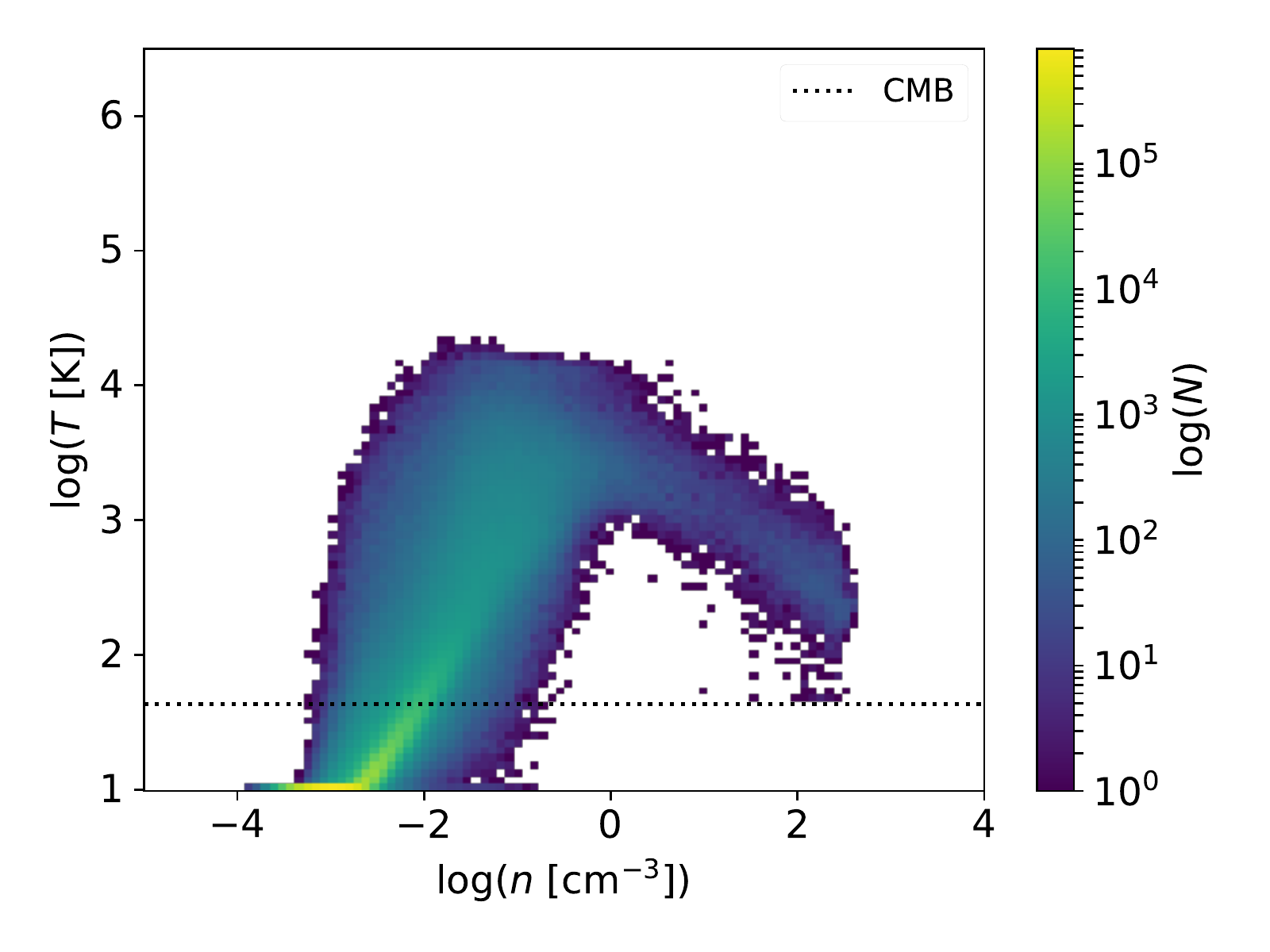}}
\vspace{1pt}
\subfloat[WDM, $z=12.8$]{\includegraphics[width= 1\columnwidth]{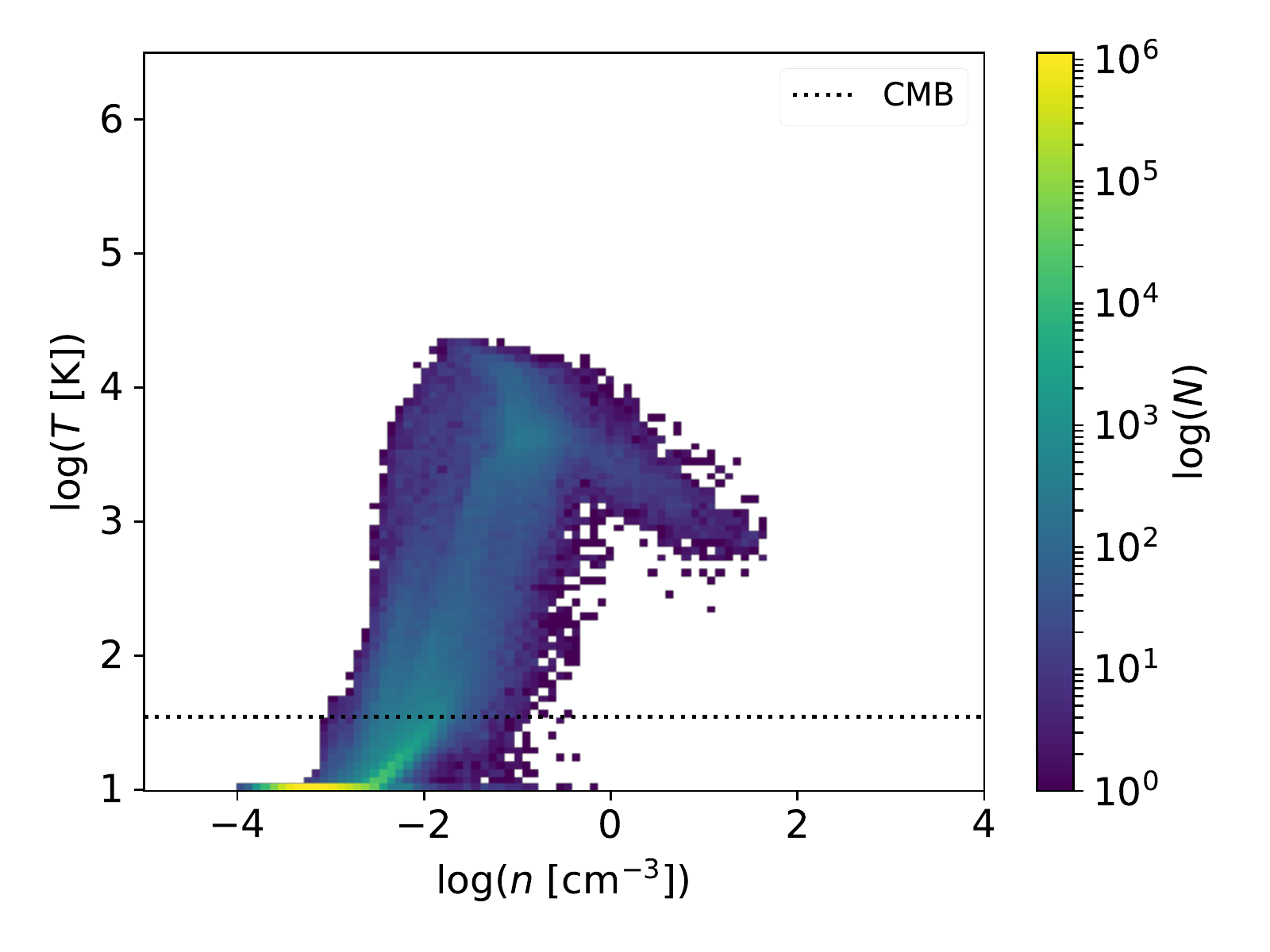}}
\hfill
\subfloat[CDM, $z=12.8$]{\includegraphics[width= 1\columnwidth]{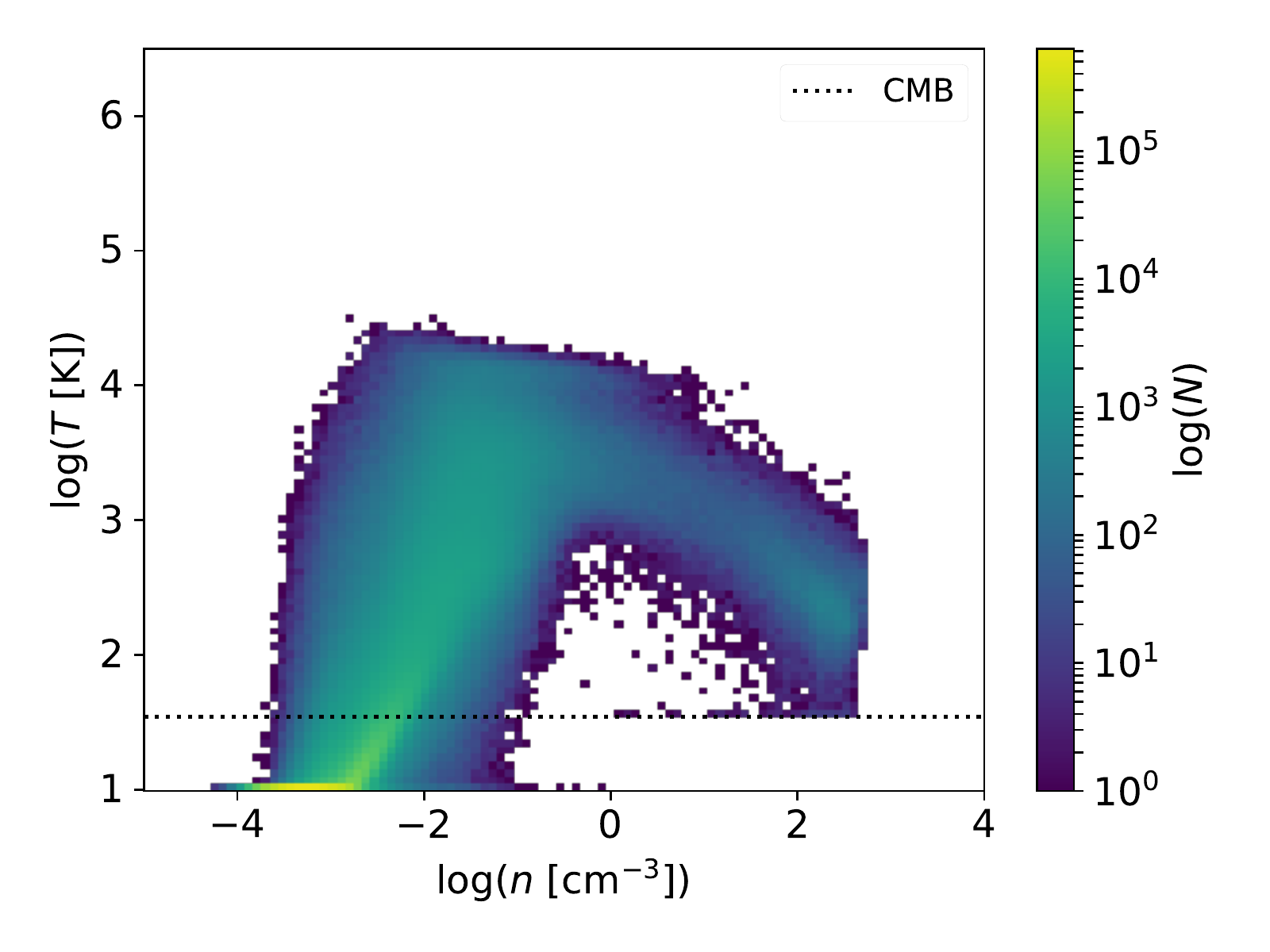}}
\vspace{1pt}
\subfloat[WDM, $z=7.7$]{\includegraphics[width= 1\columnwidth]{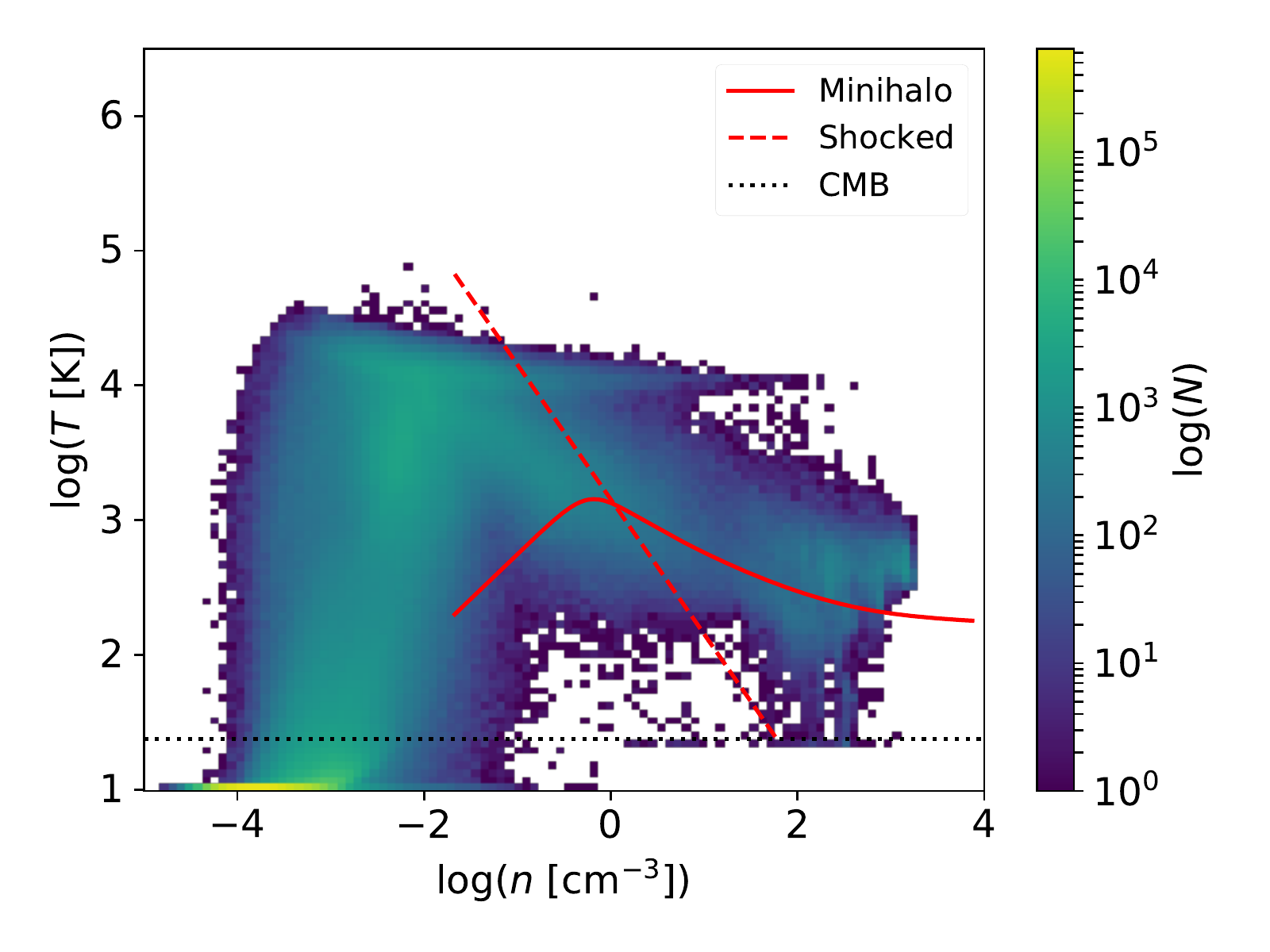}}
\hfill
\subfloat[CDM, $z=7.7$]{\includegraphics[width= 1\columnwidth]{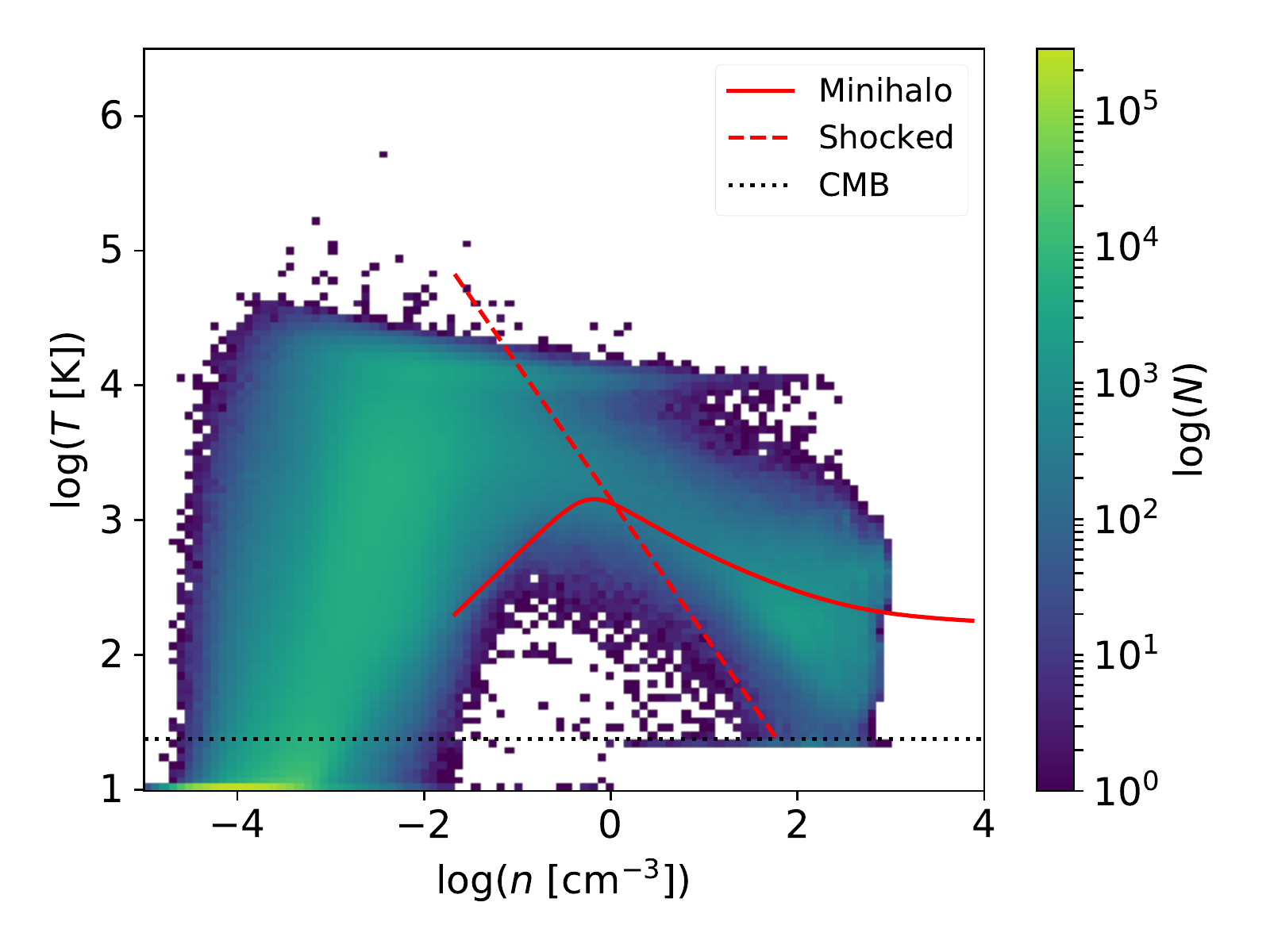}}
\caption{Thermal phase diagrams of the sample zoom-in region in the no-feedback case (Z\_Nsfdbk), at redshifts $z=$15.8 ({\it top}), 12.8 ({\it middle}), and 7.7 ({\it bottom}), for WDM ({\it left}) and CDM ({\it right}) cosmologies. Here, color indicates the number count of gas particles in each bin. Throughout, we show the CMB temperature $T_{\mathrm{CMB}}=2.73 (1+z)$~K ({\it dotted line}) for comparison. In the bottom panel, we also show the evolutionary tracks from the one-zone model \citep{lithium} for free-fall collapsing primordial gas in minihaloes ({\it solid}) and for shocked primordial gas under isobaric conditions ({\it dashed}). It is evident that the development of both collapse modes is delayed in the WDM model at early stages, while at late stages ($z\lesssim 10$) the overall phase-space distributions for the different DM models are similar.}
\label{pd}
\end{figure*}

\begin{figure*}
\hspace{-10pt}
\subfloat[WDM, $z=8.5$]{\includegraphics[width= 1.065\columnwidth]{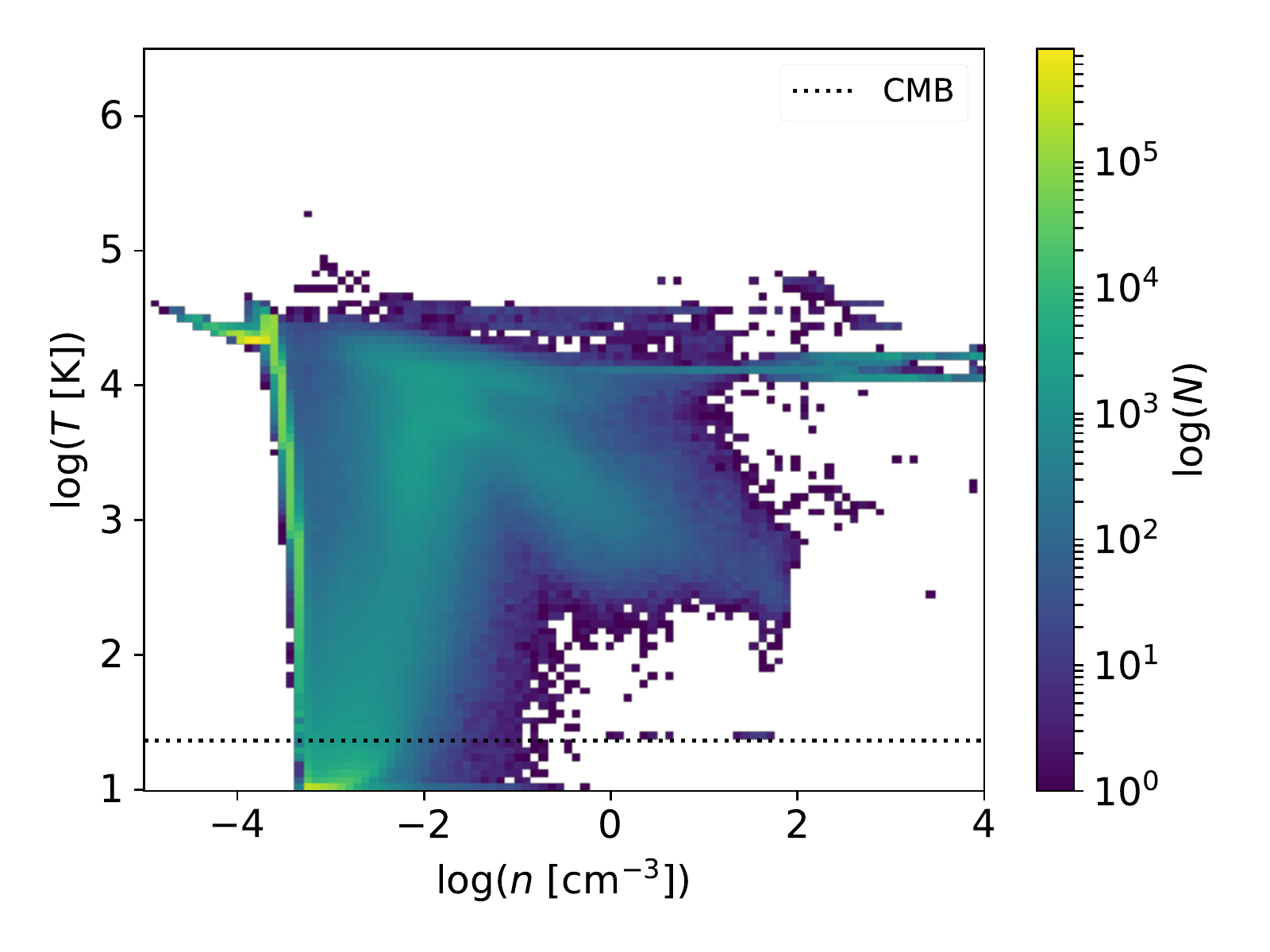}}
\subfloat[CDM, $z=8.5$]{\includegraphics[width= 1.065\columnwidth]{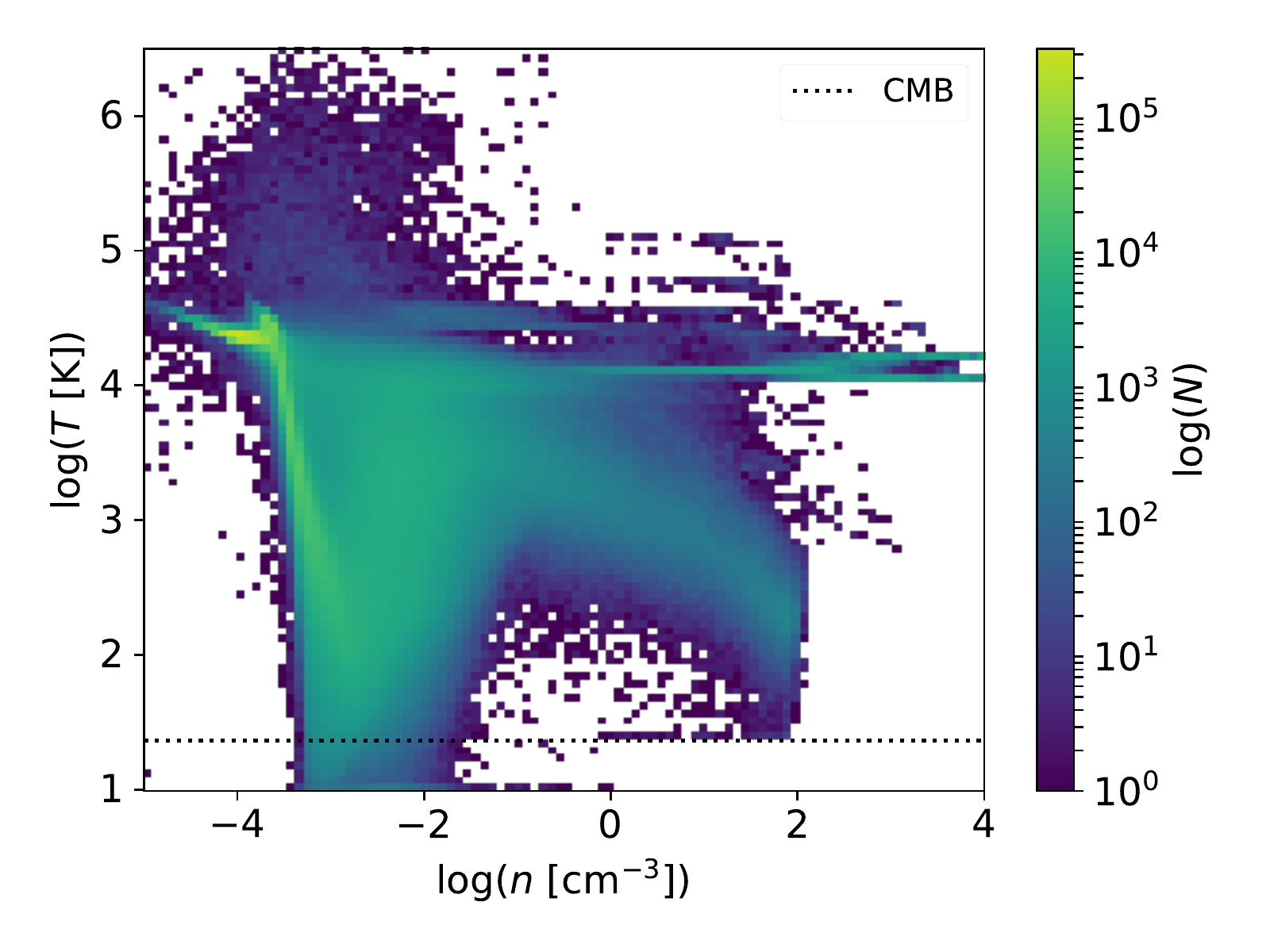}}
\vspace{-10pt}
\caption{$T$-$n$ phase diagrams of the sample zoom-in region at $z=8.5$, with stellar feedback included (Z\_sfdbk). {\it (a)} WDM model. {\it (b)} CDM model. Color again indicates the number of gas particles in each bin ($N$). 
Compared with the no-feedback case (Z\_Nsfdbk), there is a new hot dense component at $T\gtrsim 10^{4}$~K and $1\lesssim n\ [\mathrm{cm^{-3}}]\lesssim 10^{4}$, corresponding to the \HII\ regions around newly-formed stars. In addition, there is a hot diffuse component at $T\gtrsim 3\times 10^{4}$~K and $10^{-4}\lesssim n\ [\mathrm{cm^{-3}}]\lesssim 10^{-2}$ in the CDM cosmology, which does not appear in the WDM model. This component is caused by heating and ionization of the low-density circum-galactic medium (CGM) in low-mass subhaloes by stellar feedback.}
\label{pd1}
\end{figure*}

\subsection{Thermal evolution}
\label{s3.1}

It is instructive to first consider the distribution of gas particles in the temperature-density ($T$-$n$) phase diagram, investigating in particular how stellar feedback changes the thermal evolution of gas in different DM models. 

We start with the simple case of Z\_Nsfdbk to evaluate the performance of the primordial cooling model and SF criteria. Fig.~\ref{pd} shows the distribution of all gas particles (in the sample zoom-in region) in $T$-$n$ phase space for WDM and CDM cosmologies, for a sequence of redshifts. For reference, we reproduce predictions from idealized one-zone models \citep{lithium} for primordial gas collapsing into minihaloes and experiencing shocks under isobaric conditions, at redshift $z=7.7$. The one-zone models are initialized at density $n_{0}=0.3[(1+z)/21]^{3}\,\mathrm{cm^{-3}}=0.021\ \mathrm{cm^{-3}}$, which is the average density of baryons in DM haloes at the point of virialization \citep{clarke2003}. The free-fall collapsing primordial gas in minihaloes evolves from an initial temperature $T_{0}=200$~K and ionization fraction $x_{\mathrm{e}}=10^{-4}$, while the initial values for the shocked primordial gas are $T_{0}=6.7\times 10^{4}$~K ($\simeq T_{\mathrm{vir}}$) and $x_{\mathrm{e}}=0.1$. Here, $T_{\mathrm{vir}}$ is the virial temperature of the target halo, estimated as
\begin{align}
T_{\mathrm{vir}}=&\frac{GM\mu m_{\mathrm{H}}}{5k_{B} R_{\mathrm{vir}}}=9.8\times 10^{4}\ \mathrm{K}\notag\\
&\cdot \left(\frac{\Delta\Omega_{m}}{200\cdot 0.315}\right)^{1/3}\left(\frac{1+z}{10}\right)\left(\frac{M}{10^{10}\ M_{\odot}}\right)^{2/3} ,\label{e0}
\end{align}
where $R_{\mathrm{vir}}=\left[3M/(4\pi\rho_{\mathrm{crit},0}\Omega_{m}\Delta)\right]^{1/3}a$ is the (physical) virial radius, $a=1/(1+z)$ the scale factor, $\rho_{\mathrm{crit},0}=(8\pi G/3)^{-1}H_{0}^{2}$ the present-day critical density, $\Delta=200$ the virial overdensity, and $\mu=0.63$ the mean molecular weight of primordial gas with fully ionized hydrogen. 

In general, in DM haloes such as the target system with virial masses above the threshold for the onset of atomic-hydrogen cooling, $M_{\mathrm{th}}\sim 10^{8}\ M_{\odot}$, there are two modes of accretion, leading to different evolutionary paths for the primordial gas \citep{greif2008first}. For hot accretion, the gas is first heated to temperatures $T\gtrsim 10^{4}$~K by structure formation shocks, at which point cooling by atomic hydrogen becomes efficient. Then, the gas quickly cools and enters a cold dense phase ($n\gtrsim 1\ \mathrm{cm^{-3}}$, $T\lesssim 10^{3}$~K), which enables fragmentation and subsequent star formation. The second mode is cold accretion, where gas is accreted along filaments, so that it remains cold and dense without being shocked. The one-zone model for isobaric post-shock evolution represents the idealized behavior during hot accretion, while that for free-fall collapse exemplifies gas during cold accretion.

As can be seen, both modes of accretion are delayed in the WDM model at early stages ($z\gtrsim 10$). For instance, the (star-forming) cold dense component occurs in CDM cosmology at redshift $z\sim 20$, whereas for WDM, the initial heating during hot accretion just starts at $z\sim 16$, and the gas enters its cold dense phase after $z\sim 12$. 
However, at late stages, after virialization ($z\lesssim 10$)\footnote{In Z\_Nsfdbk, the distribution of gas particles in $T$-$n$ phase space remains nearly unchanged at $z\lesssim 10$, implying that the central object has reached a dynamical equilibrium. We thus conclude that the target halo virializes at $z\sim 10$.}, the thermal phase space behaviour in the two cosmologies becomes quite similar. A small difference exists in the region with $T\lesssim 200$~K and $n\gtrsim 10^{2}\ \mathrm{cm^{-3}}$, where the amount of cold dense gas is smaller in the WDM cosmology, implying that cold accretion is suppressed. These results are consistent with the trend found in \citet{hirano2017first} for FDM that the onset of Pop~III star formation is delayed, and shifted to more massive host structures, while the late-stage thermal properties of primordial gas remain asymptotically the same. We note that the slope of the idealised one-zone isobaric track ($T\propto n^{-1}$) is steeper than what is seen in the simulations. This indicates that pressure is actually increasing during post-shock evolution, due to the gas falling deeper into the gravitational potential well\footnote{The hot dense component at $T\sim 10^{4}$~K and $n\gtrsim 1\ \mathrm{cm^{-3}}$ is unphysical in the case of no feedback, caused by artificial virial heating around sink particles.}. 

For the Z\_sfdbk runs, we are only interested in the late-stage ($z\lesssim 10$) properties, from which we can better appreciate the effects of stellar feedback. As presented in Fig.~\ref{pd1}, which shows the situation at $z=8.5$, a common feature in the $T$-$n$ phase diagrams with stellar feedback for both DM models is a hot dense component at $T\gtrsim 10^{4}$~K and $1\lesssim n\ [\mathrm{cm^{-3}}]\lesssim 10^{4}$, which corresponds to the \HII\ regions around newly-formed stellar populations\footnote{Gas in this component has a typical temperature $T\sim 1.5\times 10^{4}$~K, emerging from the balance between photo-ionization heating and atomic cooling $\Gamma_{\mathrm{pi}}=\Lambda$. For the dense gas close to LTE ($n\gtrsim 10^{2}\ \mathrm{cm^{2}}$), both $\Gamma_{\mathrm{pi}}$ and $\Lambda$ are proportional to $n$, such that the equilibrium temperature is independent of density.}. This phase produces the majority of free-free emission ($\gtrsim 99$\%). However, the densest part in this component (with $n\gtrsim 10^{3}\ \mathrm{cm^{-3}}$) is unphysical due to the legacy nature of our feedback model. Actually, the P2L model for Pop~II stellar feedback in \citet{jaacks2018legacy} tends to over-predict the volumes of compact \HII\ regions, as it uses a fixed ionization front radius $R_{\mathrm{ion}}\propto n^{-2/3}$ for ionization heating, based on a typical density $n=1\ \mathrm{cm^{-3}}$, which is not valid in dense environments with $n\gg 1\ \mathrm{cm^{-3}}$, e.g. at centres of subhaloes. We have rerun the simulations under the same condition with a modified P2L model of adaptive ionization radii $R_{\mathrm{ion}}\propto n^{-2/3}$, and find that the highest density of hot gas drops to $\sim 10^{3}\ \mathrm{cm^{-3}}$. As a result, we expect that the free-free signal to be strongly overestimated if this unphysical hot dense gas is taken into account. Therefore, we place an upper bound to gas density when calculating the free-free emission in Section~\ref{s4.1}. 

Another less important common feature in the $T$-$n$ phase diagrams from Z\_sfdbk is the heating of the diffuse IGM ($n\lesssim 10^{-4}\ \mathrm{cm^{-3}}$) by the UV background.
For cold gas with $T\lesssim 200$~K, most of it resides in the low-density region ($n\lesssim 1\ \mathrm{cm^{-3}}$) for the WDM cosmology, while a significant amount of dense gas ($n\gtrsim 1\ \mathrm{cm^{-3}}$) is found in the CDM case. This results from the absence of small-scale structures and weaker stellar feedback (due to delayed Pop~II star formation, see the next subsection) in the WDM cosmology.

Interestingly, in the CDM cosmology, there is an additional hot diffuse component with $T\gtrsim 3\times 10^{4}$~K and $10^{-4}\lesssim n\ [ \mathrm{cm^{-3}}]\lesssim 10^{-2}$, which is also found in other simulations of atomic cooling haloes for standard CDM (e.g. see fig.~10 in \citealt{jeon2015first}). Since this component only emerges in CDM cosmology, it must be associated with star formation in small-scale structures, such as minihaloes. A possible scenario is that stellar feedback heats and ionizes the low-density circum-galactic medium (CGM) in low-mass subhaloes, whose gravity is not strong enough to contract and compress the heated gas, so that the CGM density remains low. This results in insufficient cooling and high temperatures, in particular when the gas is affected by multiple star formation events. 


\subsection{Star formation history}
\label{s3.2}

\begin{figure*}
\hspace{-10pt}
\subfloat[]{\includegraphics[width= 1.065\columnwidth]{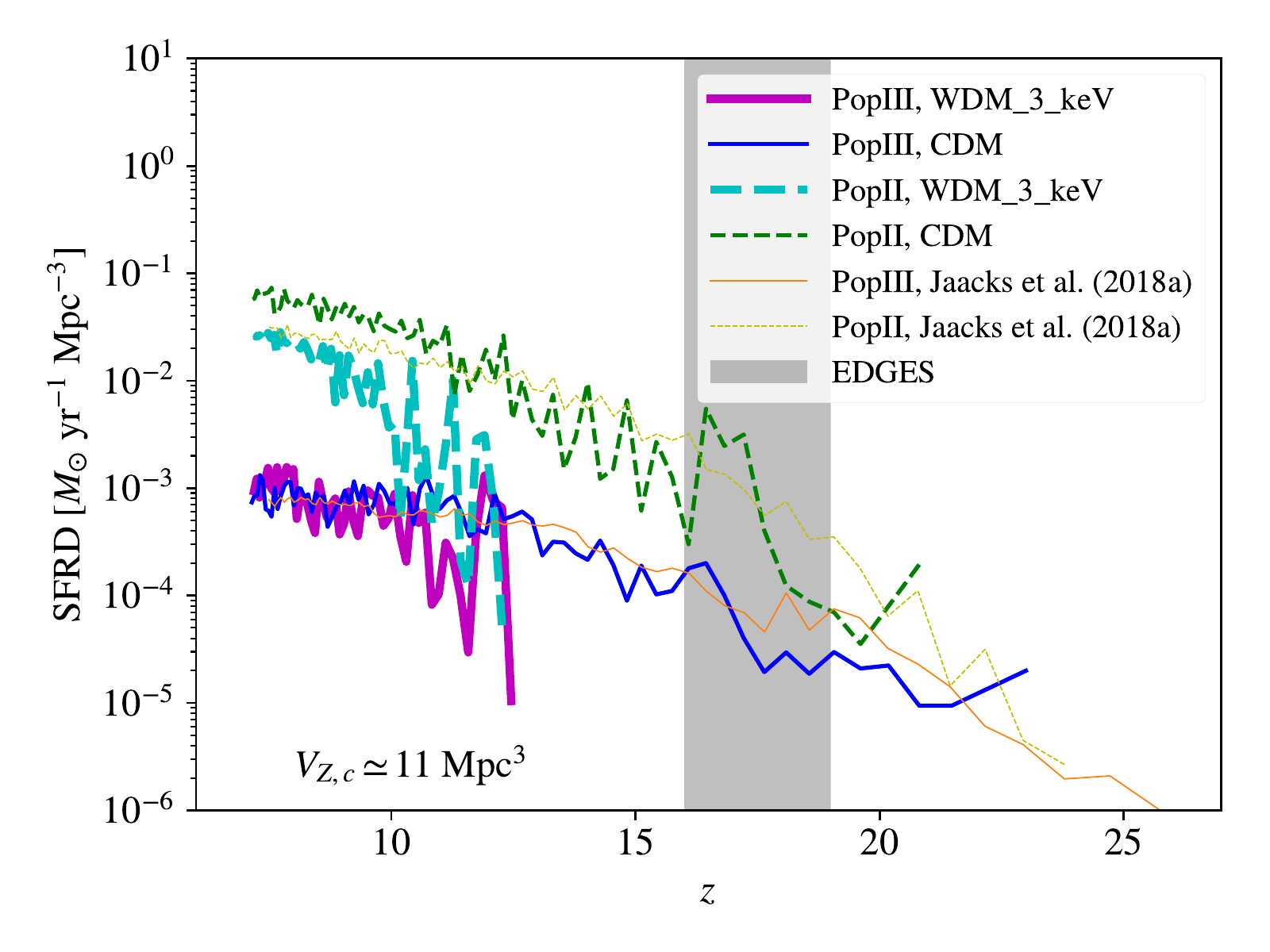}}
\subfloat[]{\includegraphics[width= 1.065\columnwidth]{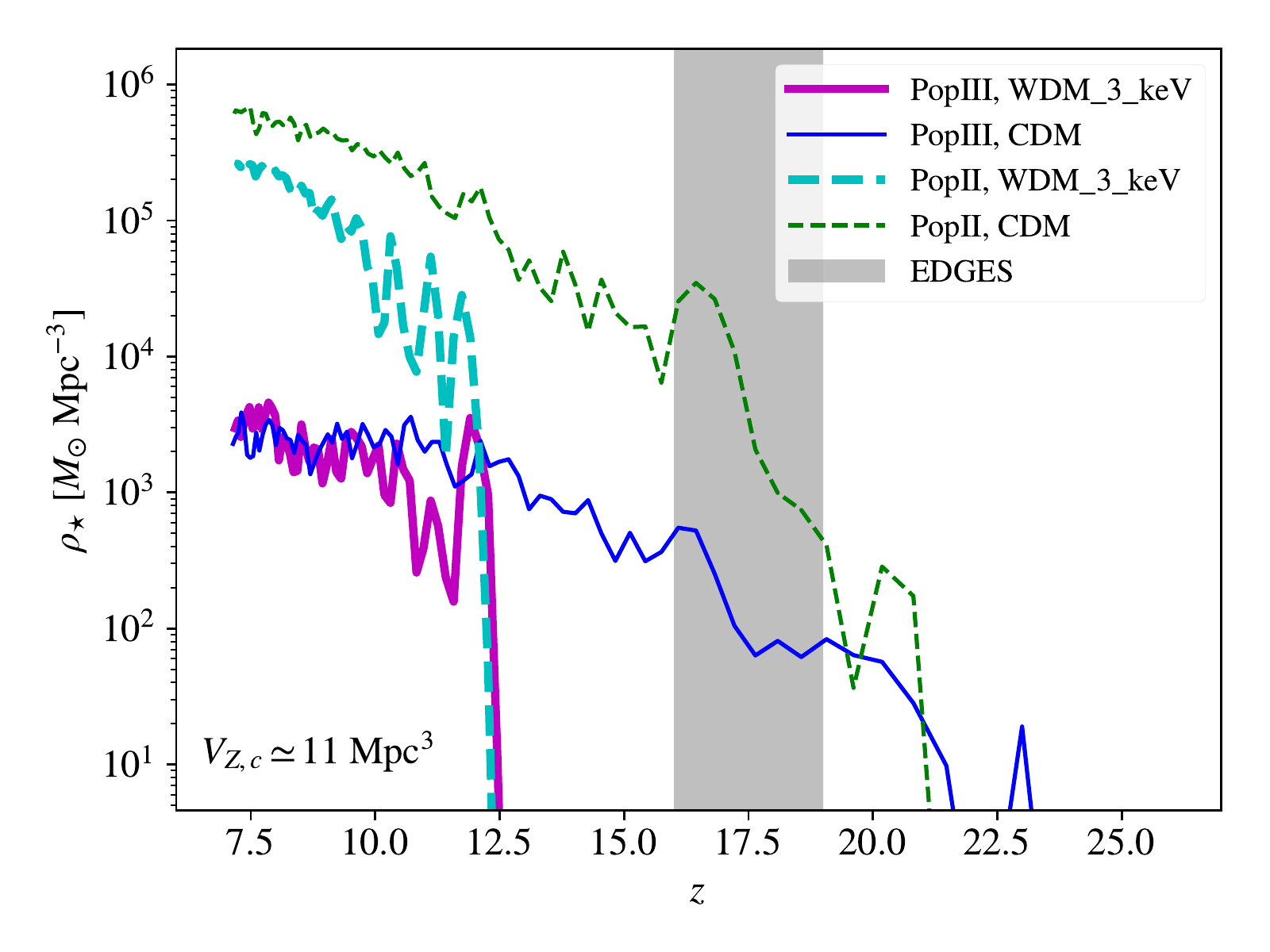}}
\vspace{-10pt}
\caption{Evolution of {\it (a)} SFRD and {\it (b)} stellar mass density with redshift, measured in the sample zoom-in region with a co-moving volume $V_{Z,c}\simeq 11\ \mathrm{Mpc^{3}}$, from Z\_sfdbk. The results for Pop~III and Pop~II stars are shown with solid and dashed curves, for the WDM ({\it thick lines}) and CDM ({\it normal lines}) models. In panel (a), we also plot the CDM results from \citet{jaacks2018legacy} with thin curves. Furthermore, we show the redshift range around $z\sim 17$ of the 21-cm absorption signal detected by EDGES \citep{nature} as the shaded region. If the EDGES signal were confirmed, the WDM model simulated here would be ruled out, since it cannot form stars before $z\sim 17$.} 
\label{sfr}
\end{figure*}

Fig.~\ref{sfr} shows the (co-moving) star formation rate density (SFRD) and stellar mass density $\rho_{\star}$ (of young stellar populations with strong stellar feedback) as functions of redshift, from Z\_sfdbk\footnote{In general, the stellar mass density estimated from Z\_Nsfdbk (with a SF efficiency $\epsilon=0.05$) is higher than that from Z\_sfdbk by one order of magnitude. The result in Z\_Nsfdbk is unphysical due to the absence of stellar feedback.}, in comparison with the results from \citet{jaacks2018legacy} for the standard CDM model. It turns out that star formation first occurs at redshift $z\sim 23$ in the CDM cosmology, while at redshift $z\sim 12.5$ in the WDM cosmology, which implies a delay of $\sim 200$~Myr. Note that the 21-cm absorption signal detected by EDGES is centered at $z\sim 17$. If this signal is confirmed, the WDM model simulated here with a DM particle mass of 3~keV will be disfavoured, because there is no star formation before $z\sim 17$ to generate the Ly$\alpha$ radiation field that couples the 21-cm spin temperature with the kinetic temperature of the IGM to produce the absorption signal. This is consistent with the result in \citet{Schneider2018} based on the timing of the EDGES signal that the mass of thermal WDM is limited to $m_{\chi}c^{2}>6.1$~keV, while previous studies obtained lower minimum WDM masses of $2-3$~keV \citep{Sitwell2014,Safar2018}, applying a similar analysis but making different approximations for the WDM transfer function and astrophysical parameters (such as star formation efficiency). It is necessary to point out that none of these semi-analytical studies, as well as this work considers the relative velocities between DM and baryons (i.e. the streaming motion), in the presence of which haloes have to be heavier than what they would be if no velocity effect was present to form stars. As a result, high-redshift star formation will be delayed/suppressed (e.g. \citealt{greif2011delay,stacy2011effect,naoz12,naoz13,Anna2018}). The recent study by \citet{Anna2019} takes into account these velocities and shows that sufficient Pop III star formation in small-scale structures at $z\gtrsim 20$ is indispensable to produce the 21-cm signal, which, however, is suppressed in WDM models. In light of this, we suspect that the constraint on WDM mass would be further tightened with the streaming motion between baryons and DM, and the model with $m_{\chi}c^{2}=3$~keV would be ruled out if the EDGES signal is real. 

For both DM models, Pop~II star formation dominates the overall SFRD once it occurs. In the CDM cosmology, Pop~II star formation commences at redshift $z\sim 20.8$, which is $\sim 22.6$~Myr after the initial Pop~III activity, whereas for WDM, the initial Pop~II stellar population is formed at redshift $z\sim 12.26$, shortly (7.5~Myr) after the appearance of Pop~III stars. 

Interestingly, the Pop~III SFRD in the WDM model is similar to the CDM case for $z\lesssim 10$, although the number density of minihaloes (with $M\sim 10^{6}\ M_{\cdot}$) is lower by one order of magnitude\footnote{Our zoom-in simulation for the WDM model actually over-predicts the abundance of low-mass ($M\lesssim 5\times 10^{6}\ M_{\odot}$) haloes by up to a factor of 5. This is caused by spurious numerical fragmentation, which is a common outcome of simulations with a power spectrum cut-off (e.g. \citealt{2007MNRAS.380...93W,angulo2013warm}). However, the abundance of low-mass haloes in the WDM model is still much lower compared with the CDM case. So this will not affect our results of star formation histories and radiation signature.}. This shows that formation of Pop~III stars in dense filaments within WDM \citep{gao2007lighting,10.1093/mnras/stz766} is as efficient as that in minihaloes within CDM. 
However, the Pop~II SFRD in the WDM model is significantly lower than the CDM counterpart even at $z\sim 7.2$, but the difference decreases toward lower redshifts. 
For $7.2\lesssim z\lesssim 12.5$, the Pop~II stellar mass density in the CDM cosmology is always higher than for WDM by at least a factor of 4, while the Pop~III stellar mass densities are almost identical in the two DM models at $z\sim 7.2$. This is explained by the suppression of star formation in small-scale structures for the WDM cosmology, which leads to less efficient metal enrichment, especially in terms of the volume filling fraction of enriched gas that can host Pop~II stellar populations, as shown below. For the CDM cosmology, our results are consistent with those in \citet{jaacks2018legacy}, acknowledging the fact that our sample zoom-in region has a much smaller volume (5.28\%) than that of the simulation box in their work. Note that our zoom-in region represents an overdense region, and is thus more efficient in creating massive DM haloes (with $M\gtrsim 10^{8}\ M_{\odot}$). As a result, it is reasonable that the Pop~II SFRD in the sample zoom-in region is higher than that from \citet{jaacks2018legacy}. On the other hand, the Pop~III SFRD predicted by our simulations is identical to that from \citet{jaacks2018legacy}, since the volume of the sample zoom-in region is large enough to produce a cosmic-mean number density of small-scale structures, such as minihaloes, where Pop~III stars are formed. 

\subsection{Metal enrichment}
\label{s3.3}

\begin{figure*}
\hspace{-10pt}
\subfloat[]{\includegraphics[width= 1.065\columnwidth]{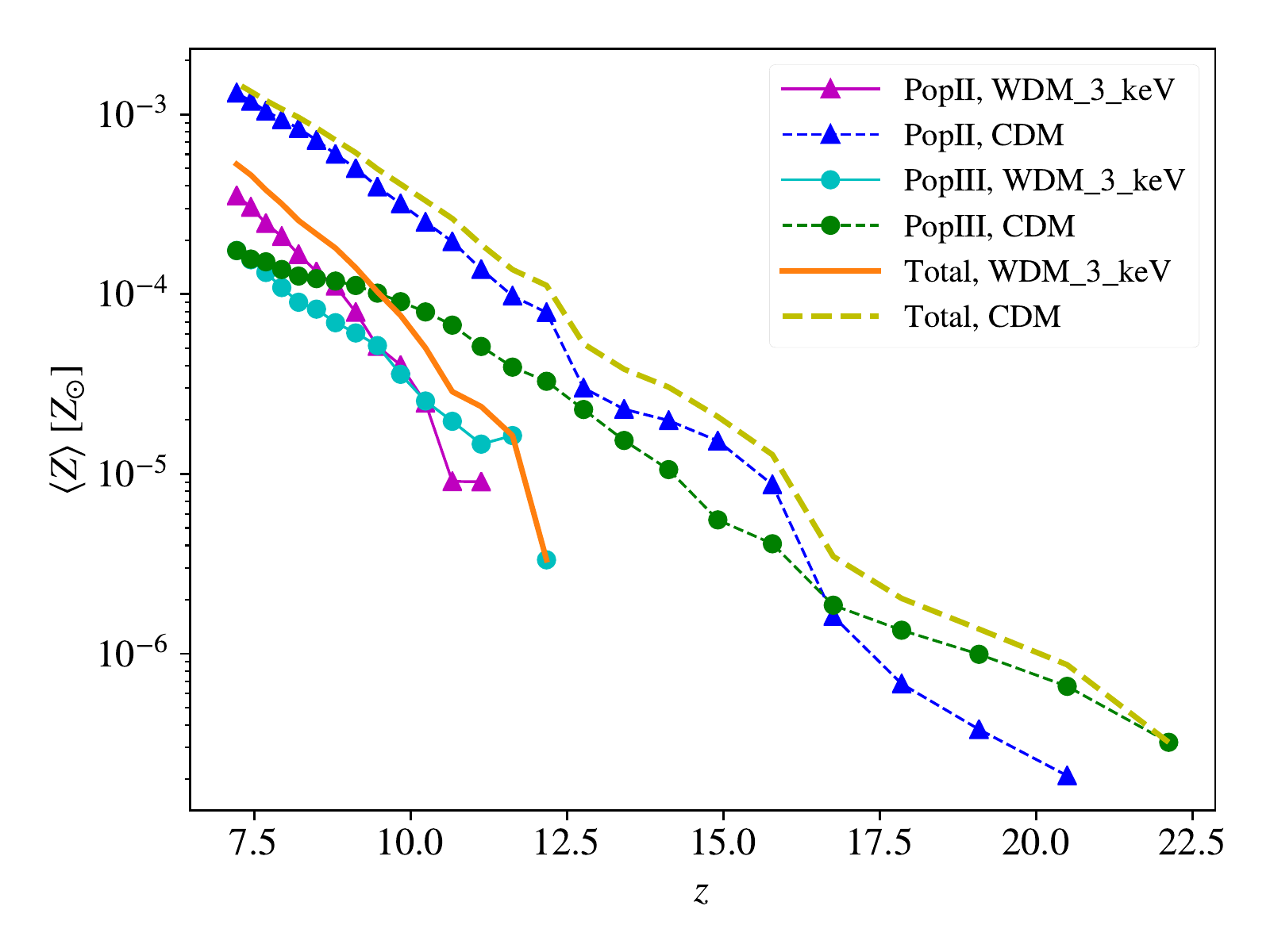}}
\subfloat[]{\includegraphics[width= 1.065\columnwidth]{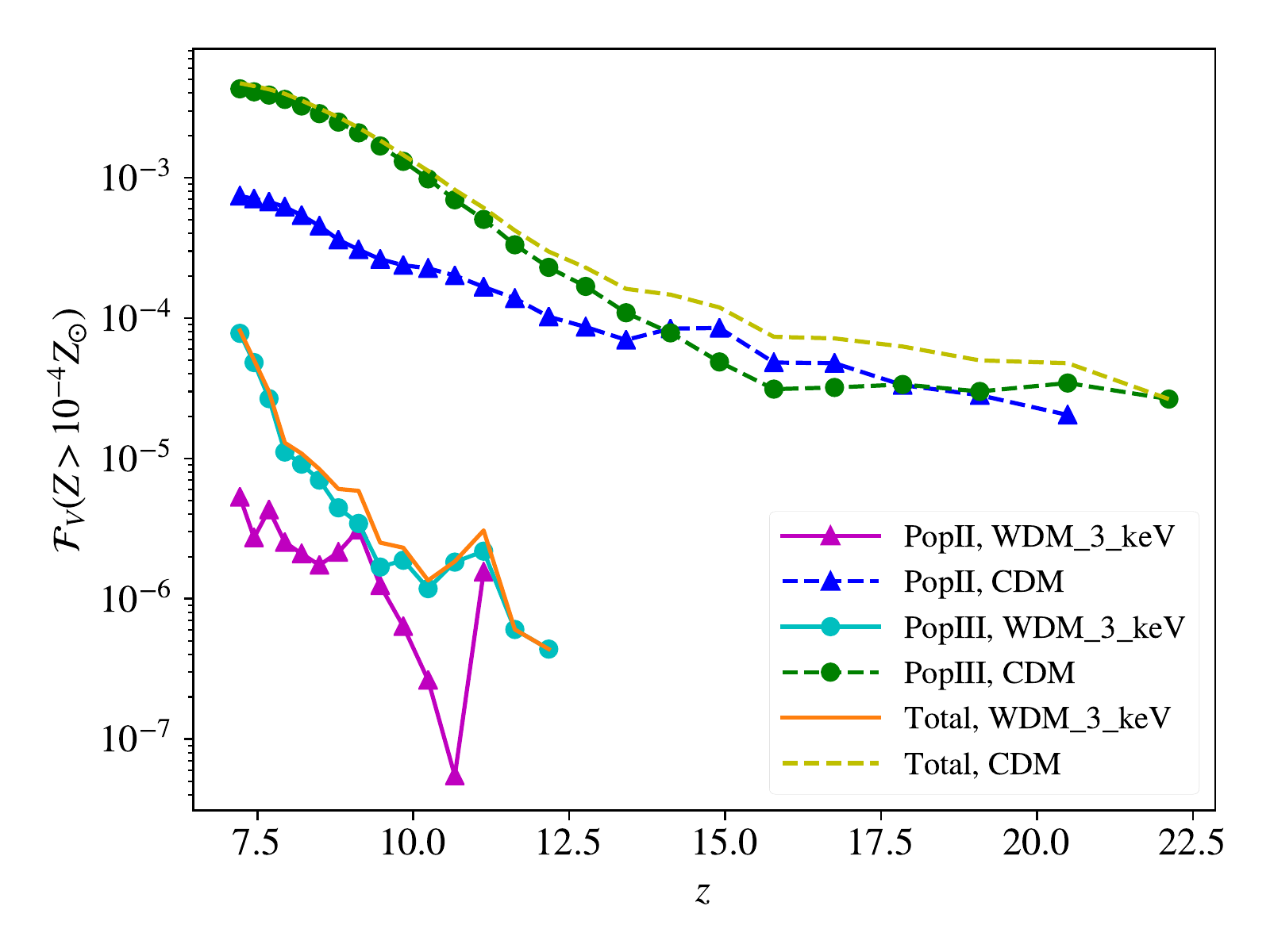}}
\vspace{-10pt}
\caption{Redshift evolution of the {\it (a)} average metallicity and {\it (b)} volume filling fraction, $\mathcal{F}_{V}$, of gas with $Z>Z_{\mathrm{crit}}=10^{-4}\ Z_{\odot}$, measured in the sample zoom-in region. Results for Pop~III, Pop~II and total metals are shown with circles, triangles and no marker, respectively, for WDM ({\it solid lines}) and CDM ({\it dashed lines}). Evidently, the volume filling fraction of Pop~II gas in the CDM cosmology is always much higher than for WDM, indicating that significant metal enrichment tends to occur only in dense environments in the WDM model, affecting a small volume, while in the CDM model, star formation in small-scale structures can enrich a large volume of gas with low densities in addition to the dense regions.}
\label{Zz}
\end{figure*}

\begin{figure*}
\hspace{-10pt}
\subfloat[WDM, $z=8.5$]{\includegraphics[width= 1.065\columnwidth]{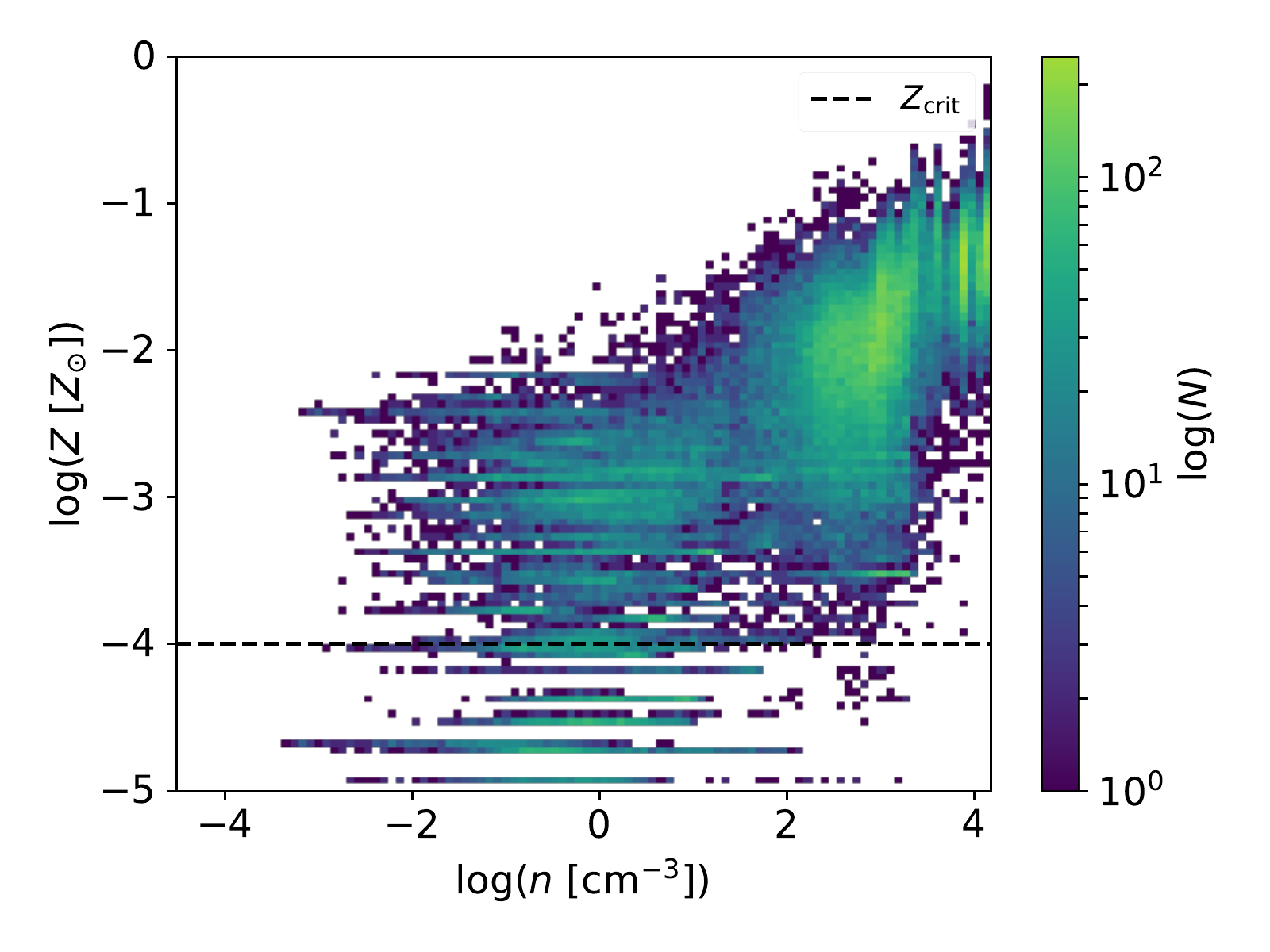}}
\hfill
\subfloat[CDM, $z=8.5$]{\includegraphics[width= 1.065\columnwidth]{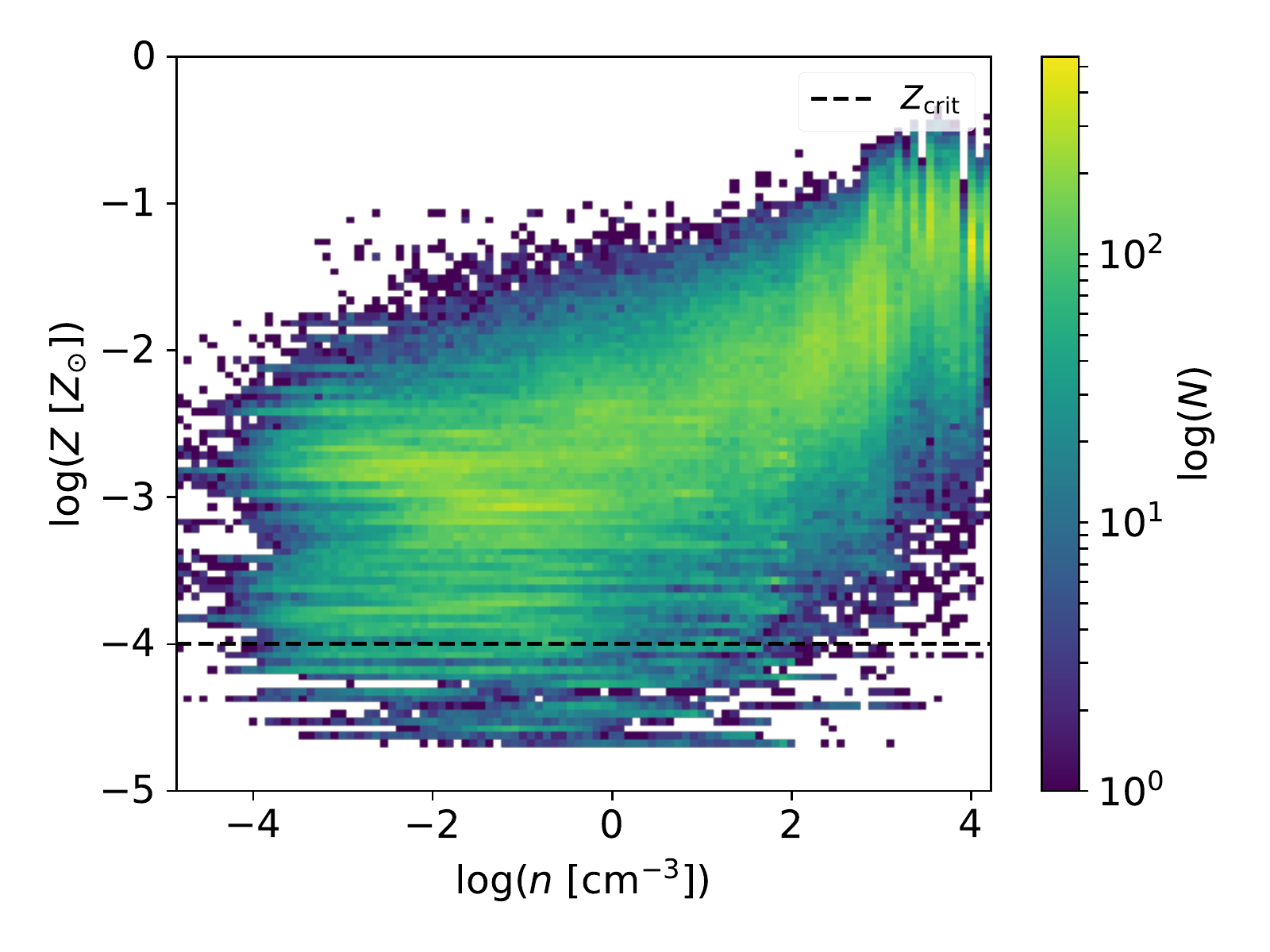}}
\vspace{-10pt}
\caption{Environment for metal enrichment. Metallicity-density ($Z$-$n$) phase diagrams of the sample zoom-in region, at $z=8.5$, from Z\_sfdbk, in {\it (a)} WDM and {\it (b)} CDM cosmologies, where color indicates the number count of enriched gas particles in each bin ($N$). For WDM, metal enrichment mostly affects the densest gas, while for CDM, enrichment is much more widespread, also reaching gas with lower densities, due to star formation in small-scale structures.}
\label{Znpd}
\end{figure*}

Fig.~\ref{Zz} illustrates the metal enrichment histories in the sample zoom-in region, in terms of the (mass-weighted) average metallicity $\langle Z\rangle$ and volume filling fraction $\mathcal{F}_{V}$ of gas with $Z>Z_{\mathrm{crit}}=10^{-4}\ Z_{\odot}$ (Pop~II gas, henceforth), for metals produced by both Pop~III and Pop~II stars, from the Z\_sfdbk runs. 
On average, metal enrichment in the WDM cosmology is delayed by $\sim 200$~Myr compared to the CDM case, which reflects the delay in the initial Pop~III activity and the rise of Pop~II star formation. 
The volume fraction of Pop~II gas in the CDM cosmology is always much higher (by a factor of $10^{2}-10^{4}$) than for WDM. This indicates that significant metal enrichment tends to occur only in dense environments in the WDM model, affecting a small volume, while in the CDM model, star formation in small-scale structures can enrich a large volume of gas with low densities in addition to the dense regions.

This trend is confirmed by the distribution of enriched gas in the metallicity-density ($Z$-$n$) phase diagram, as shown in Fig.~\ref{Znpd}. 
Actually, the difference in $\mathcal{F}_{V}$ between the two DM models for Pop~II produced metals is about two orders of magnitude and does not change much with redshift, while that for Pop~III produced metals decreases towards lower redshifts. 
Finally, for CDM cosmology, our simulations predict slightly higher $\langle Z\rangle\sim 10^{-3} Z_{\odot}$ and $\mathcal{F}_{V}\sim 10^{-3}$ at late stages ($z\sim 7.5$), compared with the results in \citet{jaacks2018legacy} and \citet{pallottini2014simulating}, with much larger volumes ($4^{3}\ h^{-3}\ \mathrm{Mpc^{3}}$ and $10^{3}\ h^{-3}\ \mathrm{Mpc^{3}}$, respectively). The reason again is that our zoom-in simulations are targeted at overdense parts of the Universe, with corresponding metal production efficiency higher than the cosmic average.



\section{Radiation signature}
\label{s4}
In this section, we present the radiation signature of the target halo derived from the zoom-in simulations. The radiative transfer calculation is only performed for the cubic central box with a co-moving volume of $500^{3}\ h^{-3}\mathrm{kpc^{3}}\sim (10R_{\mathrm{vir,c}})^{3}$ in the sample zoom-in region, to include all the emission associated with the formation of the target halo, while excluding the contribution from other DM haloes that also form in this zoom-in region. Here $R_{\mathrm{vir,c}}\sim 50\ h^{-1}\mathrm{kpc}$ is the co-moving virial radius of the target halo at redshift $z=8.8$. Note that the sample zoom-in region is defined as the smallest box that enclose the \textit{initial} distribution of particles from the target halo, which has a fixed \textit{co-moving} volume. In this way, it also includes some particles that will not belong to the target halo at late stages, and these particles can form DM haloes other than the target halo in the zoom-in region. Actually, the target halo only occupies the central part of the sample zoom-in region at $z\lesssim 12$, when the radiation is built up.  

\subsection{Free-free emission}
\label{s4.1}

\begin{figure*}
\hspace{-10pt}
\subfloat[WDM]{\includegraphics[width= 1.065\columnwidth]{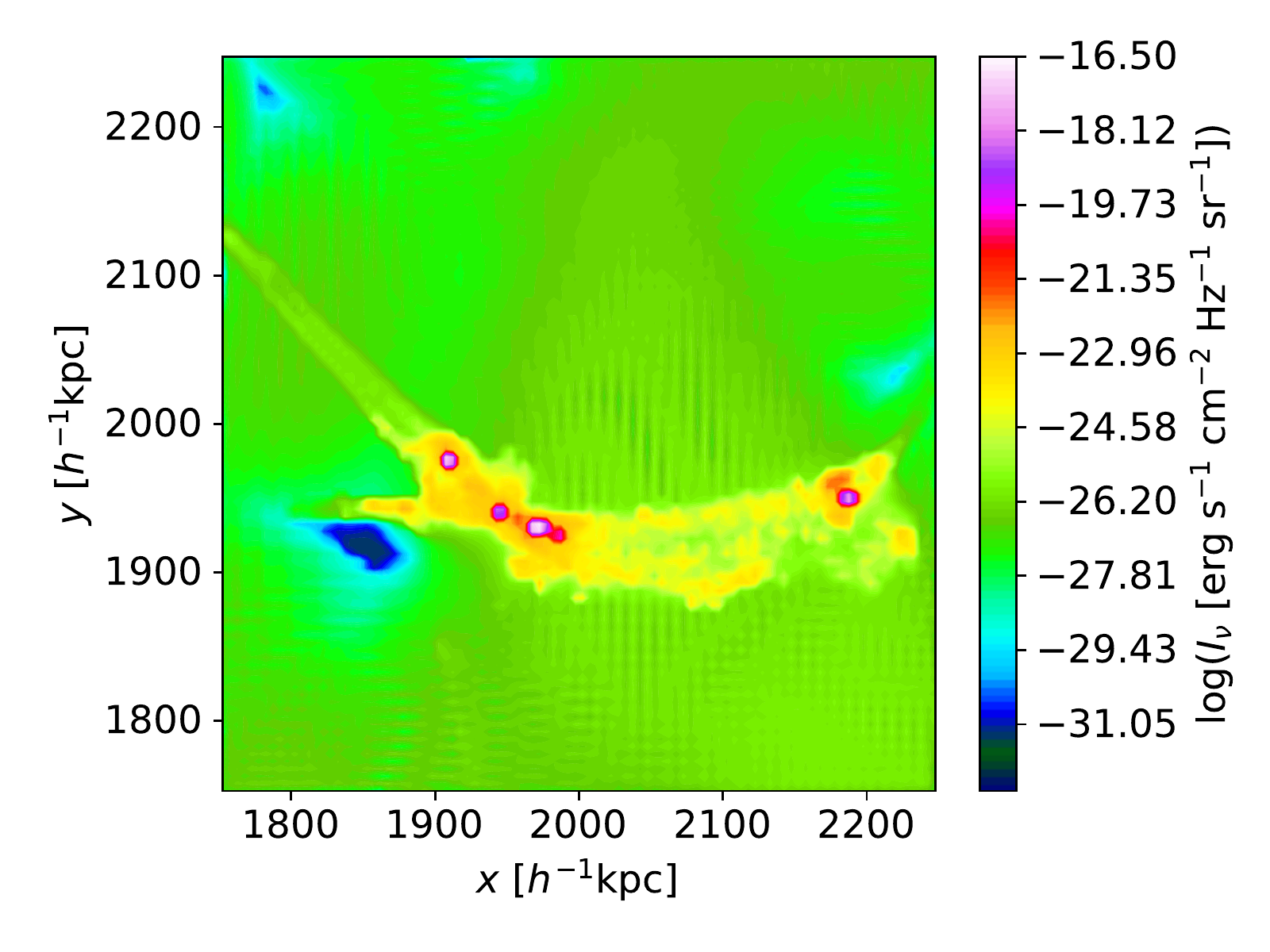}}
\hfill
\subfloat[CDM]{\includegraphics[width= 1.065\columnwidth]{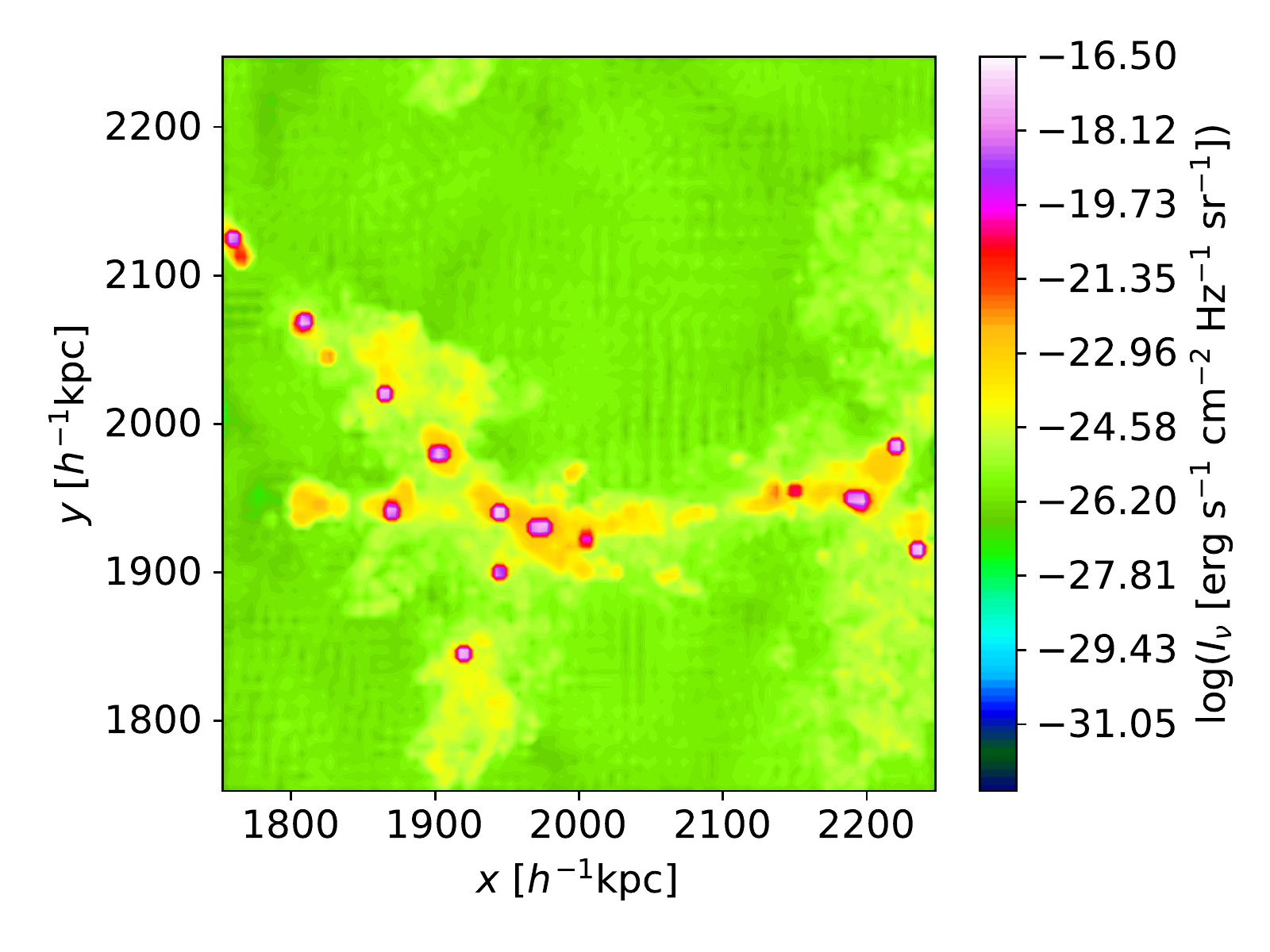}}
\vspace{-10pt}
\caption{Rest-frame specific intensity maps of free-free emission at rest-frame frequency $\nu=0.1\ \mathrm{GHz}$ and redshift $z=8.5$, in {\it (a)} WDM and {\it (b)} CDM cosmologies, for the feedback simulations (Z\_sfdbk). Significantly fewer sources are found in the WDM run, compared with the CDM case, caused by suppression of small-scale structure formation.}
\label{f00}
\end{figure*}

\begin{figure}
\hspace{-10pt}
\includegraphics[width= 1.05\columnwidth]{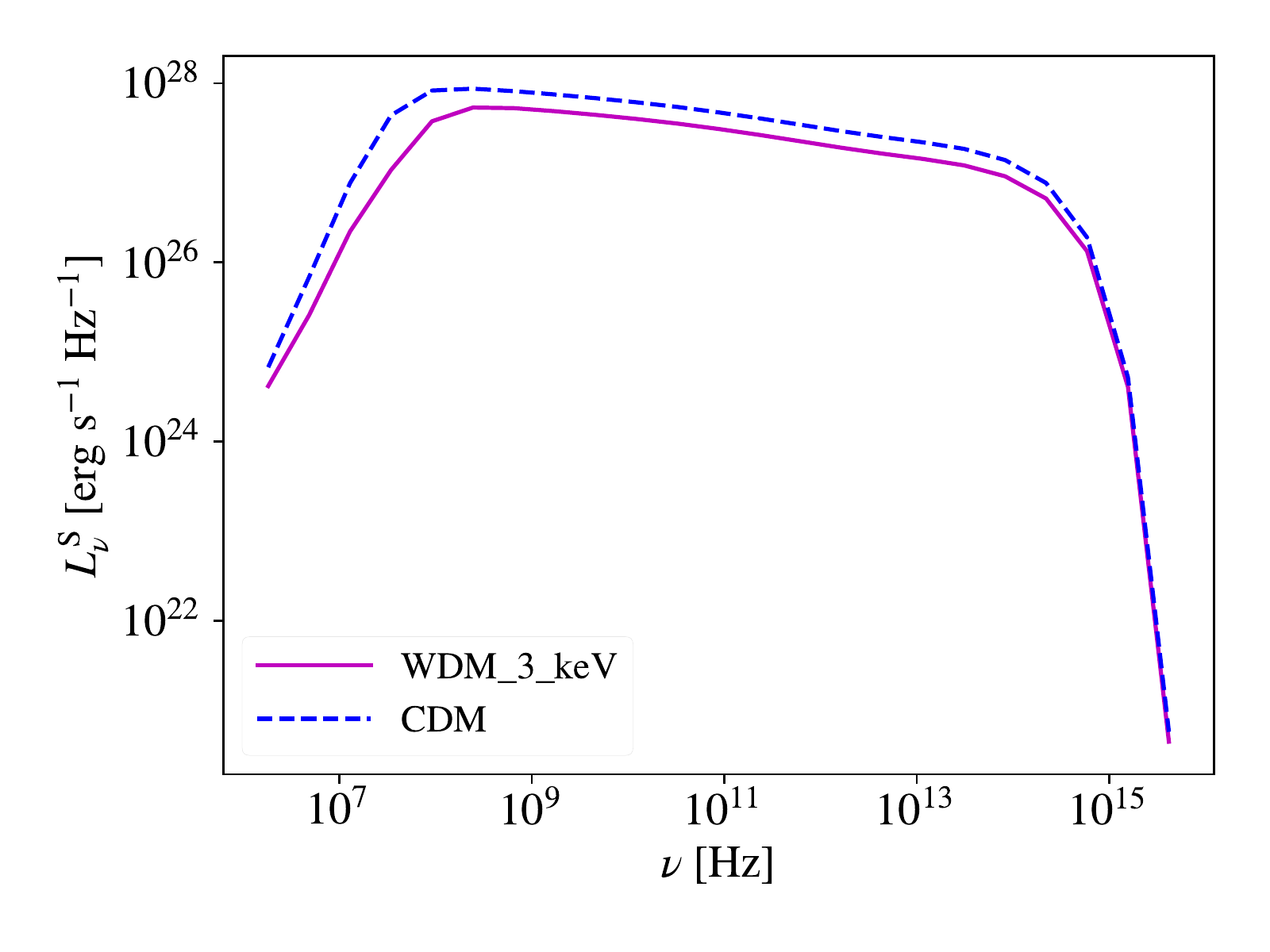}
\vspace{-15pt}
\caption{Specific luminosity of free-free emission as a function of rest-frame frequency, at redshift $z=8.5$, from Z\_sfdbk.}
\label{f0}
\end{figure}

\begin{figure}
\hspace{-10pt}
\includegraphics[width= 1.05\columnwidth]{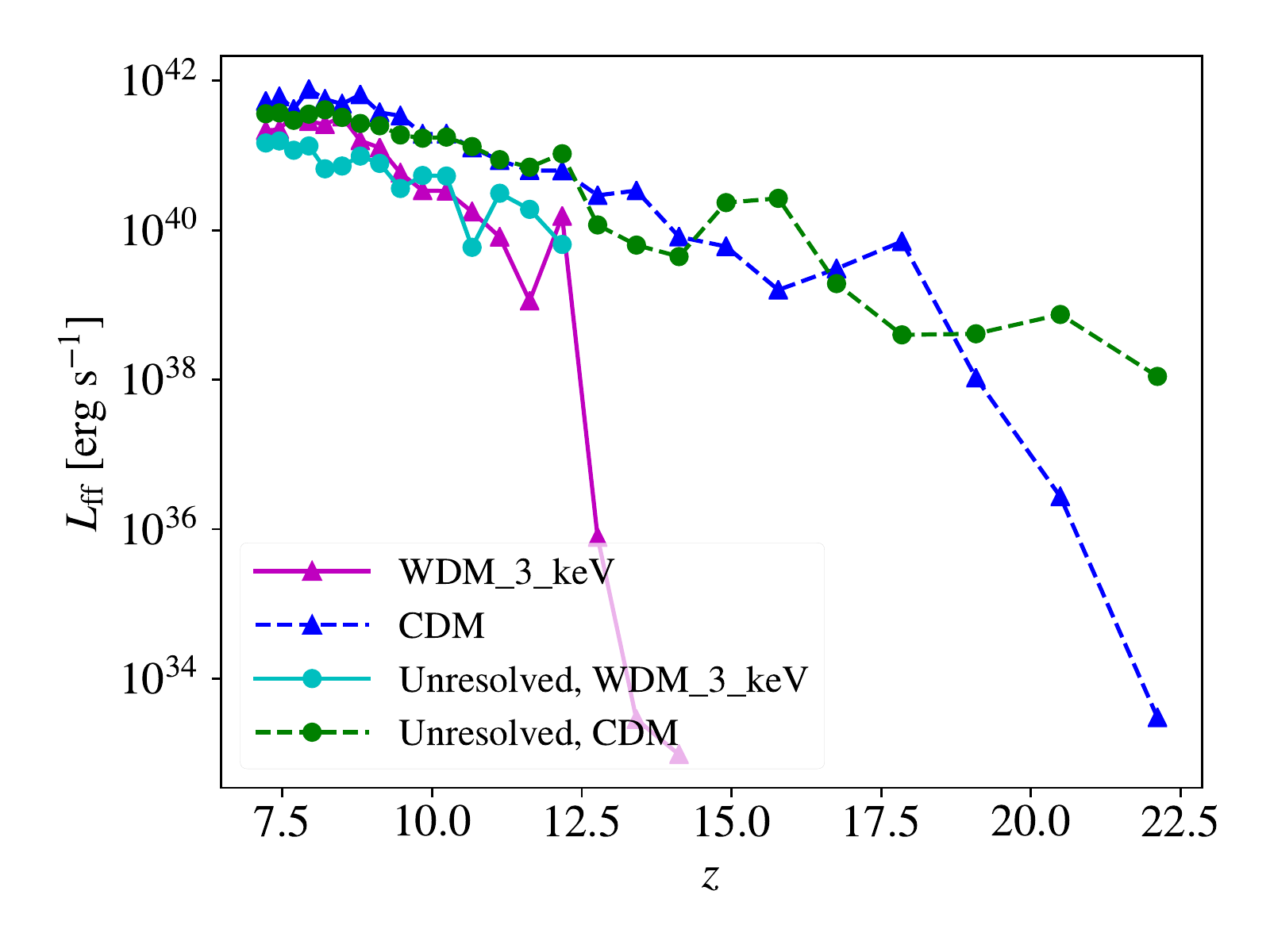}
\vspace{-15pt}
\caption{Evolution of the integrated free-free emission luminosity with redshift, in Z\_sfdbk. Results for gas particles are marked with triangles, while those for unresolved gas in newly-formed stellar particles with squares. The signal for the WDM cosmology is weaker than that for CDM by a factor of 10 (5) at redshift $z\sim 11.5$ (9.5), but catches up at $z\lesssim 8.5$. }
\label{f13}
\end{figure}

To calculate the free-free emission from the sample zoom-in region, we first {\it (i)} map the Bremsstrahlung emissivities of individual gas particles (with $T\ge 10^{4}\ \mathrm{K}$ and $n\le 500\ \mathrm{cm^{-3}}$) onto a 3-D grid of $100\times 100\times 100$ cells, covering the central zone, with the standard clouds-in-cells (CIC) method. We have verified that the contribution from gas with $T<10^{4}$~K is negligible. The upper limit of density $500\ \mathrm{cm^{-3}}$ is chosen to include the gas in typical \HII\ regions, and meanwhile exclude the unphysical hot dense gas produced by our legacy feedback model (see Section~\ref{s3.1} for details). We then {\it (ii)} perform radiative transfer along the direction of the $z$ axis, chosen to be the line-of-sight direction. Based on the emissivity and mass-weighted average gas temperature on the grid, we integrate the radiative transfer equation
\begin{align}
\frac{dI_{\nu}}{dz}=k_{\nu}I_{\nu}+j_{\nu}\ ,\label{e1}
\end{align}
for each 2-D cell in the upper surface of the cubic central zone, which is a rectangular area $A_{\mathrm{c}}$, defined by $1750<x<2250\ h^{-1}\mathrm{kpc}$ and $1750<y<2250\ h^{-1}\mathrm{kpc}$ in the $xy$ plane, at $z=2250\ h^{-1}\mathrm{kpc}$. Here $I_{\nu}$ is the (rest-frame) specific intensity, and the boundary condition is $I_{\nu}(z=1750\ h^{-1}\mathrm{kpc})\equiv 0$. For each 3-D cell for a volume $V_{\mathrm{cell}}$, the emission coefficient is approximated as $j_{\nu}=\sum_{p} w_{p}V_{p}\epsilon_{\mathrm{ff},p}\exp[-h\nu/(k_{B}T)] g_{\mathrm{ff}}(\nu,T)/(4\pi V_{\mathrm{cell}})$, while the absorption coefficient $k_{\nu}$ is obtained from Kirchhoff's law under the assumption of local thermodynamic equilibrium (LTE). Here, the index $p$ goes over all gas particles that overlap with the cell, whose mass-weighted average temperature is $T=\sum_{p} w_{p}T_{p}$, where $w_{p}$ and $V_{p}=m_{p}/\rho_{p}$ are the CIC (mass-)weight and effective volume for particle $p$. Finally, $\epsilon_{\mathrm{ff}}=\epsilon_{0}n_{e}n_{i}T^{-1/2}$ is the Bremsstrahlung emissivity, and $g_{\mathrm{ff}}(\nu,T)$ the Gaunt factor \citep{RPA}, where $\epsilon_{0}=6.84\times 10^{-38}$ in c.g.s. units. 

We can thus obtain the specific intensity map $I_{\nu}$ on $A_{\mathrm{c}}$ for any rest-frame frequency $\nu$, from which the (simulated) rest-frame specific luminosity can be derived by integrating the intensity across the projected area as $L^{\mathrm{S}}_{\nu}= 4\pi\int_{A_{\mathrm{c}}} I_{\nu}dxdy$. For instance, Fig.~\ref{f00} shows the specific intensity map at $\nu=0.1\ \mathrm{GHz}$ for both DM models, and Fig.~\ref{f0} the luminosity as a function of rest-frame frequency, at redshift $z=8.5$, for the Z\_sfdbk runs. The spectra are almost flat in the frequency range $0.1-10^{6}\ \mathrm{GHz}$. According to standard Bremsstrahlung theory, the higher-frequency cut-off arises from the exponential term $\exp[-h\nu/(k_{B}T)]$ as $\nu_{\rm max}=k_{B}T/h\sim 10^{6}$~GHz for $T\sim 2\times 10^{4}$~K, and the lower cut-off is due to optical depth effects. It is straightforward to show that for a uniform isothermal sphere of radius $R$, with a temperature $T$, an ionization fraction $x_{\mathrm{e}}$ and a hydrogen number density of $n_{\mathrm{H}}$, assuming LTE conditions, the optical depth of free-free emission exceeds unity, when $\nu<\nu_{\rm min}\simeq 2\mathrm{GHz}\cdot (R/100\ \mathrm{pc})^{1/2}(x_{\mathrm{e}}n_{\mathrm{H}}/10^{3}\ \mathrm{cm^{-3}})[T/(2\times 10^{4}\ \mathrm{K})]^{-4/3}$. For the \HII\ regions around Pop~II stellar populations that produce the majority of free-free emission, $R\sim R_{\mathrm{ion}}\sim 100$~pc, $x_{\mathrm{e}}n_{\mathrm{H}}\sim 50\ \mathrm{cm^{-3}}$, and $T\sim 2\times 10^{4}$~K, such that $\nu_{\rm min}\sim 0.1$~GHz. We also plot the evolution of the integrated luminosity $L_{\mathrm{ff}}=\int L_{\nu}^{\mathrm{S}}d\nu$ with redshift in Fig.~\ref{f13}, again for Z\_sfdbk. To estimate the contribution from the unresolved high-density gas that is locked up in sink particles, we assume that the ISM inside is fully ionized and heated to $T=2\times 10^{4}$~K by OB stars for $10$~Myr, and that it has a number density of $\sim100\ \mathrm{cm^{-3}}$. It turns out that the luminosity from such unresolved sources is close to that from gas particles at late stages ($z\lesssim 12$), implying that our choices for the temperature and density thresholds are reasonable to describe the gas in \HII\  regions.

In our simulations, the free-free signal for the WDM model is weaker than for CDM by a factor of 10 (5) at $z\sim 11.5$ (9.5), which is due to the delayed structure formation, as reflected in the star formation histories (see Fig.~\ref{sfr}). The difference becomes smaller towards lower redshifts, and converges to a factor of 2.5 for $z\lesssim 8.5$. 
Interestingly, the difference in SFRD also converges to a factor of 2.5 at lower redshifts (see Fig.~\ref{sfr}), implying that free-free emission is strongly correlated with star formation rate (SFR).

With stellar feedback included, the free-free emission luminosity is $\sim 10^{3-4}$ times larger, compared with the prediction from Z\_Nsfdbk (not shown). This indicates that free-free emission is mostly powered by stellar feedback. 
Interestingly, at $z=7.7$ in the no-feedback case (Z\_Nsfdbk), we find $L_{\mathrm{ff}}\sim 10^{37}-10^{38}\ \mathrm{erg\ s^{-1}}$, while the virial luminosity is $L_{\mathrm{vir}}\sim 10^{41} \ \mathrm{erg\ s^{-1}}$, calculated from
\begin{align}
L_{\mathrm{vir}}=& \frac{3GM^{2}}{5R_{\mathrm{vir}}t_{\mathrm{ff}}}= 2.7\times 10^{41}\ \mathrm{erg\ s^{-1}}\notag \\
&\cdot \left(\frac{\Delta\Omega_{m}}{200\cdot 0.315}\right)^{5/6}\left(\frac{1+z}{10}\right)^{5/2}\left(\frac{M}{10^{10}\ M_{\odot}}\right)^{5/3}\ ,\label{e3}
\end{align}
where $t_{\mathrm{ff}}=\left[3\pi/(32G\rho_{\mathrm{crit},0}\Omega_{m}\Delta)\right]^{1/2}a^{3/2}$ is the free-fall timescale. Therefore, only $\sim 10^{-4}-10^{-3}$ of the gravitational potential energy during collapse is carried away by free-free emission\footnote{In Z\_Nsfdbk, we have neglected free-free emission from the dense gas incorporated by sink particles. Extrapolating the properties of the unresolved gas, we assume an average temperature of $\bar{T}=1000$~K, a number density of $\bar{n}=n_{\mathrm{th}}=100\ \mathrm{cm^{-3}}$, and an average degree of ionization of $\bar{x}_{\mathrm{e}}=10^{-4}$. We thus estimate that the unresolved gas would only contribute $\sim$0.1\% of the total emission.}. 
The majority of gravitational energy is converted into atomic hydrogen and $\mathrm{H_{2}}$ line emissions. 

\subsection{Contribution to the cosmic radio background}
\label{s4.2}

\subsubsection{General formalism}
Based on the above calculations, we can infer the contribution of free-free emission from early structure formation to the cosmic radio background (CRB). This complements studies of the cosmic background radiation in the near-infrared \citep[e.g.][]{helgason2016} and the far-infrared/sub-millimeter band \citep{DeRossi2017}, which are repositories for reprocessed starlight at $z\gtrsim 7$. In general, the observed background intensity $J_{\nu_{\mathrm{obs}}}(>z)$ from sources beyond redshift $z_{\mathrm{end}}$ is calculated by integrating the cosmic radiative transfer equation
\begin{align}
J_{\nu_{\mathrm{obs}}}(>z_{\mathrm{end}})=\int_{0}^{t_{H}(z_{\mathrm{end}})}\frac{j_{\nu}}{(1+z)^{3}}cdt\ ,\label{e4}
\end{align}
where $\nu=\nu_{\mathrm{obs}}(1+z)$ is the rest-frame frequency, $j_{\nu}\equiv j_{\nu}(z)$ the cosmic-average emission coefficient, and $t_{H}(z_{\mathrm{end}})$ the age of the Universe at redshift $z_{\mathrm{end}}$. 

In evaluating this integral, we for simplicity only consider the emission from DM haloes, even though shocks in the IGM can also produce free-free emission. Later, we will show that the IGM contribution is negligible. We divide $j_{\nu}$ into contributions from DM haloes of different masses as
\begin{align}
&j_{\nu}=\int_{M_{\rm min}}^{M_{\rm max}}\frac{dj_{\nu}}{dM}dM\ ,\label{e5}\\
&\frac{dj_{\nu}}{dM}=\frac{1}{4\pi} L_{\nu}t_{\star}\max\left[0,\frac{d \mathcal{N}}{d t}({z})\right](1+z)^{3}\ .\label{e6}
\end{align}
Here $L_{\nu}\equiv L_{\nu}(M, z)$ and $t_{\star}\equiv t_{\star}(M,z)$ are the typical (specific) luminosity and timescale of free-free emission for DM haloes with a virial mass $M$ at redshift $z$, whereas $\mathcal{N}=dn_{h}/dM$ is the halo mass function (the number of DM haloes per unit \textit{co-moving} volume per unit mass). Equation~(\ref{e6}) implies that the radiation energy from individual newly-born DM haloes is distributed across space and time to produce an effective emission coefficient, describing the time- and spatially-averaged state of radiation. Note that we neglect the effect of accretion and mergers on shaping the halo mass function to obtain Equ.~(\ref{e6}), where we assume that
\begin{displaymath}
\frac{\partial\mathcal{N}}{\partial t}=\frac{d\mathcal{N}}{d t}-\dot{M}\frac{\partial \mathcal{N}}{\partial M}\simeq \max\left[0,\frac{d\mathcal{N}}{d t}\right]\mbox{\,.}
\end{displaymath}
We can rewrite the cosmic radiative transfer equation (Equ.~\ref{e4}) in the form
\begin{align}
&J_{\nu_{\mathrm{obs}}}(>z)=\int_{\infty}^{z_{\mathrm{end}}}dz\int_{M_{\rm min}}^{M_{\rm max}}dM\frac{L_{\nu}t_{\star} c}{4\pi}\notag\\
&\hspace{60pt}\cdot \max\left[0,\frac{d\mathcal{N}}{d t}\right]\frac{dt}{dz}\notag\\
&=\int_{z_{\mathrm{end}}}^{\infty}dz\int_{M_{\rm min}}^{M_{\rm max}}dM\frac{L_{\nu}t_{\star} c}{4\pi}\max\left[0,-\frac{d\mathcal{N}}{dz}\right] \ ,\label{e7}
\end{align}
where the time evolution of the halo mass function $d\mathcal{N}/dz$ is evaluated with the \textsc{python} package \href{http://hmf.readthedocs.io/en/latest/index.html}{\texttt{hmf}} \citep{murray2013hmfcalc}, given the default fitting model from \citet{tinker2008toward} and WDM model from \citet{bode2001halo,viel2005constraining}. The minus sign in the second line comes from $dt/dz<0$. 

Now our task is to derive $L_{\nu}(M, z)$ and $t_{\star}(M,z)$, and to determine the mass range ($M_{\rm min}$ and $M_{\rm max}$), which may vary with $z$. From our zoom-in simulations, we obtain the free-free luminosity $L_{\nu}^{\mathrm{S}}$ of the target halo with a virial mass $M_{\mathrm{ref}}\simeq 10^{10}\ M_{\odot}$, formed at redshift $z\sim 10$ (see Fig.~\ref{f0} for an example at $z=8.5$). The free-free luminosity reaches and stays at a high level in the snapshots with $7.22\le z\le 10.24$. Therefore, we choose the time-averaged luminosity of the target halo in this redshift range as the reference luminosity $L_{\nu}^{\mathrm{ref}}=\eta\int_{t_{H}(z=10.24)}^{t_{H}(z=7.22)}L_{\nu}^{\mathrm{S}}dt/t_{\star,\mathrm{ref}}$, where $t_{\mathrm{\star,ref}}=\left[t_{H}(z=7.22)-t_{H}(z=10.24)\right]=272$~Myr is the reference timescale, and $\eta=2$ a boosting factor to take into account the emission from unresolved sources (see Fig.~\ref{f13}). 
For simplicity, we obtain the free-free luminosity and timescale for atomic cooling haloes from these reference values with simple power-law scalings (see below). We further assume that this normalization is independent of formation redshift, such that $L_{\nu}(M=M_{\mathrm{ref}},z)=L_{\nu}^{\mathrm{ref}}$ and $t_{\star}(M=M_{\mathrm{ref}},z)=t_{\star,\mathrm{ref}}$.

In the next three subsections, we construct $L_{\nu}(M, z)$ and $t_{\star}(M,z)$ for three groups of DM haloes, utilizing existing results in the literature, as well as our reference luminosity $L_{\nu}^{\mathrm{ref}}$ and timescale $t_{\mathrm{\star,ref}}$.

\subsubsection{Minihaloes}
\label{s4.2.2}
In minihaloes, free-free emission originates in the \HII\ regions around Pop~III stars, which can expand into the diffuse IGM and cool rapidly to a temperature $T\simeq 10^{3}$~K \citep{greif2009}. The corresponding free-free luminosity is
\begin{align}
L_{\nu}^{\mathrm{mini}}&=4\pi \epsilon_{0}n_{e}n_{i}V_{\text{\HII}}T^{-1/2}\simeq 4\pi \epsilon_{0}n_{\mathrm{bg}}^{2}V_{\text{\HII}}T^{-1/2}\notag\\
&=4\pi\epsilon_{0}N_{\mathrm{ion}}n_{\mathrm{bg}}T^{-1/2}\ .\label{e8}
\end{align}
Here $n_{\mathrm{bg}}=\rho_{\mathrm{crit, 0}}a^{-3}/(\mu m_{\mathrm{H}})$ is the background (IGM) number density of baryons ($\mu=1.22$ for neutral gas), $V_{\text{\HII}}\simeq N_{\mathrm{ion}}/n_{\mathrm{bg}}$ the \HII\ region volume, $N_{\mathrm{ion}}\propto M_{\star}=\epsilon_{\star}M$ the number of ionizing photons produced by Pop~III stars in the minihalo, $M_{\star}$ the stellar mass, and $\epsilon_{\star}$ the star formation efficiency. For simplicity, we assume that the star formation efficiency is a constant in minihaloes, so that $N_{\mathrm{ion}}\propto M$. Then, calibrating to the results of \citet{greif2009}, such that $N_{\mathrm{ion}}\simeq 2\times 10^{64}$ for $M\simeq 2\times 10^{6}\ M_{\odot}$, assuming $T\simeq 10^{3}$~K, the expression above can be rewritten as
\begin{align}
L_{\nu}^{\mathrm{mini}}(z,M)=&2.8\times 10^{22}\ \mathrm{erg\ s^{-1}\ Hz^{-1}}\notag\\
&\times \left(\frac{M}{10^{6}\ M_{\odot}}\right)\left(\frac{1+z}{10}\right)^{3}\ .\label{e9}
\end{align}
Note that this expression is only valid for minihaloes in the mass range $M_{1}\le M\le M_{2}$, where (e.g. \citealt{barkana2001beginning,yoshida2003simulations,trenti2009formation})
\begin{align}
M_{1}\equiv M_{1}(z)&\simeq 10^{6}\ M_{\odot}\left(\frac{1+z}{10}\right)^{-2}\ ,\\
M_{2}\equiv M_{2}(z)&\simeq 2.5\times 10^{7}\ M_{\odot}\left(\frac{1+z}{10}\right)^{-3/2}\ ,\label{e11}
\end{align}
as it assumes that the \HII\ regions are produced by Pop~III stars and can expand into the diffuse IGM. In more massive DM haloes with stronger gravity and higher virial temperatures, star formation is dominated by Pop~II stars and the \HII\ regions can be confined to have much higher electron/ion densities (see the $T$-$n$ phase diagrams in Fig.~\ref{pd1}). As a result, formula~(\ref{e9}) will generally underestimate the free-free luminosity for DM haloes with $M>M_{2}$. We further impose an exponential cut-off
\begin{align}
L_{\nu}(z,M)=L_{\nu}^{\mathrm{mini}}(z,M)\exp[-h\nu/(k_{B}T)]\ , \label{e12}
\end{align} 
to model the high-frequency truncation for $T=10^{3}$~K. The timescale for free-free emission in minihaloes is the recombination time in the associated \HII\ regions:
\begin{align}
t_{\star}(z, M)&=t_{\mathrm{rec}}(z)=\frac{1}{\alpha_{B}n_{\mathrm{bg}}}\notag\\
&=92\ \mathrm{Myr}\cdot\left(\frac{1+z}{10}\right)^{-3}\ ,\label{e13}
\end{align}
where $\alpha_{B}=2.6\times 10^{-13}\ \mathrm{cm^{3}\ s^{-1}}$ is the case B recombination coefficient for hydrogen \citep{greif2009}.

In our simulations, the number density of minihaloes in the WDM model is lower than that for CDM by a factor of $10-100$. However, it turns out that minihaloes only contribute $\sim 0.5$\% (1\%) of the total free-free signal in the WDM (CDM) model for $z_{\mathrm{end}}=6$. Thus, the huge difference in small-scale structures is not reflected in the CRB.

\subsubsection{Low-mass atomic cooling haloes}
\label{s4.2.3}
We define DM haloes with $M_{2}<M\le 10^{10}\ M_{\odot}$ as low-mass atomic cooling haloes. As mentioned above, minihaloes and haloes with $M\sim 10^{10}\ M_{\odot}$, simulated here, behave rather differently, due to the different conditions in their \HII\ regions. In the former case, \HII\ regions are unconfined, while in the latter case they remain confined, and the transition between them can be complex. For simplicity, we model this transition with power-law expressions, such that for $M_{2}<M\le 10^{10}\ M_{\odot}$, we have
\begin{align}
L_{\nu}(M, z)&=L_{\nu}^{\mathrm{ref}}\cdot\left(\frac{M}{M_{\mathrm{ref}}}\right)^{\beta_{M}}\cdot\exp[-h\nu/(k_{B}T)]\ ,\label{e14}\\
t_{\star}(M,z)&=t_{\star,\mathrm{ref}}\cdot\left(\frac{M}{M_{\mathrm{ref}}}\right)^{\beta_{t}}\ .\label{e15}
\end{align}
Here, $T=\max\{10^{3}\ \mathrm{K}\cdot T_{\mathrm{vir}}(M,z)/T_{\mathrm{vir}}[M_{2}(z),z],2\times 10^{4}\ \mathrm{K}\}$ is the estimated electron temperature, with the upper bound, $2\times 10^{4}$\,K, given by the typical temperature of \HII\ regions in our zoom-in simulations. The power-law indexes $\beta_{M}\equiv\beta_{M}(z)$ and $\beta_{t}\equiv\beta_{t}(z)$ are determined by continuity of $L_{\nu}(z,M)$ and $t_{\star}(M,z)$ as functions of $M$ at $M=M_{2}$, evaluated at $\nu=100$~GHz, such that
\begin{align}
L_{\nu=100\ \mathrm{GHz}}^{\mathrm{ref}}\cdot\left[\frac{M_{2}(z)}{M_{\mathrm{ref}}}\right]^{\beta_{M}(z)}&=L_{\nu}^{\mathrm{mini}}[ M=M_{2}(z),z]\ ,\notag\\
t_{\star,\mathrm{ref}}\cdot\left[\frac{M_{2}(z)}{M_{\mathrm{ref}}}\right]^{\beta_{t}(z)}&=t_{\mathrm{rec}}(z)\ .\label{e16}
\end{align}

\subsubsection{Massive haloes}
\label{s4.2.4}
For more massive haloes, with virial masses $M>10^{10}\ M_{\odot}$, we expect the free-free luminosity to be higher, but the spectrum will also be shifted to higher frequencies. For instance, the free-free emission from galaxy clusters takes the form of X-rays. 
Given that the evolution of $L_{\mathrm{ff}}$ closely mirrors that of the SFRD (see Fig.~\ref{sfr} and \ref{f13}), we assume that the free-free luminosity of these massive haloes ($M>10^{10}\ M_{\odot}$) is proportional to the SFR $\dot{M}_{\star}$. We again normalize to the reference spectrum $L_{\nu}^{\mathrm{ref}}$ at $M=M_{\mathrm{ref}}=10^{10}\ M_{\odot}$. \citet{Mirocha2018} argue that the SFR in the mass range $10^{10}-10^{12}\ M_{\odot}$ satisfies $\dot{M}_{\star}\propto M^{5/3}$. We then have
\begin{align}
L_{\nu}(z,M)=\begin{cases} L_{\nu}^{\mathrm{ref}}\cdot\left(\frac{10^{10}\ M_{\odot}}{M_{\mathrm{ref}}}\right)^{5/3}\ ,\quad \hspace{2pt}\nu\ge\nu_{\rm min} \\
L_{\nu=\nu_{\rm min}}\cdot\left(\frac{\nu}{\nu_{\rm min}}\right)^{-2}\ ,\quad \nu<\nu_{\rm min}\ ,\end{cases}\label{e17}
\end{align}
valid for haloes with $M>10^{10}\ M_{\odot}$. 
Here we have truncated the spectrum at low frequencies to model the spectral shift. Below the truncation frequency $\nu_{\rm min}\equiv\nu_{\rm min}(M)=\nu_{\rm min,\mathrm{ref}}\left[R_{\mathrm{vir}}(M)/R_{\mathrm{vir}}(M_{\mathrm{ref}})\right]^{1/2}=\nu_{\rm min,\mathrm{ref}}\left(M/M_{\mathrm{ref}}\right)^{1/6}$, the system is optically thick, and the spectrum approaches the black-body form, with $\propto \nu^{-2}$ under the Rayleigh-Jeans approximation. Specifically, $\nu_{\rm min,\mathrm{ref}}\simeq 0.1$~GHz is the truncation frequency for the reference spectrum $L_{\nu}^{\mathrm{ref}}$. The expression for $\nu_{\rm min}(M)$ derives from the fact that $\nu_{\rm min}\propto R^{1/2}T^{-4/3}(n_{e}n_{i})^{1/2}$ \citep{RPA}, where $R$ is the characteristic size of the system, which in our case is assumed to be proportional to the virial radius of the DM halo. We here further assume that the \HII\ region properties (their $T$, $n_{e}$ and $n_{i}$) are approximately the same for these massive haloes. Since the free-free emission is predominantly powered by stellar feedback, we can estimate its timescale with the star formation timescale, such that $t_{\star}=\epsilon_{\star} M_{\mathrm{baryon}}/\dot{M}_{\star}=\epsilon_{\star}(\Omega_{b}/\Omega_{m})M/\dot{M}_{\star}$. According to \citet{Mirocha2018}, for DM haloes in the range $10^{10}-10^{12}\ M_{\odot}$, $\epsilon_{\star}\propto M^{2/3}$ and $\dot{M}_{\star}\propto M^{5/3}$, so that $t_{\star}$ is a constant. Therefore, we set
\begin{align}
t_{\star}(M,z)=t_{\star}(M=10^{10}\ M_{\odot},z)\ ,\quad M>10^{10}\ M_{\odot}\ .\label{e18}
\end{align}
Actually, under the above assumptions and approximations, the final $J_{\nu_{\mathrm{obs}}}$ result is not sensitive to $M_{\rm max}$, as long as it is sufficiently large, since massive haloes are rare in the early Universe. For definiteness, we choose $M_{\rm max}=10^{12}\ M_{\odot}$, and have verified that the contribution from more massive haloes ($M>10^{12}\ M_{\odot}$) is indeed negligible ($<4$\%) for $z_{\mathrm{end}}=6$.

\subsubsection{Comparison with other models}

\begin{figure}
\hspace{-10pt}
\includegraphics[width=1.05\columnwidth]{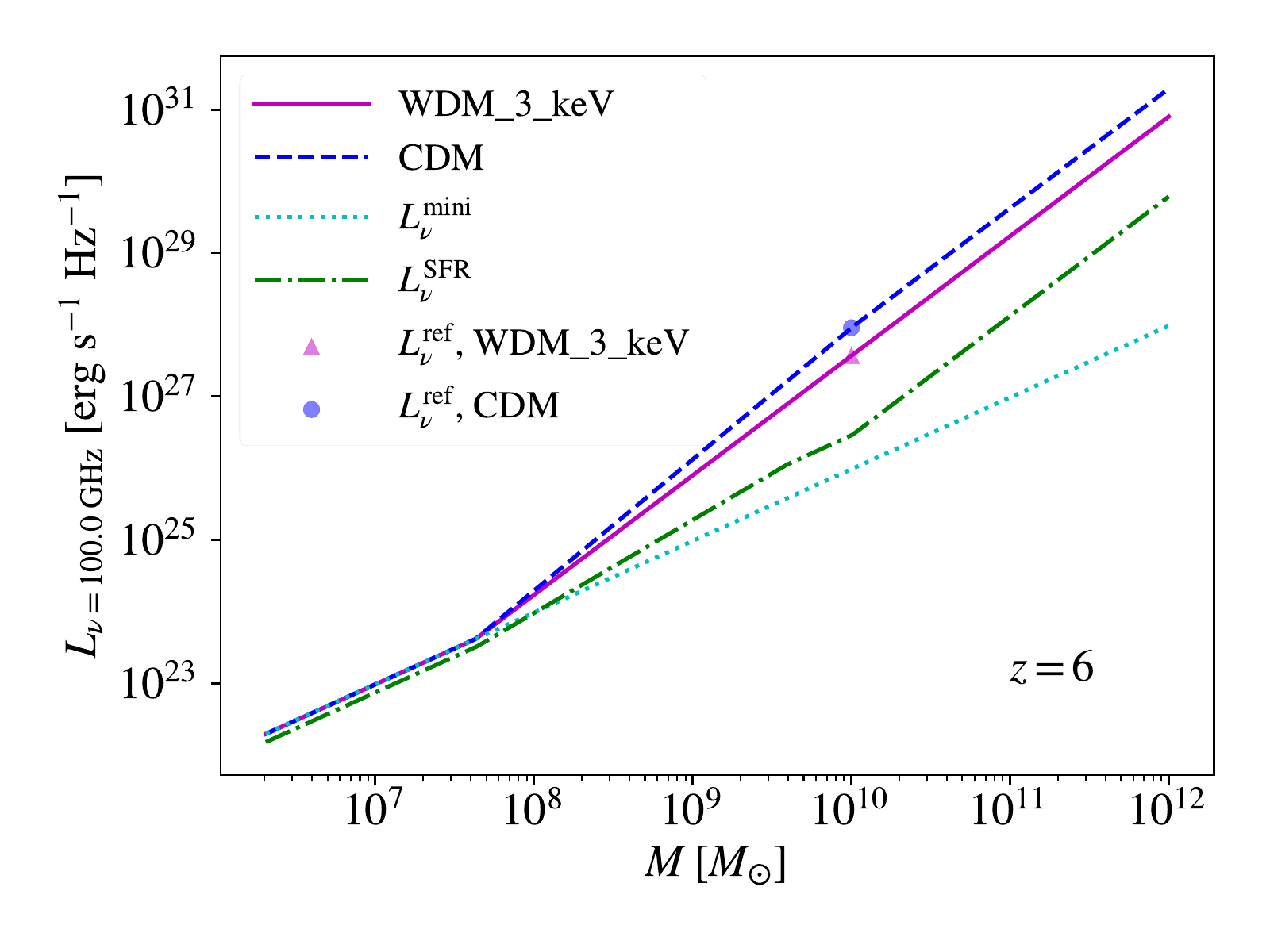}
\vspace{-25pt}
\caption{Typical free-free luminosity from star-forming haloes as a function of virial mass at $z=6$, for $\nu=100$~GHz, in the WDM (solid) and CDM (dashed) models. The respective reference luminosities are shown with a triangle and a square. The extrapolated minihalo and SFR-based luminosities (Equ.~\ref{e9} and \ref{e19}) are also plotted with dotted and dashed-dotted lines, respectively.}
\label{Lnu_M}
\end{figure}

Finally, we obtain the typical luminosity of free-free emission $L_{\nu}(M, z)$ by combining formulae~(\ref{e12}), (\ref{e14}) and (\ref{e17}), as well as the typical timescale $t_{\star}(M,z)$ from formulae~(\ref{e13}), (\ref{e15}) and (\ref{e18}). 
An example of $L_{\nu}(M, z)$ is shown in Fig.~\ref{Lnu_M} 
for $z=6$, in comparison with an extrapolation of the minihalo luminosity $L_{\nu}^{\mathrm{mini}}$, and a model for present-day galaxies, based on the relation between radio luminosity and SFR. Assuming solar metallicity and continuous star formation, this relation is \citep{murphy2011calibrating}:
\begin{align}
L_{\nu}^{\mathrm{SFR}}=& 2.2\times 10^{27}\ \mathrm{erg\ s^{-1}\ Hz^{-1}}\notag\\
&\cdot \left(\frac{T}{10^{4}\ \mathrm{K}}\right)^{0.45}\left(\frac{\nu}{1\ \mathrm{GHz}}\right)^{-0.1}\left(\frac{\mathrm{SFR}}{M_{\odot}\ \mathrm{yr^{-1}}}\right)\ ,\label{e19}
\end{align}
where ${\rm SFR} = f(M,z)$, and $T$ is the electron temperature (see Subsection~\ref{s4.2.3}). For the SFR, we here adopt the result from \citet{Mirocha2018} based on the EDGES signal (see their Fig.~4): 
\begin{align}
{\rm SFR} = \begin{cases} 3\times 10^{-3}\cdot \left(\frac{1+z}{11}\right)^{3/2}\left(\frac{M}{10^{8}M_{\odot}}\right),\quad M\le 10^{10} M_{\odot}\\
0.3\cdot \left(\frac{1+z}{11}\right)^{3/2}\left(\frac{M}{10^{10}M_{\odot}}\right)^{5/3},\ 10^{10}<\frac{M}{M_{\odot}}\le 10^{12}   \ ,\end{cases}\label{e20}
\end{align}
given in units of $M_{\odot}\ \mathrm{yr^{-1}}$. The SFR-based model~(Equ.~\ref{e19}) generally underestimates the free-free luminosity, compared with our model at $z=6$, by about an order of magnitude, implying that high-redshift ($z\gtrsim 6$) galaxies are more efficient at producing free-free emission than present-day galaxies. 

\subsubsection{Results}

\begin{figure}
\hspace{-10pt}
\includegraphics[width=1.05\columnwidth]{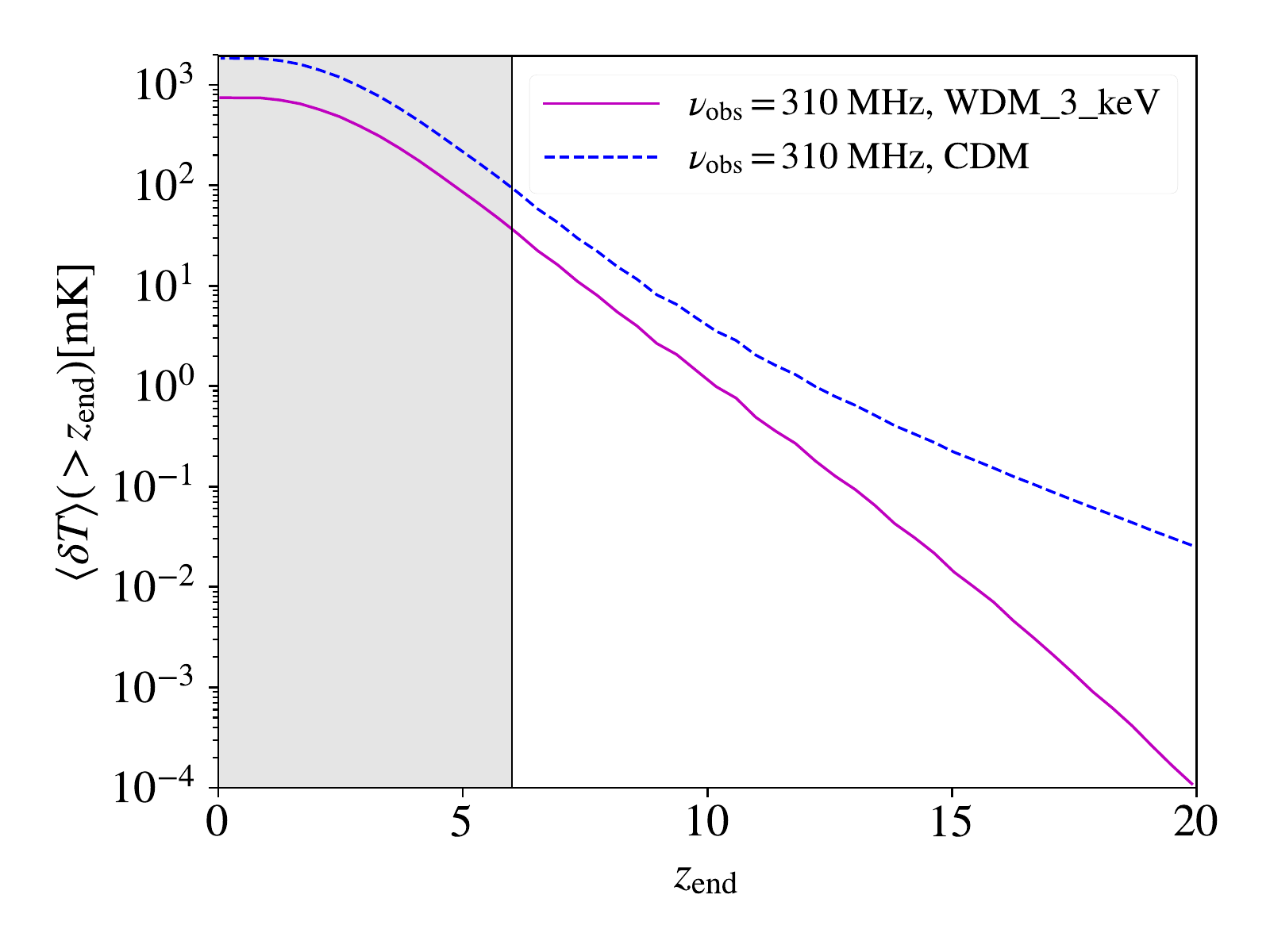}
\vspace{-25pt}
\caption{Contribution of free-free emission from structure formation to the CRB brightness temperature at $\nu_{\mathrm{obs}}=310$~MHz, as a function of the termination redshift of Pop~II star formation, from Z\_sfdbk. The shaded region denotes the post-reionization epoch.}
\label{f20}
\end{figure}

\begin{figure*}
\hspace{-10pt}
\subfloat[]{\includegraphics[width= 1.065\columnwidth]{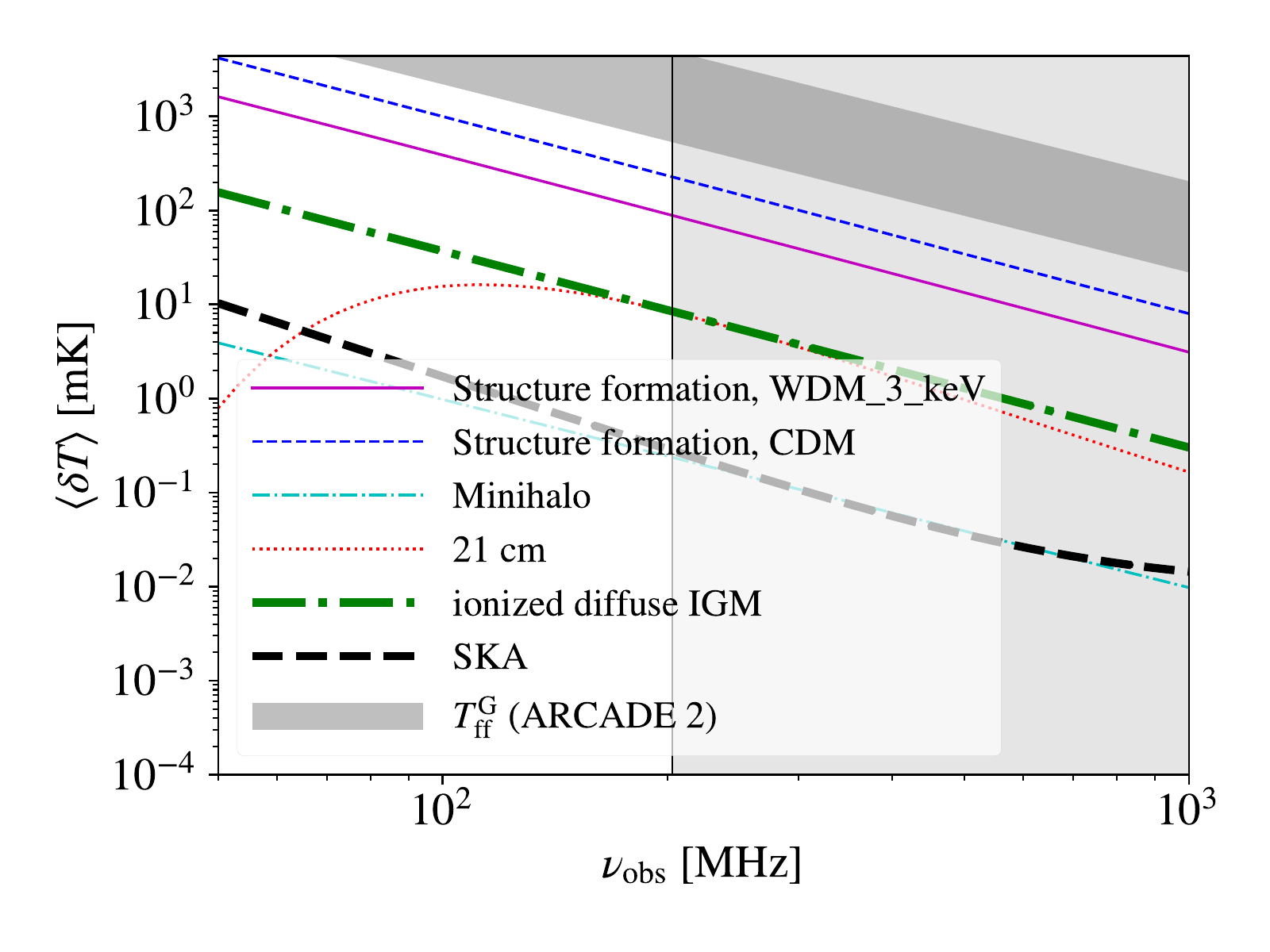}}
\hfill
\subfloat[]{\includegraphics[width= 1.065\columnwidth]{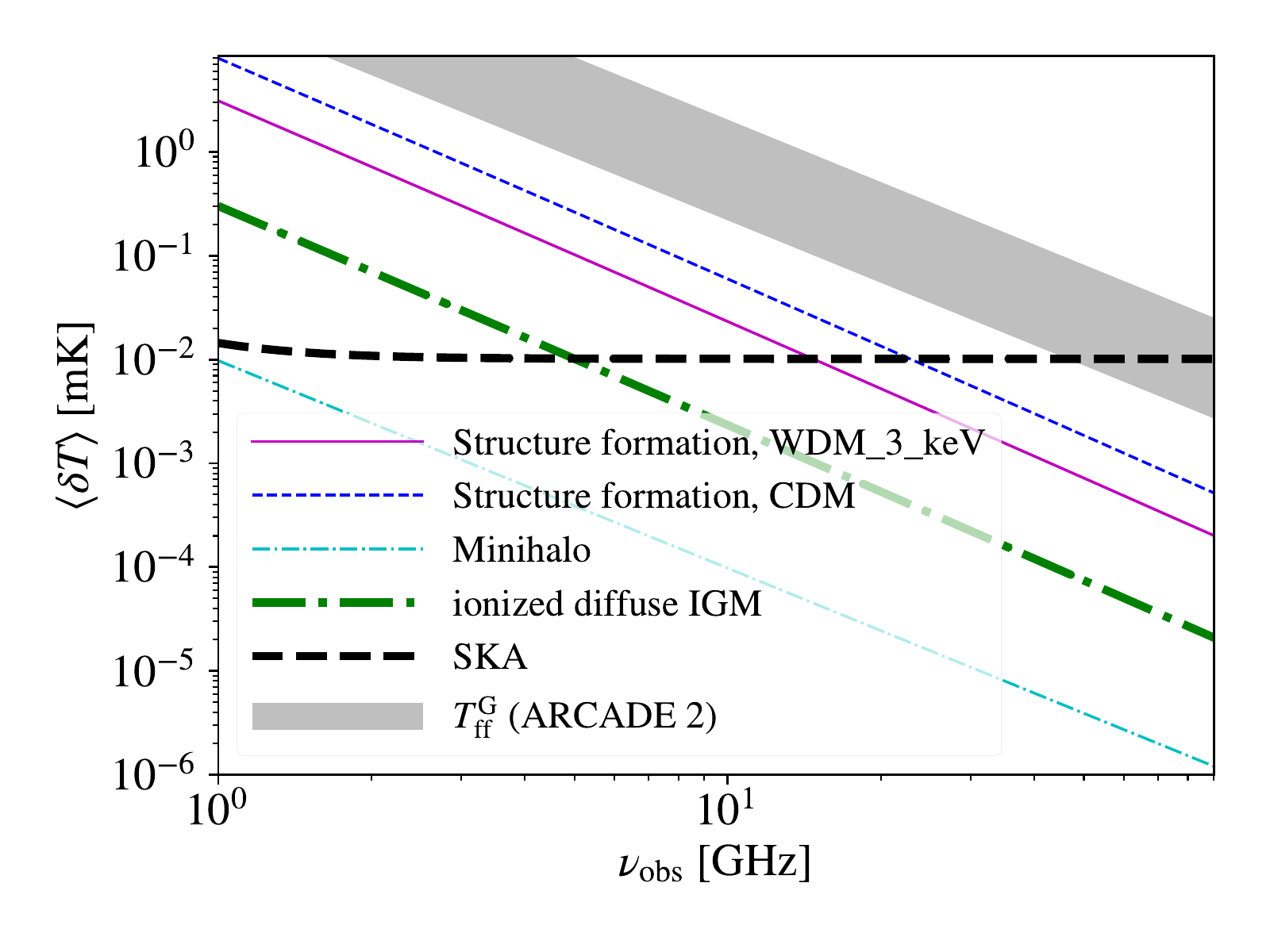}}
\vspace{-10pt}
\caption{Contribution of free-free emission from early structure formation to the CRB brightness temperature (thin solid and dashed lines for the WDM and CDM model, respectively). Our Z\_sfdbk results for $z_{\mathrm{end}}=6$ are  compared with contributions from minihaloes and the diffuse IGM after reionization (dashed-dotted and dotted, respectively). In panel (a), we show the low-frequency range, and in (b) the high-frequency one. The thick dashed curves represent the $10\sigma$ SKA detection limit for a $1000$~h integration and a bandwidth of $\Delta\nu_{\mathrm{obs}}=1$~MHz. The darker shaded regions denote the $\pm 1\sigma$ range of the CRB free-free component, $T^{\mathrm{G}}_{\mathrm{ff}}[\mathrm{mK}]=(1180\pm 952)(\nu_{\mathrm{obs}}/310\ \mathrm{MHz})^{-2}$, inferred from the data in \citet{arcade}. The curve for 21-cm line emission from relic \HII\ regions around first stars in minihaloes is not meaningful in the post-reionization epoch (lighter shaded), as gas in minihaloes will be evaporated by ionizing photons. The contribution from early structure formation dominates among the radio sources considered here. The free-free signal from early structure formation equals $3_{-1.5}^{+13}$\% ($8_{-3.5}^{+33}$\%) of the free-free component in the CRB, within the WDM (CDM) model.}
\label{f15}
\end{figure*}

In practice, we carry out the integration (Equ.~\ref{e7}) over the redshift range $z_{\rm end}\le z\le 30$, and evaluate the overall contribution from the sources considered above in terms of the enhancement in the background brightness temperature 
\begin{align}
\langle\delta T\rangle(>z_{\mathrm{end}})=\frac{c^{2}}{2k_{B}\nu_{\mathrm{obs}}^{2}}J_{\nu_{\mathrm{obs}}}(>z_{\mathrm{end}})\ . 
\end{align}
The total brightness temperature is dominated by the cosmic microwave background (CMB) and the Galactic/total synchrotron component $T_{\mathrm{sky}}\ [\mathrm{mK}]=2725\pm 1 + (24100\pm 2100)(\nu_{\mathrm{obs}}/310\ \mathrm{MHz})^{-2.599\pm 0.036}$, measured by experiments such as ARCADE 2 \citep{arcade}. 
We fit the same dataset (with 14 data points) and covariance matrix from \citet{arcade} with a new model including the contribution of free-free emission from high-$z$ sources as $T_{\mathrm{ff}}^{\mathrm{G}}\propto \nu_{\mathrm{obs}}^{-2}$, which leads to $T'_{\mathrm{sky}}\ [\mathrm{mK}]=2725.3\pm 0.1 + (21737 \pm 2809)(\nu_{\mathrm{obs}}/310\ \mathrm{MHz})^{-2.642\pm 0.037}+(1180\pm 952)(\nu_{\mathrm{obs}}/310\ \mathrm{MHz})^{-2}$, roughly consistent with the result $T_{\mathrm{ff}}^{\mathrm{G}}(310\ \mathrm{MHz})= 600\pm 60\ \mathrm{mK}$ in \citet{kogut2011}\footnote{\citet{kogut2011} found that free-free emission contributes $0.10\pm 0.01$ of the total signal in the lowest ARCADE 2 band at 3.15~GHz, while the total signal at 3.15~GHz is 58.2~mK in terms of brightness temperature \citep{arcade}. Therefore, the measured free-free emission is $T^{\mathrm{G}}_{\mathrm{ff}}[\mathrm{mK}]=(5.82\pm 0.582)(\nu_{\mathrm{obs}}/3.15\ \mathrm{GHz})^{-2}=(600\pm 60)(\nu_{\mathrm{obs}}/310\ \mathrm{MHz})^{-2}$.}. The errors denote the range of values for which the $\chi^{2}_{\mathrm{dof}}=\chi^{2}/\mathrm{dof}$ is within one $\sigma(\chi^{2}_{\mathrm{dof}})=\sqrt{2/\mathrm{dof}}$ from the (smallest) $\chi^{2}_{\mathrm{dof}}$ of the best fit. 
However, for this new model with free-free emission, the best-fit $\chi^{2}_{\mathrm{dof}}=16.87/(14-4)=1.687\sim 1+ 1.54\sigma(\chi^{2}_{\mathrm{dof}})$, while for the original model without a free-free component, the best-fit $\chi^{2}_{\mathrm{dof}}=17.4/(14-3)=1.582\sim 1+ 1.36\sigma(\chi^{2}_{\mathrm{dof}})$, which implies that the fitting becomes worse when free-free emission is considered. 
Therefore, it is necessary to point out that the free-free component in the CRB inferred from current observational data is highly uncertain (with a relative uncertainty of at least 80\%). %

Note that our estimation of $\langle\delta T\rangle=\langle\delta T\rangle(>z_{\mathrm{end}})$ is sensitive to the lower integration limit, $z_{\mathrm{end}}$, as shown in Fig.~\ref{f20}. For $\nu_{\mathrm{obs}}=310\ \mathrm{MHz}$, the signal becomes $\langle\delta T\rangle \simeq 1885\ (763)\ \mathrm{mK}$ at $z_{\mathrm{end}}\simeq 0$, within the CDM (WDM) cosmology, which is of the same order of magnitude with the observed best-fit CRB free-free signal of $1180$~mK. We find that the difference between the dark models decreases from above two orders of magnitude at $z_{\mathrm{end}}\gtrsim 20$ to less than a factor of 10 for $z_{\mathrm{end}}\lesssim 13$, with an almost constant ratio $\simeq$2.5 at $z_{\mathrm{end}}\lesssim 7$. The reason is that such global radiation signature is dominated by massive structures with $M>10^{8}\ M_{\odot}$, where deviations between WDM and CDM become less important. Thus the difference can only be significant at early epochs, when small-scale structures start to form in the CDM cosmology, while haloes have not yet collapsed in the WDM model. Henceforth, we assume $z_{\mathrm{end}}=6$, approximating the redshift below which reionization will significantly suppress star formation in the low-mass haloes considered here\footnote{This suppression is not directly captured in our simulations, as we do not self-consistently model the reionization process.}.

We compare our results with the contributions of free-free and 21-cm line emissions from (relic) \HII\ regions around the first stars, with a mass of $M_{\star}\simeq 100\ M_{\odot}$, formed in minihaloes \citep{greif2009}. In addition, we consider the free-free emission from the diffuse, ionized IGM after reionization ($z<6$). Note that our calculation of this latter contribution assumes that the IGM is uniform and isothermal with a temperature $T_{\mathrm{IGM}}=2\times 10^{4}$~K, and that reionization is instantaneous at $z=6$, which renders it a rough estimation. According to the detailed calculations from \citet{cooray2004} at $\nu_{\mathrm{obs}}=2\ \mathrm{GHz}$, our results are lower by up to a factor of 10. 
We note that among the radio sources considered here, the contribution from early structure formation is the dominant one.

In Fig.~\ref{f15}, we display the results from Z\_sfdbk for $z_{\mathrm{end}}=6$, together with the $10\sigma$ SKA detection limit for an integration time of $1000$~h and a bandwidth $\Delta\nu_{\mathrm{obs}}=1$~MHz, assuming that the foreground signal, $T_{\mathrm{sky}}$, dominates the antenna temperature, for (a) the low-frequency range ($50-1000\ \mathrm{MHz}$), and (b) the high-frequency range ($1-90\ \mathrm{GHz}$). Generally speaking, all the signals that we consider can be detected by SKA, except for the free-free emission from \HII\ regions around the first-stars in minihaloes, which can only be marginally detected at $\nu_{\mathrm{obs}}\sim 400\ \mathrm{MHz}$. However, it is not trivial to separate them, especially for the signals from structure formation and the diffuse ionized IGM, given their similar spectra. In the CDM model with $z_{\mathrm{end}}=6$, the predicted signal from early structure formation, $\langle\delta T\rangle(310\ \mathrm{MHz})=93.7$~mK, accounts for $8_{-3.5}^{+33}$\% of the observationally inferred signal of $T_{\mathrm{ff}}^{\mathrm{G}}(310\ \mathrm{MHz})$.
For WDM, on the other hand, we find $\langle\delta T\rangle(310\ \mathrm{MHz})=36.5\ \mathrm{mk}$, corresponding to  $\sim 3_{-1.5}^{+13}$\% of the measured value, lower by a factor of 2.5 in comparison to the CDM case.

\subsection{Molecular hydrogen emission}
\label{s4.3}

\begin{figure}
\hspace{-10pt}
\includegraphics[width= 1.05\columnwidth]{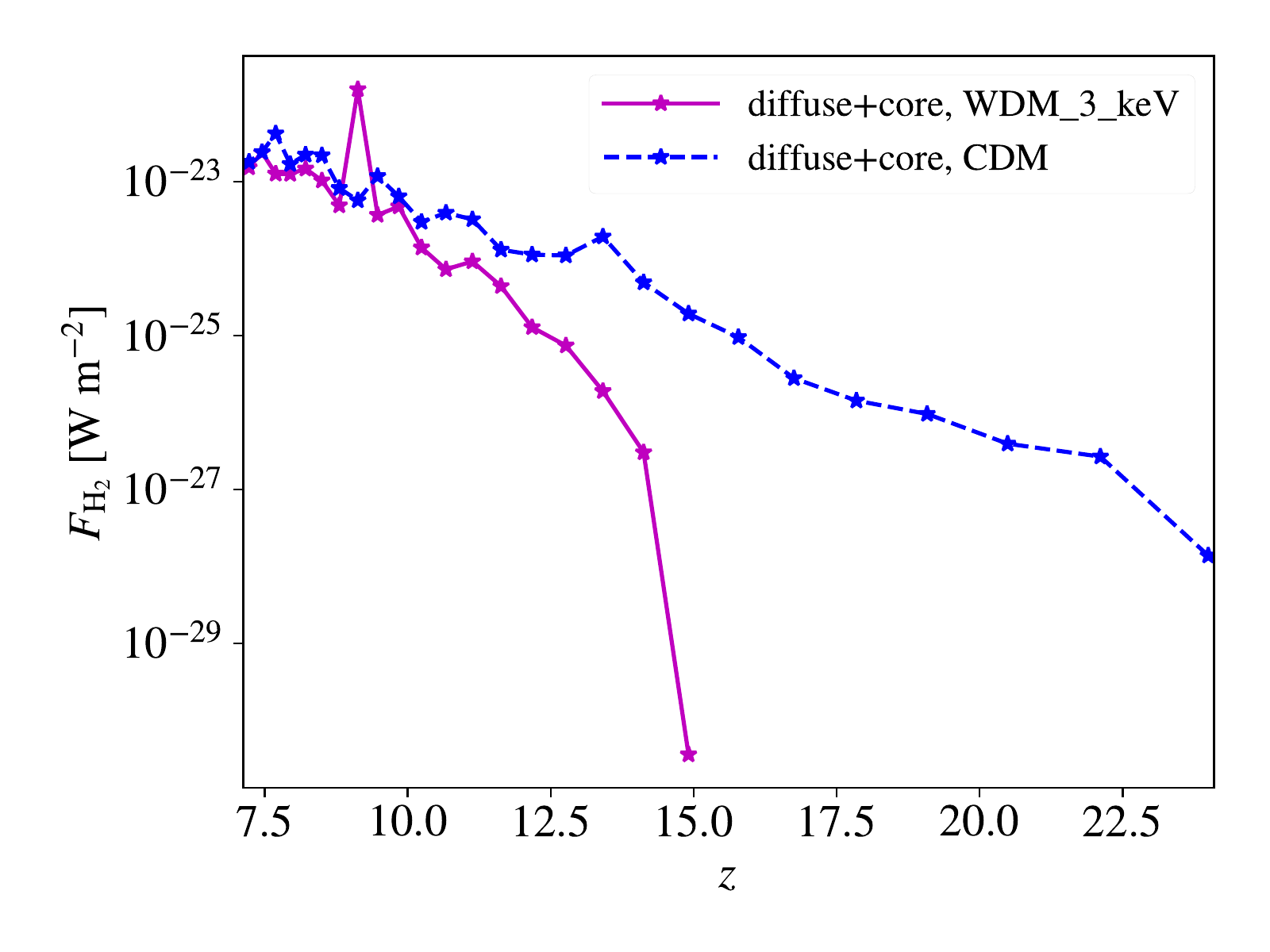}
\vspace{-15pt}
\caption{Overall $\mathrm{H_{2}}$ flux as a function of redshift for Z\_sfdbk. The overall flux in the WDM cosmology is lower than that in the CDM model by at least one order of magnitude for $z\gtrsim 12$, but reaches almost the same level at late stages ($z\lesssim 8$). For both models, the overall flux is $F_{{\rm H}_2}\sim 10^{-23} \mbox{\,W\,m}^{-2}$ at redshifts $7.2\lesssim  z\lesssim 8$, below the projected 5$\sigma$ 10-h sensitivity for 10m-class, cooled telescopes, $F_{\rm th}\simeq 10^{-22} \mbox{\,W\,m}^{-2}$, currently under development (see the text for further discussion).}
\label{f17}
\end{figure}

\begin{figure*}
\hspace{-15pt}
\subfloat[Z\_sfdbk]{
\includegraphics[width= 1.05\columnwidth]{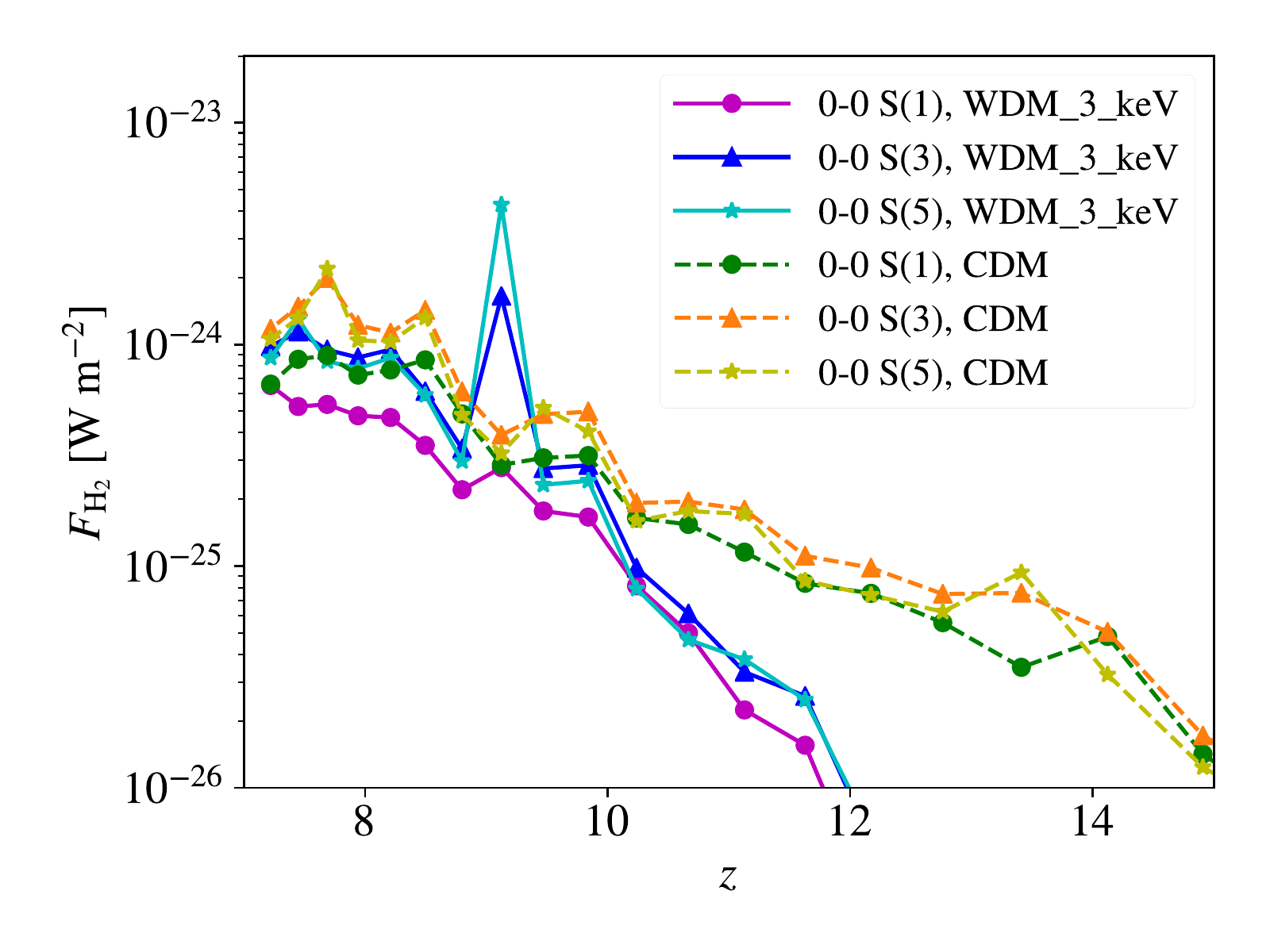}}
\vspace{15pt}
\subfloat[Z\_Nsfdbk]{
\includegraphics[width= 1.05\columnwidth]{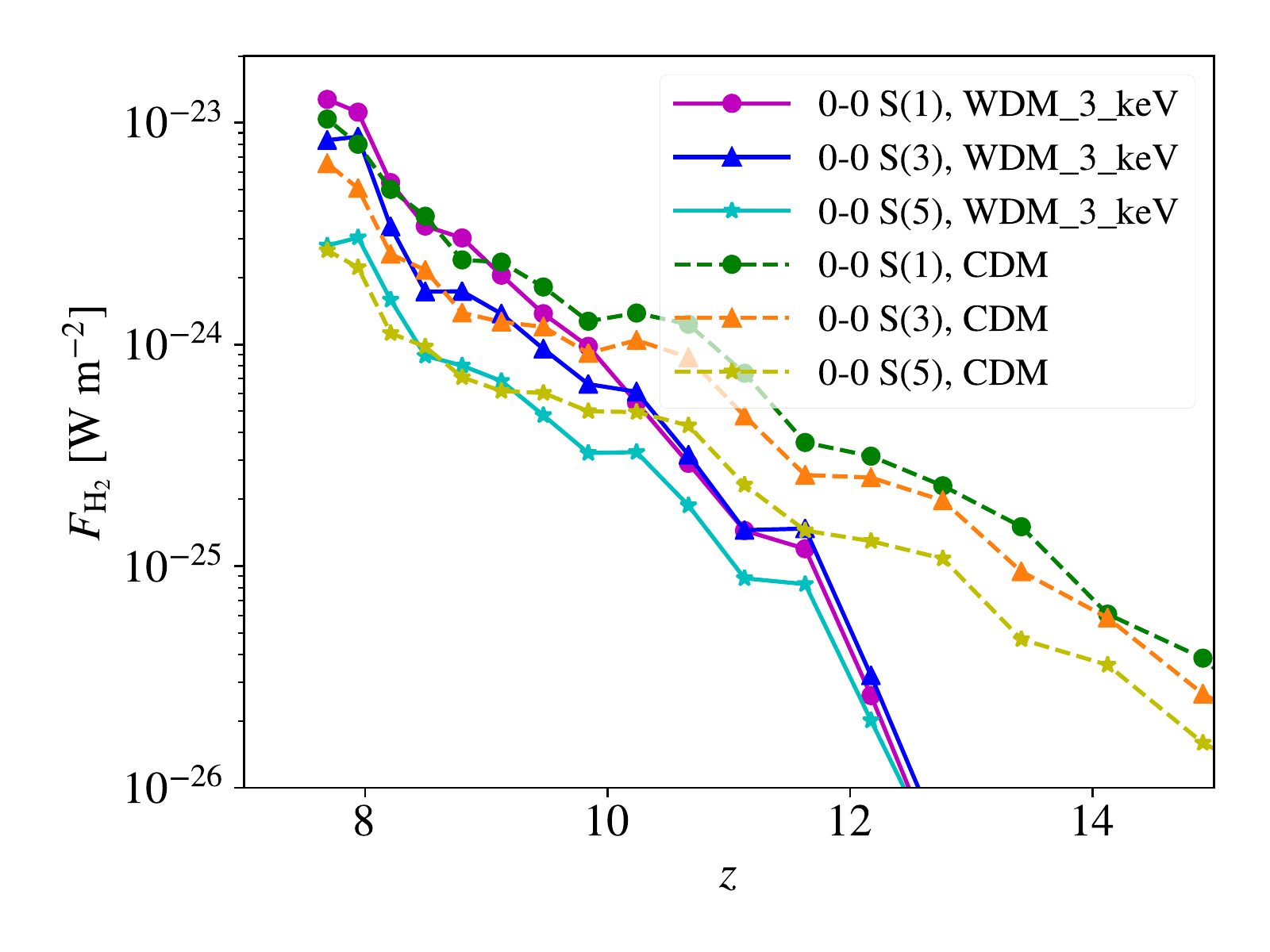}}
\vspace{-20pt}
\caption{Evolution of $\mathrm{H_{2}}$ fluxes from diffuse gas with redshift, for pure rotational lines 0-0 S(1) (solid circles), 0-0 S(3) (triangles) and 0-0 S(5) (stars), in (a) Z\_sfdbk and (b) Z\_Nsfdbk. Compared with Fig.~\ref{f17}, the fluxes of these pure rotational lines are much lower (by at least a factor of 10) than the overall flux in Z\_sfdbk, and remain two orders of magnitude below the detection limit at $z= 7.2$, indicating that they are not likely to be detected at such high redshifts, unless some extreme events (e.g. starbursts in major mergers) occur. Whereas for Z\_Nsfdbk, the flux of the strongest line 0-0 S(1) reaches $F_{\mathrm{H_{2}}}[\text{0-0 S(1)}]\sim 10^{-23}\ \mathrm{W\ m^{-2}}$ at $z=7.2$, which may be detectable by gravitational lensing or shocks with more than a factor of 10 enhancement. }
\label{f18}
\end{figure*}

For $\mathrm{H_{2}}$ emission, we only take into account the cold molecule-rich component with $[\mathrm{H_{2}/H}]>10^{-5}$ and $T<10^{4}$~K, as hotter gas usually resides in the \HII\ regions, where $\mathrm{H_{2}}$ will be photo-dissociated by photons from nearby stars. Since this local photo-dissociation effect is not explicitly included in our simulations, we indirectly model it by imposing the selection criterion $T<10^{4}$~K for gas to contribute to the H$_2$ emission. Note that this prescription is only important for Z\_sfdbk, where \HII\ regions are generated around newly-born stellar populations (see the $T$-$n$ diagrams in Section~\ref{s3.1}). 

We infer the luminosities of different $\mathrm{H_{2}}$ lines from their respective cooling rates. For transition (line) $k$, the cooling rate per $\mathrm{H_{2}}$ molecule at temperature $T$ and neutral hydrogen (collider) number density $n_{\mathrm{H}}$ can be written in the form
\begin{align}
W_{k}(T,n_{\mathrm{H}})=\frac{g_{k}\exp(-\beta E_{k})}{Z(T)(1+n_{\mathrm{cr},k}/n_{\mathrm{H}})}A_{k}\Delta E_{k}\ ,\label{e24}
\end{align}
where $E_{k}$ and $g_{k}$ are the energy and degeneracy of the upper level, $\Delta E_{k}=hc/\lambda_{k}$ the energy change, $A_{k}$ the Einstein spontaneous emission coefficient, $Z(T)$ the partition function, $n_{\mathrm{cr},k}$ the critical density for transition $k$, and $\beta=1/(k_{B}T)$. As an approximation, we assume that 
\begin{align}
n_{\mathrm{cr},k}\equiv n_{\mathrm{cr},k}(T)=\frac{A_{k}}{\sigma_{T}v_{T}}\ ,\label{e21}
\end{align}
where $\sigma_{T}$ and $v_{T}$ are the average cross section and relative velocity for collision of $\mathrm{H_{2}}$ molecules and neutral hydrogen atoms, which only depend on temperature. The total cooling rate is simply the sum over different lines as $W=\sum_{k}W_{k}$. When the density is low, $n_{\mathrm{H}}\rightarrow 0$, we define
\begin{align}
w(T)=\lim_{n_{\mathrm{H}}\rightarrow 0} \frac{W}{n_{\mathrm{H}}}&=\sum_{k}\frac{g_{k}\exp(-\beta E_{k})}{Z(T)n_{\mathrm{cr},k}}A_{k}\Delta E_{k}\notag\\
&=\sigma_{T}v_{T}\sum_{k}\frac{g_{k}\exp(-\beta E_{k})}{Z(T)}\Delta E_{k}\ ,\label{e22}
\end{align}
where in the last line we have used the assumption in Equ.~(\ref{e21}). Once $w(T)$ is known, one can easily solve for $\sigma_{T}v_{T}$, and obtain  $n_{\mathrm{cr},k}$ for any line $k$ with Equ.~(\ref{e21}). Here, we adopt the fitting formula from \citet{galli1998} (in c.g.s. units) as
\begin{align}
\log w(T)=&-103.0+97.59\log T-48.05(\log T)^{2}\notag\\
&+10.8(\log T)^{3}-0.9032(\log T)^{4}\ .\label{e23}
\end{align}

We apply the above formalism to 42 lines\footnote{The first part of the notation used here for molecular lines denotes the vibrational transition, e.g. `1-0' indicates from $v=1$ to $v=0$, while the second part denotes the rotational quantum number after transition, which is presented in the bracket, and the change of rotational quantum number $\Delta J$. O, P, Q, R and S correspond to $\Delta J=2,\ 1,\ 0,\ -1$ and $-2$, respectively.}, including the pure rotational lines 0-0 S($j$), $j=0,1,...,15$, and the ro-vibrational lines with energies of the upper level $E_{k}/k_{B}\lesssim 2\times 10^{4}$~K, e.g. 1-0 Q(1), 1-0 O(3) and 1-0 O(5). The properties of select lines are summarized in Table~\ref{t1}. We compare the overall cooling rate calculated in this way with the one used in \citet{lithium}, finding that any deviation is within a factor of 1.5, in the temperature range $30\ \mathrm{K}-10^{4}$~K and $n_{\mathrm{H}}\lesssim 100\ \mathrm{cm^{-3}}$. We have assumed optically thin conditions for the $\mathrm{H_{2}}$ emission, as only diffuse gas ($n\lesssim 100\ \mathrm{cm^{-3}}$) is considered here.
Then, for each gas particle $p$, the temperature $T^{p}$, $\mathrm{H_{2}}$ number density $n_{\mathrm{H_{2}}}^{p}$, and that of neutral hydrogen atoms $n_{\mathrm{H}}^{p}$ are known from the simulation, with which the cooling rate for any line $k$ is derived by formulae~(\ref{e24})-(\ref{e23}) as $W^{p}_{k}=W_{k}(T^{p},n_{\mathrm{H}}^{p})$, and the corresponding luminosity is $L^{p}_{k}=W_{k}^{p}n_{\mathrm{H_{2}}}^{p}V_{p}$, where $V_{p}=m_{p}/\rho_{p}$ is the physical volume associated with particle $p$. Finally, for each line, we sum up the luminosities from all gas particles to obtain the total luminosity and flux. The overall $\mathrm{H_{2}}$ luminosity (flux) is just the summation of those from the 42 lines. We also calculate the overall $\mathrm{HD}$ emission with similar methods, and find that $L_{\mathrm{H_{2}}}\sim 10^{3}-10^{4} L_{\mathrm{HD}}$. 

Fig.~\ref{f17} shows the evolution of overall $\mathrm{H_{2}}$ flux in Z\_sfdbk, where we also include the contribution from collapsing star-forming cores in newly-created stellar particles, whose overall $\mathrm{H_{2}}$ luminosity is estimated with a star formation efficiency $\epsilon=0.05$, a typical core mass $M_{\star}=10\ M_{\odot}$, a duration of the core collapse $\Delta t_{\mathrm{core}}=0.1$~Myr, and a time-averaged luminosity $L_{\mathrm{core}}=5\times 10^{33}\ \mathrm{erg\ s^{-1}}$ per core \citep{mizusawa2005}. The overall $\mathrm{H_{2}}$ flux in the WDM cosmology is lower than that in the CDM case by at least one order of magnitude for $z\gtrsim 12$, but reaches almost the same level at late stages ($z\lesssim 8$). 

Fig.~\ref{f18} shows the evolution of $\mathrm{H_{2}}$ fluxes for the pure rotational lines 0-0 S(1), 0-0 S(3) and 0-0 S(5), from diffuse gas, in (a) the Z\_sfdbk, and (b) the Z\_Nsfdbk models. With stellar feedback included (Z\_sfdbk), 0-0 S(1) contributes 4.3\% (2.1\%) of the overall flux, 0-0 S(3) 7.5\% (4.7\%), and 0-0 S(5) 6.7\% (5.2\%), at redshift $z=7.7$, in the WDM (CDM) cosmology. For the Z\_Nsfdbk case, 0-0 S(1) contributes 25.6\% (24.0\%) of the overall flux, 0-0 S(3) 16.8\% (15.1\%), and 0-0 S(5) 5.6\% (6.1\%), for the WDM (CDM) cosmology. As shown in Table~\ref{t1}, the distribution of $\mathrm{H_{2}}$ luminosity among different lines in Z\_Nsfdbk implies that most of the luminosity originates from the gas in the temperature range $T\sim 1000-2000\ \mathrm{K}$, which represents the typical condition before runaway collapse (see Fig.~\ref{pd}). For Z\_sfdbk, on the other hand, the overall luminosity shows a significant contribution from $\mathrm{H_{2}}$ molecules at higher temperatures ($T\sim 2000-5000\ \mathrm{K}$). This indicates that the state of $\mathrm{H_{2}}$ is regulated by the stellar UV and LW radiation fields. In the presence of this radiation, the $\mathrm{H_{2}}$ molecules become generally hotter, implying higher cooling rates, but meanwhile the amount of molecule-rich gas will be reduced due to photo- and collisional dissociation in \HII\ regions. In our simulations these two competing effects roughly cancel each other out, in terms of the overall $\mathrm{H_{2}}$ emission. However, with stellar feedback, a larger fraction of radiation energy is carried by higher-frequency (mid-IR) photons from ro-vibrational transitions, so that the far-IR (FIR) emission from (low-energy) pure rotational lines 0-0 S(1), 0-0 S(3) and 0-0 S(5) will be reduced.


\begin{figure}
\hspace{-10pt}
\includegraphics[width=1.05\columnwidth]{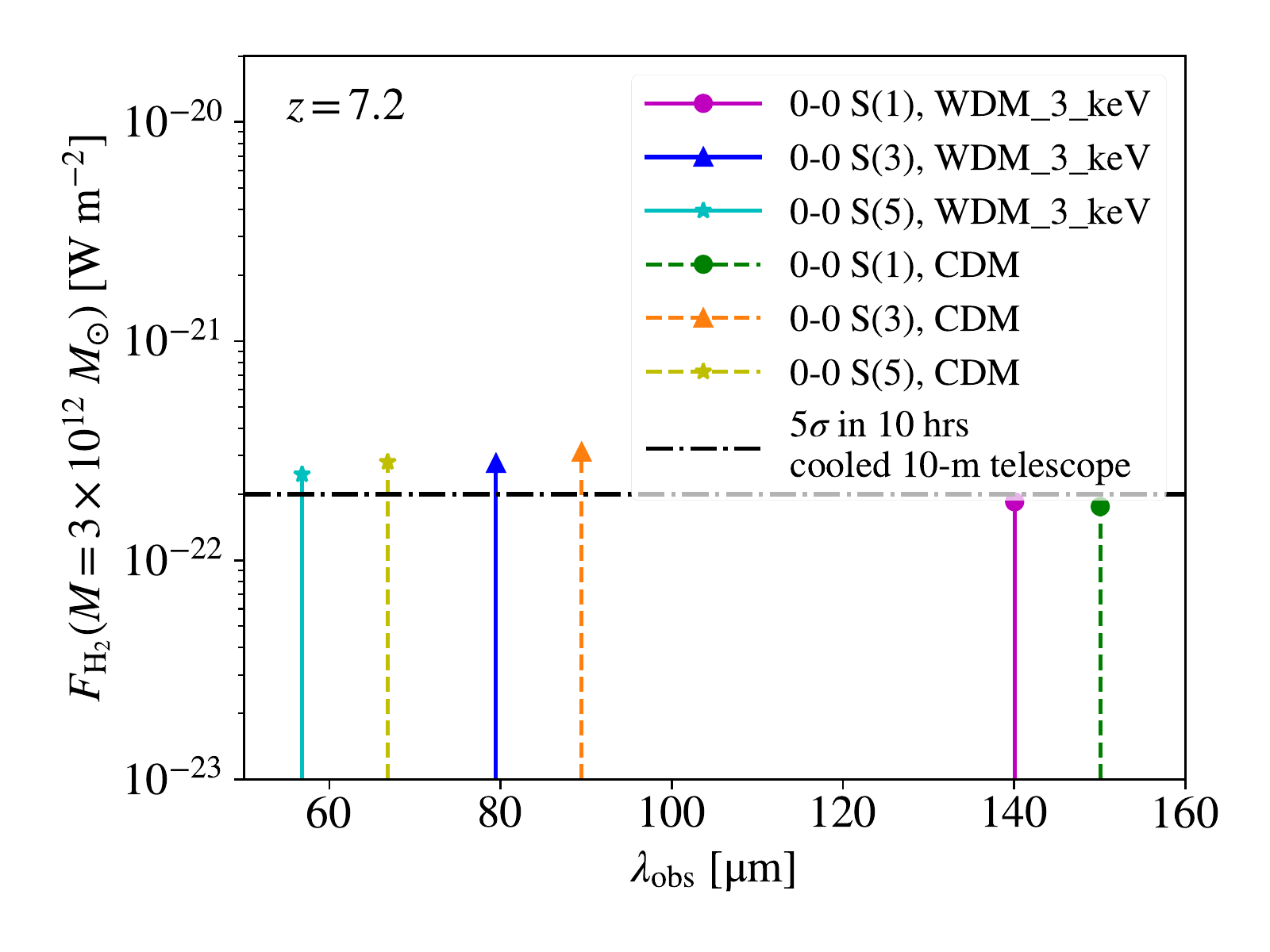}
\vspace{-20pt}
\caption{Extrapolated fluxes of the pure rotational lines 0-0 S(1) (solid circles), 0-0 S(3) (triangles) and 0-0 S(5) (stars) at $z=7.2$, in Z\_sfdbk, for a halo mass of $3\times 10^{12}\ M_{\odot}$. The dashed-dotted line shows the projected 5$\sigma$ 10-h sensitivity for future 10m-class, cooled telescopes, where $F_{\mathrm{th}}\simeq 2\times 10^{-22}\ \mathrm{W\ m^{-2}}$. For illustration, the wavelengths of lines in the CDM cosmology are shifted by $10\ \mathrm{\mu m}$. In both the WDM (solid) and CDM (dashed) models, these lines are marginally detectable in FIR bands.}
\label{FH2_extra}
\end{figure}

\begin{figure}
\hspace{-10pt}
\includegraphics[width=1.05\columnwidth]{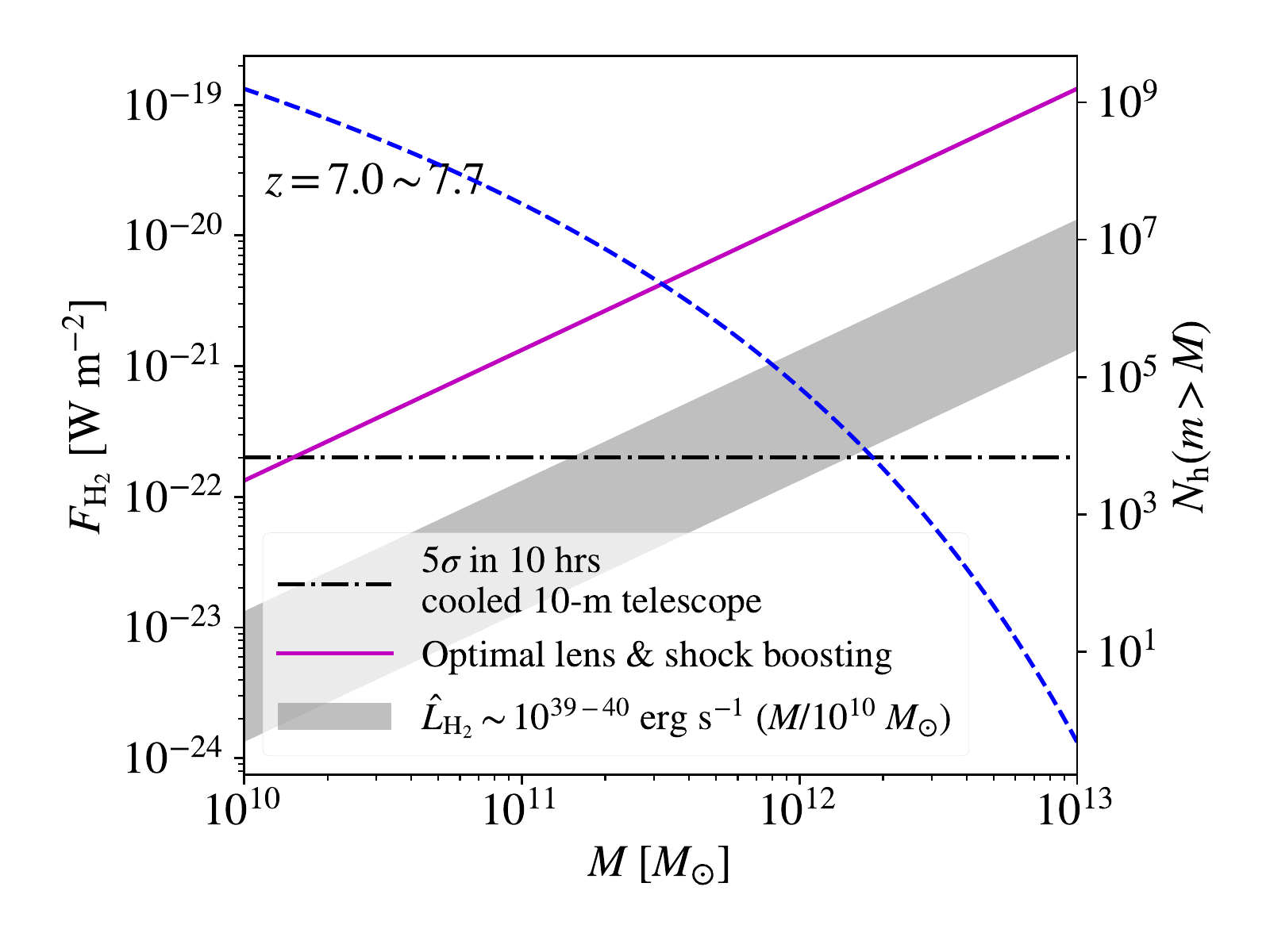}
\vspace{-20pt}
\caption{Flux estimates for the strongest $\mathrm{H_{2}}$ line at $z\sim 7$ as a function of halo mass $M$ (left axis, shaded region), and number of (star-forming) haloes with masses above $M$ and ages less than 100~Myr at $z\sim 7$, formed over the entire sky (right axis, dashed curve). The dashed-dotted line shows again the expected 5$\sigma$ 10-h sensitivity for a future 10m-class FIR telescope in space. 
An upper limit for the flux is also plotted, with boosting from lensing or shocks by a factor of 10 (right axis, solid curve). }
\label{FH2_M}
\end{figure}

To summarize, for Z\_sfdbk, in both the WDM and CDM models, the overall $\mathrm{H_{2}}$ flux is $F_{\mathrm{H_{2}}}\sim 10^{-23}\ \mathrm{W\ m^{-2}}$ at redshifts $7.2\lesssim  z\lesssim 8$, below the expected 5$\sigma$ 10-h sensitivity for projected 10m-class, cooled FIR telescopes of $F_{\mathrm{th}}\simeq 10^{-22}\ \mathrm{W\ m^{-2}}$, such as the OST\footnote{\url{https://asd.gsfc.nasa.gov/firs/science/history.html}}. 
The signal from Z\_Nsfdbk without stellar feedback is larger by a factor of 3, but still below the detection limit. 
In the Z\_sfdbk simulation, the fluxes for the pure rotational lines 0-0 S(1), 0-0 S(3) and 0-0 S(5) are much lower (by at least a factor of 10) than the overall flux, and remain two orders of magnitude below the detection limit at $z= 7.2$, indicating that their detection at such high redshifts will be challenging, requiring some unusual event, such as a starburst in a major merger. For Z\_Nsfdbk, on the other hand, the flux of the strongest line 0-0 S(1) reaches $F_{\mathrm{H_{2}}}[\text{0-0 S(1)}]\sim 10^{-23}\ \mathrm{W\ m^{-2}}$ at $z=7.2$, which is possible to detect with a factor of 10 enhancement from (strong) gravitational lensing \citep{appleton2009} or shocks, as observed in Stephan's Quintet \citep{appleton2017}. Note that the simulated haloes with $M\sim 10^{10}\ M_{\odot}$ are not the most massive ones at $z\sim 7$. There are more massive haloes with correspondingly stronger signals. Assuming that the properties of star forming clouds are roughly independent of halo mass, we extrapolate the $\mathrm{H_{2}}$ flux to higher halo masses by simply assuming $F_{\mathrm{H_{2}}}\propto M$. We find that the pure rotational lines 0-0 S(1), 0-0 S(3) and 0-0 S(5) become marginally detectable for $M=3\times 10^{12}\ M_{\odot}$ at $z=7.2$, even for Z\_sfdbk, as shown in Fig.~\ref{FH2_extra}. 

As mentioned in Section~\ref{s3.1}, the Pop~II feedback model used in our simulations may over-predict the effect of ionization. As a result, the strength of $\mathrm{H}_{2}$ pure rotational lines may be underestimated in Z\_sfdbk\footnote{We have rerun the simulations under the same condition with a modified P2L model of adaptive ionization radii and weaker feedback, from which we find that the overall flux of $\mathrm{H_{2}}$ is enhanced by one order of magnitude, while the fluxes of pure rotation lines 0-0 S(3) and 0-0 S(5) are increased by up to a factor of 5, compared with the results shown here for the P2L model in \citet{jaacks2018legacy}. }. While stellar feedback certainly will destroy and heat up the $\mathrm{H}_{2}$ molecules in \HII\ regions, lowering the luminosity of pure rotational lines that comes mainly from cold/warm gas, so that Z\_Nsfdbk may overestimate the luminosity of pure rotational lines. Generally speaking, we can regard the results from Z\_sfdbk and those from Z\_Nsfdbk as lower bounds and upper bounds, respectively. In light of this, we estimate the luminosity of the strongest $\mathrm{H_{2}}$ line from massive DM haloes as $\hat{L}_{\mathrm{H_{2}}}\sim 10^{39-40}\ \mathrm{erg\ s^{-1}}\ (M/10^{10}\ M_{\odot})$, based on our simulations. We plot the corresponding flux at $z\sim 7$ as a function of halo mass in Fig.~\ref{FH2_M}, together with the number of (star-forming) haloes, with masses above a given threshold and ages less than 100~Myr, corresponding to the $z=7-7.7$ range\footnote{This is the timescale in which $\mathrm{H_{2}}$ emission is maintained at a high level in our simulations, as shown in Fig.~\ref{f17}.}. In Fig.~\ref{FH2_M}, we also show an upper flux limit, assuming boosting from lensing or shocks by a factor of 10. 
As can be seen, even with the most conservative assumptions, massive haloes with $M\gtrsim 10^{12}\ M_{\odot}$ are detectable without boosts at $z\sim 7$, and there are $\sim 10^{5}$ such massive haloes with ages less than 100~Myr over the entire sky. Therefore, a few massive star-forming haloes are expected to be detected via $\mathrm{H_{2}}$ emission in a survey area of a few square degrees. Furthermore, with lens and shock boosting, haloes with $M\gtrsim 10^{10}\ M_{\odot}$ become detectable at $z\sim 7$, and there are roughly $10^{9}$ such sources in the entire sky, indicating that observation of such lower-mass haloes is also feasible. 
However, at higher redshifts $z\gtrsim 12$, even with lensing and shock boosting, there are only $\lesssim 10^{4}$ (star-forming) haloes that are massive enough ($M\gtrsim 6\times 10^{10}\ M_{\odot}$) to be detectable, making observation of their $\mathrm{H_{2}}$ lines challenging.

\begin{table*}
\centering
\caption{Properties of select $\mathrm{H_{2}}$ lines. The first section (a) lists the physical properties of these lines, including the wavelength $\lambda=hc/\Delta E_{k}$, Einstein spontaneous emission coefficient $A_{k}$, energy and degree of degeneracy of the upper level $E_{k}$ and $g_{k}$ (from \url{http://www.astronomy.ohio-state.edu/~depoy/research/observing/molhyd.htm}). In section (b), fractions of the luminosity from these lines are summarized for different temperatures at $n_{\mathrm{H}}=100\ \mathrm{cm^{-3}}$, followed by the line strengths implied by our simulations at $z=7.7$. }
\begin{tabular}{cccccccccc}
\hline
(a) Physical properties & 0-0 S(0) & 0-0 S(1) & 0-0 S(2) & 0-0 S(3) & 0-0 S(4) & 0-0 S(5) & 1-0 Q(1) & 1-0 O(3) & 1-0 O(5)\\
\hline
$\lambda\ [\mu\mathrm{m}]$ & 28.221 & 17.035, & 12.279 & 9.6649 & 8.0258 & 6.9091 & 2.4066 & 2.8025 & 3.235\\
$A_{k}\ [\mathrm{s^{-1}}]$ & 2.94e-11 & 4.76e-10 & 2.76e-9 & 9.84e-9 & 2.64e-8 & 5.88e-8 & 4.29e-7 & 4.23e-7 & 2.09e-7\\
$E_{k}/k_{B}\ [\mathrm{K}]$ & 510 & 1015 & 1682 & 2504 & 3474 & 4586 & 6149 & 6149 & 6956\\
$g_{k}$ & 5 & 21 & 9 & 33 & 13 & 45 & 9 & 9 & 21\\
\hline
(b) Temperature/DM model & 0-0 S(0) & 0-0 S(1) & 0-0 S(2) & 0-0 S(3) & 0-0 S(4) & 0-0 S(5) & 1-0 Q(1) & 1-0 O(3) & 1-0 O(5)\\
\hline
500 K & 0.1012 & 0.6685 & 0.1148 & 0.1048 & 0.0072 & 0.0031 & $\sim 10^{-5}$ & $\sim 10^{-5}$ & $\sim 10^{-5}$\\
1000 K & 0.0087 & 0.2903 & 0.1424 & 0.3208 & 0.0591 & 0.0789 & 0.0095 & 0.0082 & 0.0074\\
1500 K & 0.0014 & 0.0804 & 0.0734 & 0.2499 & 0.0664 & 0.1302 & 0.0268 & 0.0230 & 0.0271\\
2000 K & 0.0003 & 0.0244 & 0.0329 & 0.1505 & 0.0498 & 0.1201 & 0.0326 & 0.0280 & 0.0376\\
3500 K & $10^{-5}$ & 0.0020 & 0.0047 & 0.0389 & 0.0201 & 0.0687 & 0.0286 & 0.0246 & 0.0386\\
5000 K & $10^{-6}$ & 0.0005 & 0.0013 & 0.0146 & 0.0101 & 0.0433 & 0.0237 & 0.0204 & 0.0334\\

\hline
WDM\_3\_keV (Z\_Nsfdbk) & 0.0171 & 0.2562 & 0.0902 & 0.1677 & 0.0313 & 0.0561 & 0.0145 & 0.0124 & 0.0176\\
CDM (Z\_Nsfdbk) & 0.0267 & 0.2402 & 0.0746 & 0.1514 & 0.0320 & 0.0614 & 0.0159 & 0.0137 & 0.0192\\
\hline
WDM\_3\_keV (Z\_sfdbk) & 0.0052 & 0.0425 & 0.0213 & 0.0749 & 0.0247 & 0.0665 & 0.0232 & 0.0200 & 0.0312\\
CDM (Z\_sfdbk) & 0.0019 & 0.0211 & 0.0122 & 0.0471 & 0.0173 & 0.0520 & 0.0216 & 0.0186 & 0.0302\\
\hline
\end{tabular}
\label{t1}
\end{table*}


\begin{figure}
\hspace{-10pt}
\includegraphics[width= 1.05\columnwidth]{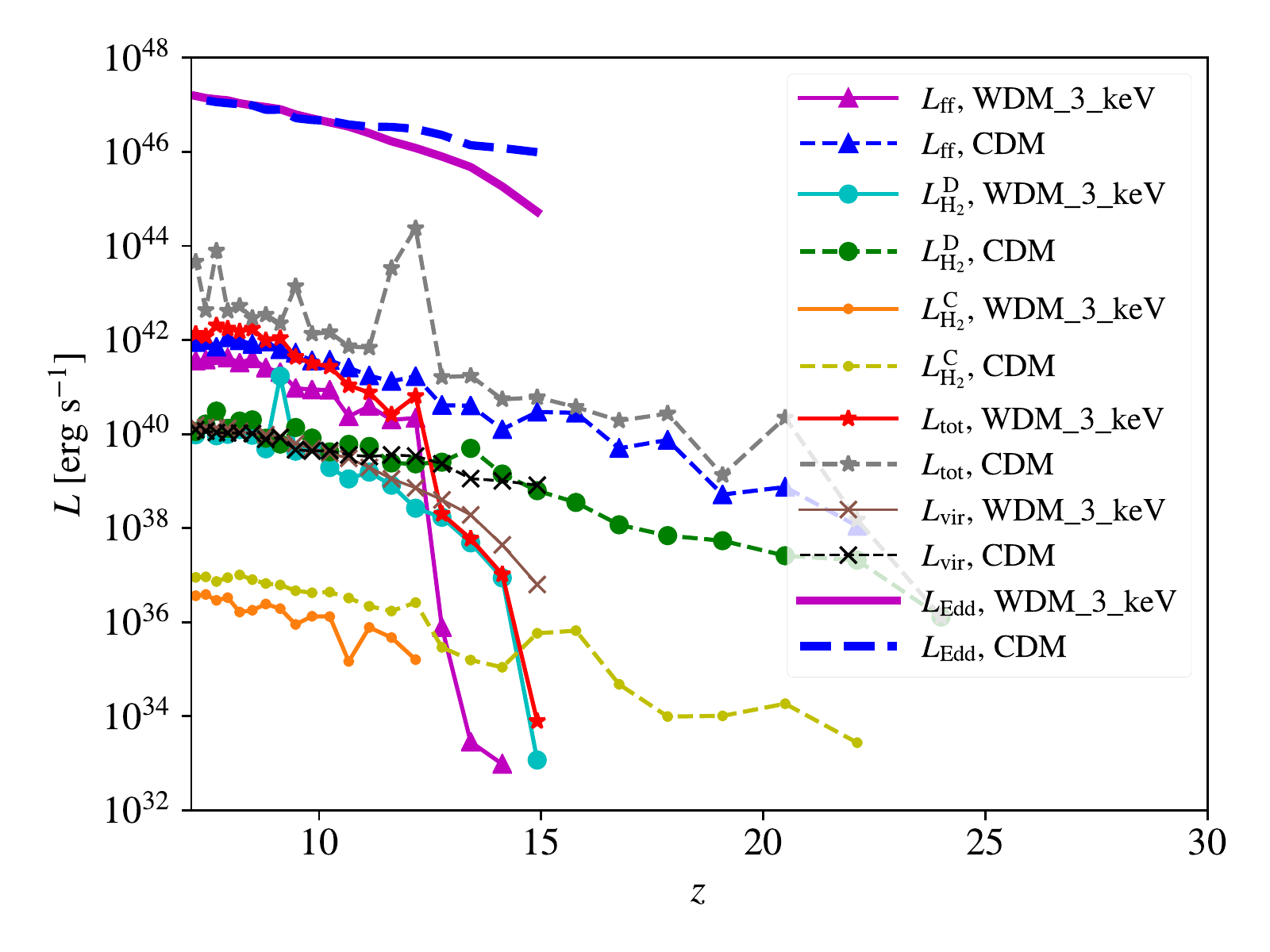}
\vspace{-10pt}
\caption{Luminosities from the target halo as functions of redshift, in the WDM (solid) and CDM (dashed) models, from Z\_sfdbk. The total luminosity $L_{\mathrm{tot}}$ is marked with stars, the $\mathrm{H_{2}}$ luminosity from diffuse gas $L_{\mathrm{H_{2}}}^{\mathrm{D}}$ with squares, the $\mathrm{H_{2}}$ luminosity from proto-stellar cores $L_{\mathrm{H_{2}}}^{\mathrm{C}}$ with dots, the free-free luminosity $L_{\mathrm{ff}}$ with triangles, and the virial luminosity $L_{\mathrm{vir}}$ with crosses. The Eddington limit $L_{\mathrm{Edd}}$ is also plotted to provide a strong upper limit (thick curves).}
\label{f16}
\end{figure}

Finally, in Fig.~\ref{f16} we show the general radiation signature of the target halo (assuming optically thin conditions), in terms of total (integrated) luminosity $L_{\mathrm{tot}}$, overall $\mathrm{H_{2}}$ luminosity from diffuse gas and protostellar cores, $L_{\mathrm{H_{2}}}^{\mathrm{D}}$ and $L_{\mathrm{H_{2}}}^{\mathrm{C}}$, free-free luminosity $L_{\mathrm{ff}}$, virial luminosity $L_{\mathrm{vir}}$ from Equ.~(\ref{e3}), and the Eddington luminosity as an extreme upper limit
\begin{align}
L_{\mathrm{Edd}}=3.2\times 10^{4}\left(\frac{M_{b}}{M_{\odot}}\right)L_{\odot}\ ,
\end{align}
where $M_{b}=(\Omega_{b}/\Omega_{m})M_{\mathrm{{vir}}}$ is the total baryon mass in the target halo. The luminosity from star-forming cores is negligible (lower by $3-4$ orders of magnitude), compared with that from diffuse gas, so that the total $\mathrm{H_{2}}$ luminosity $L_{\mathrm{H_{2}}}\approx L_{\mathrm{H_{2}}}^{\mathrm{D}}$. We find that $\mathrm{H_{2}}$ emission contributes only 1\% of the total luminosity at $z\lesssim 12$ (17) in WDM (CDM) cosmology, although it dominates in the early era before the first star formation event. The luminosity of free-free emission counts for about 10\% of the total luminosity in late stages ($z\lesssim 12$), when most radiation comes from atomic hydrogen emission. 
The total luminosity is far below ($\sim 10^{-5}$) the Eddington limit, but above the virial luminosity by a factor of 10, for $z\lesssim 12$, showing that radiation is mostly powered by stellar feedback. 

\section{Summary and Discussion}
We have carried out a series of cosmological hydrodynamic zoom-in simulations of DM haloes with virial masses $M\sim 10^{10}\ M_{\odot}$, for two DM models, a standard CDM cosmology and a WDM model with a particle mass of $m_{\chi}c^{2}=3$~keV. We investigate how the nature of DM correlates with high-redshift ($z\gtrsim 7$) structure formation and the corresponding radiation signature. 
The simulations include primordial chemistry and cooling, and adopt two idealized schemes for star formation, a simple one without stellar feedback (Z\_Nsfdbk), and another one with the model for Pop~III and Pop~II star formation and feedback from \citet{jaacks2018legacy,jaacks2018baseline} (Z\_sfdbk). Free-free and $\mathrm{H_{2}}$ $(\mathrm{HD}$) emissions are calculated by post-processing radiative transfer. We summarize the main findings for Z\_sfdbk below, which is the physically more realistic case. 

We find different early structure formation histories in the two DM models, consistent with the trend found in previous work \citep[e.g.][]{yoshida2003simulations,gao2007lighting,dayal2017reionization,hirano2017first,10.1093/mnras/stz766}:
\begin{itemize}
\item The initial Pop~III star formation event is delayed by $\sim 200$~Myr in the WDM cosmology, compared to the CDM case. However, Pop~III star formation in the WDM cosmology quickly reaches the same level as for CDM, once it occurs at $z\sim 12$.
\item Metal enrichment and Pop~II star formation are also delayed by $\sim 200$~Myr in the WDM cosmology. The difference between the (mass-weighted) average metallicities in the two DM models decreases at lower redshifts ($z\lesssim 10$), but still does not vanish by $z\sim 7.2$.
\item Significant metal enrichment ($Z>Z_{\mathrm{crit}}=10^{-4}\ Z_{\odot}$) tends to be restricted in dense environments in the WDM model, affecting a small volume, while in the CDM model, metals can also break out of low-mass haloes to enrich a large volume of low-density gas.
\end{itemize}
These results can be interpreted as follows. For early structure formation before reionization, when neutral gas is abundant, metal enrichment and Pop~II star formation tend to enhance each other. Higher metallicities lead to more efficient cooling, thus facilitating Pop~II star formation, while an abundance of newly-formed massive stars will enrich the ISM more strongly with metals, when they die in SN events. This establishes a positive feedback cycle, turned on initially by Pop~III star formation. Since formation of small-scale structures is suppressed in the WDM cosmology, the initial Pop~III stellar population is formed at lower redshifts, in more massive DM haloes. The onset of this feedback cycle is thus also delayed. However, when initiated in the more massive structures encountered in the WDM cosmology, star formation and metal enrichment are more efficient in the more vigorous gravitational collapse, such that the difference between CDM and WDM models in terms of mass-weighted average metallicity will be reduced when the DM halo evolves to lower redshifts. Besides, massive structures only occupy a small volume and can more easily confine the metals generated within, while in the CDM model, star formation and the relevant feedback in small-scale structures can enrich a large volume of gas with lower densities.

For the corresponding free-free and $\mathrm{H_{2}}$ emissions originating from high-$z$ structure formation, we identify these trends: 
\begin{itemize}
\item The free-free signal, derived from our simulations of early structure formation at $z>6$, is $3_{-1.5}^{+13}$\% ($8_{-3.5}^{+33}$\%) of the free-free component in the cosmic radio background (CRB), measured by radio experiments such as ARCADE 2 \citep{arcade}, in the WDM (CDM) model. 
\item The overall $\mathrm{H_{2}}$ flux from individual DM haloes with virial masses $M\sim 10^{10}\ M_{\odot}$ is typically below the detection limit of the next generation of FIR space telescopes, $F_{\mathrm{th}}\sim 10^{-22}\ \mathrm{W\ m^{-2}}$, for both DM models at $z\gtrsim 7.2$. Direct detection of the $\mathrm{H_{2}}$ emission, especially for individual lines, from non-lensed high-redshift galaxies is only possible for more massive haloes with $M\gtrsim 10^{12}\ M_{\odot}$. With further boosting by a factor of 10 from gravitational lensing or shocks, $M\sim 10^{10}\ M_{\odot}$ haloes may also be observable.
\end{itemize}
Note that the free-free component in the CRB, inferred from observational data, is highly uncertain (with a relative uncertainty of at least 80\%). 
Moreover, our theoretical calculations are based on a single simulated halo, with multiple approximations and assumptions, such as the redshift when Pop~II star formation terminates ($z_{\mathrm{end}}=6$), and how the free-free luminosity and timescale correlate with halo virial mass and formation redshift. These uncertainties imply that our results should be considered as first explorations, to be followed-up with improved, more complete studies. Nevertheless, extrapolation of our calculations to lower redshifts produces total CRB signals quite close to (within 0.7$\sigma$) the free-free component inferred from observation. 
Specifically, we find that early structure formation only contributes a small fraction (<10\%) of the total free-free component in the CRB. The difference between the two DM models is not significant (similar within a factor of 2.5) in terms of the global (background) free-free signal, comparable to the halo-halo scatter of free-free luminosity. We expect the difference in the global free-free signal between the two models to be even smaller, when a statistically representative halo sample is used for such calculations. The reason is that massive (atomic cooling) haloes ($M\gtrsim 10^{8}\ M_{\odot}$) contribute the majority ($>95\%$) of the emission, and the abundance of these haloes is almost the same in the two models. For instance, in an extreme case where the free-free luminosity model $L_{\nu}(z,M)$ is identical in the WDM and CDM models, the difference in the global free-free signal is only $\sim 10$\% for $z_{\mathrm{end}}=6$, and drops to $\sim 1$\% for $z_{\mathrm{end}}\sim 0$. 
Therefore, the global free-free signal may not be a good diagnostic of the underlying DM model.

Actually, synchrotron emission dominates the CRB in the low-frequency range ($\nu_{\mathrm{obs}}\lesssim 700\ \mathrm{MHz}$), which cannot be fully understood in terms of known radio sources \citep{singal2010}. It would be interesting to study the synchrotron emission from early structure formation, to assess its contribution to the CRB. Our simulations are not suitable for such an investigation, as they do not include the relevant relativistic magneto-hydrodynamics (MHD). 

For determining the $\mathrm{H}_{2}$ emission, our feedback model is likely to be too aggressive, thus underestimating the signals of the FIR pure rotational lines. Note that the haloes simulated here, with virial masses $\sim 10^{10}\ \mathrm{M_{\odot}}$, are rather low-mass and  abundant at $z\lesssim 10$. It may be possible to observe their $\mathrm{H}_{2}$ emission in FIR bands with gravitational lensing, extending what has already been done in radio bands for molecular lines from lensed, dusty star forming galaxies \citep{spilker2018fast}. Furthermore, we estimate the luminosity of the strongest $\mathrm{H_{2}}$ line from more massive DM haloes ($M\gtrsim 10^{10}\ M_{\odot}$) as $\hat{L}_{\mathrm{H_{2}}}\sim 10^{39-40}\ \mathrm{erg\ s^{-1}}\ (M/10^{10}\ M_{\odot})$ by linear extrapolation from the signals of the $M\sim 10^{10}\ M_{\odot}$ haloes simulated here. This implies that the more massive haloes, with $M\gtrsim 10^{12}\ M_{\odot}$, are detectable even without lensing at $z\sim 7$, and there are $\sim 10^{5}$ such massive haloes with ages less than 100~Myr over the entire sky. In general, it is possible to observe a few massive star-forming haloes via $\mathrm{H_{2}}$ emission per square degree with the next generation of FIR space telescopes for $z\lesssim 10$. With the boosting provided by lensing or shocks, haloes with $M\gtrsim 10^{10}\ M_{\odot}$ may also be detectable at $z\sim 7$, and there are roughly $10^{9}$ such haloes over the entire sky, suggesting that FIR observation of lower-mass haloes is feasible, as well. 

This work presents an exploratory step in the direction of probing the nature of DM with the first stars and galaxies in early structure formation. Hopefully, more precise and complete future radio surveys can significantly reduce the uncertainty in the measurement of the free-free component in the CRB. On the theory side, it is important to run similar simulations for different DM cosmologies, such as models with non-gravitational DM-baryon interactions. We need larger simulation boxes, while still maintaining high resolution, with more realistic feedback models (e.g. \citealt{wise2014birth,sarmento2018following} for CDM cosmology). 
It is also interesting to study the correlations among the radio ($\mathrm{H_{2}}$) luminosity and other properties of the DM halo, such as mass, concentration and formation time, given a large sample of haloes. 
The confluence of ever more realistic simulations with the upcoming suite of frontier observations promises unprecedented insight into the formative era of star and galaxy formation, in the process providing us with novel hints on the elusive nature of dark matter.


\section*{Acknowledgements}
This work was supported by National Science Foundation (NSF) grant AST-1413501. The authors acknowledge the Texas Advanced Computing Center (TACC) for providing HPC resources under XSEDE allocation TG-AST120024.

\label{s5}

\bibliographystyle{mnras}
\bibliography{ref} 

\label{lastpage}
\end{document}